\documentclass[journal, onecolumn]{IEEEtran} %

\usepackage{ragged2e}
\usepackage{amsmath}
\usepackage{array}
\usepackage{textcomp}
\usepackage{stfloats}
\usepackage{verbatim}
\usepackage{graphicx}

\usepackage{psfrag,epsfig,subfigure}
\usepackage{enumerate}
\usepackage{amsfonts,amssymb,bm,paralist,theorem,cite,ifthen,color}
\usepackage{listings}

\usepackage{algorithm}
\usepackage{algpseudocode}
\usepackage{makecell}
\usepackage{dsfont}
\usepackage{url}

\usepackage{xcolor}
\usepackage[subject={Top1},author={YanmengWang},version=1]{pdfcomment}
\hypersetup{hidelinks}

\usepackage{threeparttable}
\usepackage{multirow}
\usepackage{pifont}
\usepackage{subcaption}

\usepackage{caption}
\usepackage[font=small,labelfont=small]{caption}
\renewcommand{\thetable}{\arabic{table}}

\usepackage[acronym]{glossaries}
\makeglossaries
\newacronym{fl}{FL}{} 
\newacronym{lora}{LoRA}{} 
\newacronym{fft}{FFT}{Federated Fine-Tuning}
\newacronym{rhs}{RHS}{right-hand side}
\newacronym{lhs}{LHS}{left-hand side}
\newacronym{dnn}{DNN}{Deep Neural Network}
\newacronym{ap}{AP}{Access Point}
\newacronym{gn}{GN}{Group Normalization}
\newacronym{fdma}{FDMA}{Frequency Division Multiple Access}
\newacronym{psd}{PSD}{Power Spectrum Density}
\newacronym{fspl}{FSPL}{Free Space Path Loss}
\newacronym{los}{LOS}{Line of Sight}
\newacronym{nlos}{NLOS}{Non-Line of Sight}
\newacronym{gan}{GAN}{Generative Adversarial Network}
\glsdisablehyper

\newtheorem{theorem}{Theorem}
\newtheorem{lemma}{Lemma}
\newtheorem{proposition}{Proposition}
\newtheorem{corollary}{Corollary}
\newtheorem{remark}{Remark}
\newtheorem{assumption}{Assumption}

\makeatletter
\renewcommand{\maketag@@@}[1]{\hbox{\m@th\normalsize\normalfont#1}}
\renewcommand{\fnum@algorithm}{\small Algorithm~\thealgorithm}
\makeatother

\allowdisplaybreaks[4]

\begin{document}

\title{Robust Federated Fine-Tuning in Heterogeneous Networks with Unreliable Connections: \\
An Aggregation View}

\author{
Yanmeng~Wang,~\IEEEmembership{Member,~IEEE,}
Zhiwen~Dai,
Shuai~Wang,~\IEEEmembership{Member,~IEEE,}
Jian~Zhou,~\IEEEmembership{Member,~IEEE,} \\
Fu~Xiao,~\IEEEmembership{Senior Member,~IEEE,}
Tony~Q.~S.~Quek,~\IEEEmembership{Fellow,~IEEE,}
Tsung-Hui~Chang,~\IEEEmembership{Fellow,~IEEE}
\thanks{
Y. Wang, Z. Dai, J. Zhou, and F. Xiao are with the School of Computer Science, Nanjing University of Posts and Telecommunications, Nanjing 210023, China
(e-mail: hiwangym@gmail.com, daizhiwen177@foxmail.com, zhoujian@njupt.edu.cn, xiaof@njupt.edu.cn).
(Corresponding author: Fu Xiao)
}
\thanks{
S. Wang are with the National Key Laboratory of Wireless Communications, University of Electronic Science and Technology of China, Chengdu 611731, China
(e-mail: shuaiwang@uestc.edu.cn).
}
\thanks{
Tony Q. S. Quek is with the Singapore University of Technology and Design, Singapore 487372, and also with the Department of Electronic Engineering, Kyung Hee University, Yongin 17104, South Korea (e-mail: tonyquek@sutd.edu.sg).
}
\thanks{
T.-H. Chang is with the School of Artificial Intelligence, The Chinese University of Hong Kong, Shenzhen 518172, China
(e-mail: tsunghui.chang@ieee.org).
}
\thanks{
\textit{(Preprint. Under review.)}
}
}



\maketitle

\begin{abstract}
Federated Fine-Tuning (FFT) has attracted growing interest as it leverages both server- and client-side data to enhance global model generalization while preserving privacy, and significantly reduces the computational burden on edge devices by avoiding training from scratch.
Despite these advantages, FFT performance is often degraded by unreliable server-client connections and heterogeneous client data distributions.
Most existing methods assume homogeneous network conditions or require prior knowledge of connection failures.
However, these assumptions are impractical in real-world networks characterized by diverse communication standards (e.g., wired, Wi-Fi, 4G, and 5G) and heterogeneous failure patterns.
To address these limitations, we propose \texttt{FedAuto}, a novel FFT framework that mitigates the combined effects of connection failures and data heterogeneity via adaptive aggregation.
\texttt{FedAuto} operates without prior knowledge of network conditions or modifications to existing infrastructure, enabling seamless plug-and-play deployment.
Moreover, we establish a rigorous convergence guarantee for \texttt{FedAuto}.
By adopting a novel per-round aggregation perspective, our analysis removes the need for assumptions on connection failures probabilities or client selection strategies commonly imposed in prior work, and guarantees convergence of \texttt{FedAuto} for each individual realization, providing a stronger theoretical assurance.
Extensive experiments demonstrate that \texttt{FedAuto} consistently outperforms state-of-the-art baselines under diverse connection failure scenarios for both full-parameter and partial-parameter fine-tuning (e.g., LoRA), and even surpasses strategies that rely on complex communication resource optimization.

\end{abstract}


\section{Introduction}\label{sec:introduction}

\IEEEPARstart{F}{ederated} Learning (FL) has emerged as a powerful paradigm for collaboratively training \gls{dnn} across distributed edge clients while preserving data privacy.
Owing to its privacy-preserving nature, FL has been widely adopted in data-sensitive domains such as healthcare, finance, and edge intelligence~\cite{wang2023batch, wang2023beyond, wang2022federated}.
In conventional \gls{fl} frameworks, such as \texttt{FedAvg}~\cite{mcmahan2017communication}, the server is responsible only for aggregating client updates, while all training is performed at resource-constrained edge devices.
This design imposes substantial computational and communication burdens on clients and fails to leverage the abundant computation and storage resources available at the server.

To better exploit the data and computational resources across both server and client tiers, \gls{fft} has recently attracted growing attention~\cite{zhang2024towards, kuang2024federatedscope}. 
\gls{fft} typically follows a two-stage paradigm: 
a global model is first pre-trained on large-scale public datasets at a resource-rich server, and is then collaboratively fine-tuned across distributed clients using their private data~\cite{zhang2023fedpetuning}.
During the fine-tuning stage, only model updates are exchanged between the server and clients, enabling the incorporation of private knowledge while preserving data privacy \cite{zhang2022fine}.
Benefiting from server-side pre-training, \gls{fft} significantly reduces both the computational and communication overhead on edge clients compared with conventional \gls{fl}.
Moreover, by integrating domain-specific private data, \gls{fft} improves the generalization capability of pre-trained models beyond what can be achieved using public data alone~\cite{sutskever2024seq2seq}.

\begin{figure}[t]
\centering
\includegraphics[width= 3.5 in ]{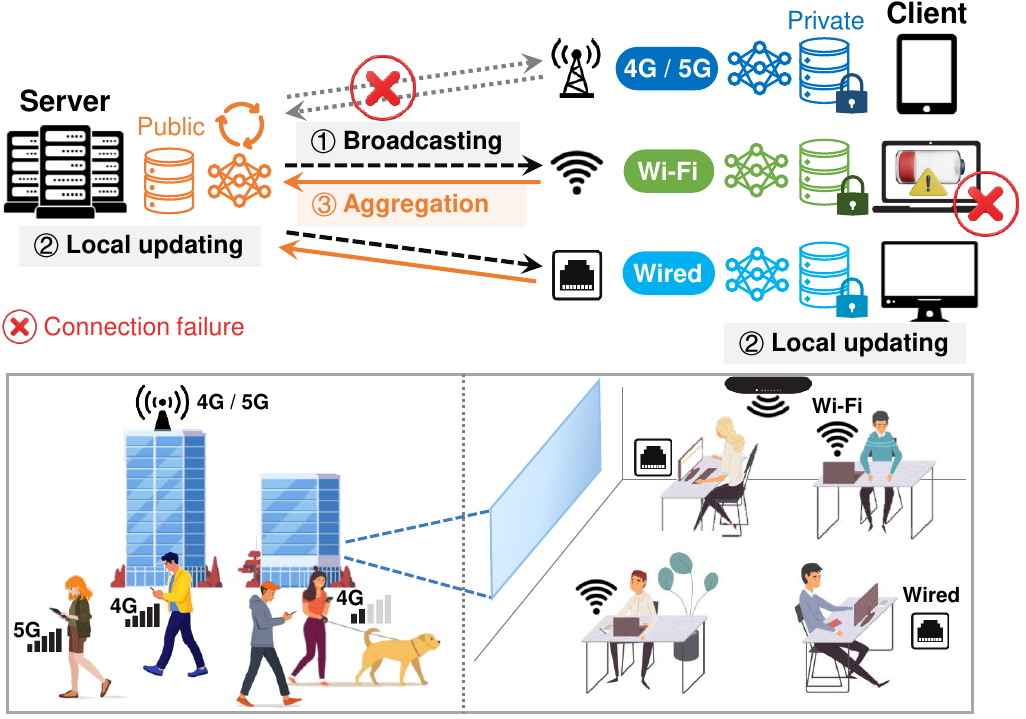}
\caption{
\gls{fft} over heterogeneous and unreliable networks.
}
\label{fig:FL unreliable networks}
\end{figure}

Despite these advantages, deploying \gls{fft} in real-world heterogeneous networks remains challenging.
First, \gls{fft} inherits the long-standing issue of \textit{data heterogeneity}, which arises not only between public and private datasets but also across private datasets held by different clients.
Second, \textit{communication heterogeneity}, a fundamental yet underexplored challenge, further complicates \gls{fft}.
In practice, as illustrated in Fig.~\ref{fig:FL unreliable networks}, clients may connect to the server via diverse communication infrastructures (e.g., wired, Wi-Fi, 4G, and 5G), resulting in heterogeneous link capacities and reliability.
Connection disruptions can occur unpredictably due to device-side factors (e.g., battery depletion or hardware failures) or channel-side conditions (e.g., weak coverage and interference)~\cite{zheng2023federated, yang2020power, {xiang2023towards}}. 
Consequently, clients experience highly heterogeneous disconnection patterns in both frequency and duration.
When combined with data heterogeneity, such communication unreliability can severely bias the global optimization process, as clients with more stable connections may disproportionately influence model aggregation.
This bias degrades convergence and generalization performance of \gls{fft}, and the empirical evidence supporting this can be found in Section~\ref{sec:aggregation unreliable networks}.

Existing approaches for mitigating data heterogeneity primarily focus on conventional \gls{fl} settings, where all training data reside exclusively on edge clients \cite{wang2020tackling, mendieta2022local, dai2023tackling}.
As such methods do not account for the mixed public-private data structure inherent in \gls{fft}, directly applying them may underutilize \gls{fft}'s unique data-distribution characteristics.
Meanwhile, most studies addressing communication unreliability in \gls{fl} either assume homogeneous network standards or rely on accurate prior knowledge of connection failure probabilities.
Physical-layer communication resource optimization approaches, such as transmit power and bandwidth allocation \cite{chen2021joint, wang2022quantized, mahmoud2023federated, zheng2023federated}, 
are often impractical in heterogeneous commercial networks spanning diverse communication standards.
In contrast, client selection strategies \cite{ribero2022federated, salehi2021federated, wang2025robust} heavily  depend on accurate estimation of time-varying failure probabilities, which is difficult to achieve in practice.
These limitations highlight the need for an \gls{fft} framework that is robust to both data heterogeneity and unpredictable communication unreliability, while remaining easy to deploy over existing heterogeneous commercial networks.

\subsection{Related work}

Since the distributed fine-tuning stage of \gls{fft} closely resembles conventional \gls{fl}, consisting of distributed training and global aggregation, we review related work in both areas.

\subsubsection{Data Heterogeneity}

Data heterogeneity is a long-standing challenge in \gls{fl} and has been extensively studied \cite{li2020federated, karimireddy2020scaffold, wang2023beyond, wang2020tackling, mendieta2022local, dai2023tackling}.
Representative methods include \texttt{FedProx}~\cite{li2020federated}, which introduces a proximal regularization term to stabilize local updates, and \texttt{SCAFFOLD}~\cite{karimireddy2020scaffold}, which mitigates client drift using control variates.
However, these methods are designed for conventional \gls{fl} and assume that all training data reside on edge clients, making them ill-suited for \gls{fft}'s mixed public-private data setting.
Recent \gls{fft}-specific methods, such as \texttt{FedIT}\cite{zhang2024towards}, \texttt{FLORA}\cite{wang2024flora}, and \texttt{FedEx-LoRA}\cite{singhal2024fedex}, consider data heterogeneity during aggregation of fine-tuned parameters.
Nevertheless, they still overlook the role of server-side public data in the overall \gls{fft} process and assume ideal, lossless communication between the server and clients. 
These assumptions limit their applicability in practical unreliable network environments.

\subsubsection{Connection Unreliability}

To mitigate the effects of communication unreliability in \gls{fl}, one line of work focuses on optimizing physical-layer resources such as transmit power and bandwidth.
For example,
\cite{chen2021joint} optimizes transmit power and bandwidth allocation in \gls{fdma} systems, which \cite{wang2022quantized}, \cite{mahmoud2023federated}, and \cite{wang2025federated} jointly optimize transmit power and bandwidth together with quantization bit allocation, communication latency, and model adaptation, respectively.
Although effective, these approaches require centralized control over communication resources, which is often infeasible in heterogeneous commercial networks with diverse communication standards and device-specific configurations.

Another line of research addresses unreliability through client selection without modifying network infrastructure.
\texttt{F3AST}\cite{ribero2022federated} balances long-term client participation by assuming that clients' successful connection probabilities follow a homogeneous Markov chain, while \cite{salehi2021federated} and \texttt{FedCote}\cite{wang2025robust} explicitly incorporate connection failure probabilities into the optimization of client selection policies.
However, these methods rely on accurate estimation of connection failure probabilities, which is challenging in practice due to time-varying channels and diverse failure modes. 
More recently, \texttt{FedAWE}~\cite{xiang2024efficient} removes this requirement but adjusts local step sizes  based on connection failure counts, rendering it ineffective under prolonged disconnections.

\subsubsection{Personalized \gls{fl}}

Beyond \gls{fft}, personalized \gls{fl} aims to improve each client’s individual model by adapting to its local data distribution, rather than learning a single global model \cite{tan2022towards, wang2023toward}.
Similar to \gls{fft}, communication reliability also plays a critical role in personalized \gls{fl}, as clients with more stable connections may dominate the aggregation of globally shared parameters.
Despite its practical importance, robustness to heterogeneous and unreliable communication remains underexplored in personalized \gls{fl}.

\subsection{Contributions}

This paper aims to enhance the robustness of \gls{fft} in heterogeneous and unreliable network environments that reflect real-world commercial communication systems, characterized by diverse network standards and unpredictable connection failures.
To this end, we propose \texttt{FedAuto}, a novel \gls{fft} framework that mitigates the joint effects of data heterogeneity and connection unreliability through adaptive aggregation alone, without requiring network infrastructure modification or prior knowledge of failure probabilities.

Our main contributions are summarized as follows:

\begin{enumerate}[1)]
\item 
\textbf{\gls{fft} under heterogeneous and unreliable networks}:
We systematically study \gls{fft} under realistic network conditions with heterogeneous communication standards and diverse connection failure patterns.
To the best of our knowledge, this is the first work to jointly consider these challenges in \gls{fft}.

\item 
\textbf{\texttt{FedAuto}}:
We propose \texttt{FedAuto}, an adaptive aggregation strategy that alleviates the adverse effects of both data heterogeneity and connection unreliability.
\texttt{FedAuto} operates solely at the aggregation level and enables plug-and-play deployment over existing commercial networks.

\item
\textbf{Convergence guarantee}:
We theoretically characterize the joint effects of aggregation weights, data heterogeneity, and connection failures on \gls{fft} convergence.
To the best of our knowledge, this is the first analysis to explicitly reveal the impact of aggregation weights under unreliable networks.
Moreover, by adopting a novel per-round aggregation perspective, our analysis eliminates the need for assumptions on connection failures probabilities or client selection strategies that are commonly imposed in conventional analyses.
Consequently, it guarantees convergence of the proposed \texttt{FedAuto} for each individual realization, providing a stronger theoretical assurance.

\item 
\textbf{Extensive experiments}:
We conduct comprehensive experiments on fine-tuning pre-trained models of varying scales (CNNs, ResNets, and ViTs) across multiple datasets and diverse connection failure scenarios.
The results demonstrate that \texttt{FedAuto} consistently outperforms existing baselines and even surpasses communication resource allocation-based methods without any network-level intervention.

\end{enumerate}

\textbf{Synopsis:}
The remainder of this paper is organized as follows.
Section~\ref{sec:system model} introduces the \gls{fft} framework under unreliable network connections.
Section~\ref{Sec:Adaptive Aggregation} presents the proposed adaptive aggregation strategy, \texttt{FedAuto}.
Section~\ref{sec:convergence analysis} provides the theoretical convergence analysis.
Section~\ref{sec:experimental results} reports the experimental results.
Finally, Section~\ref{section:Conclusion} concludes the paper and discusses future research directions.

\section{Federated fine-tuning under unreliable connections}\label{sec:system model}

\subsection{Learning Objective for \gls{fft}}\label{sec:Distributed Learning Public Private}

We consider a practical distributed learning setting, illustrated in Fig. \ref{fig:FL unreliable networks}, in which an edge server maintains a public dataset and each client holds a private dataset.
The server’s public dataset typically provides broad class coverage but contains only a limited number of samples per class.
In contrast, each client’s private dataset is domain-specific, with narrower class coverage but substantially higher sample density for the categories it contains.
Such a heterogeneous data distribution is common in real-world deployments.
Public datasets are often constructed from diverse online sources and thus lack domain specificity.
Meanwhile, clients---such as hospitals or banks---possess rich, well-curated data that are restricted to specific domains (e.g., X-ray images for a particular disease or financial documents of a specific type), but these data cannot be shared due to privacy or regulatory constraints.

To leverage both the general knowledge contained in the server-side public dataset and the domain-specific information in the client-side private datasets, the server collaborates with $N$ clients to solve the following global learning objective:
\begin{equation}\label{eq:distributed objective function}
\min_{{\mathbf{w}}} 
\,\,
F_g({\mathbf{w}}; \mathcal{D}_g)
=
p_s F_s({\mathbf{w}}; \mathcal{D}_s)
+
\sum_{i = 1}^N p_i F_i({\mathbf{w}}; \mathcal{D}_i)
,
\end{equation}
where $\mathcal{D}_s$ is the server's public dataset, $\mathcal{D}_i$ is the private dataset of client $i \in [N]$, and $\mathcal{D}_g \triangleq \mathcal{D}_s \bigcup (\cup_{i = 1}^N \mathcal{D}_i)$ denotes the global dataset combining both public and private data.
$\mathbf{w}$ represents the model parameters to be learned.
The weight coefficients $p_s = {|\mathcal{D}_s|}/{|\mathcal{D}_g|}$ and $p_i = |\mathcal{D}_i|/|\mathcal{D}_g|$ are proportional to dataset sizes and satisfy $p_s + \sum_{i=1}^N p_i = 1$.
$F_g$ denotes the global cost function, while $F_s$ and $F_i$ represent the local ones for the server and client~$i$, respectively.

Directly solving \eqref{eq:distributed objective function} through centralized learning would require all clients to upload their private datasets to the server, which is infeasible in privacy-sensitive settings.
\gls{fft} addresses this limitation by enabling collaborative training that integrates knowledge from both public and private datasets without accessing or transmitting any raw client data.
The detailed \gls{fft} procedure is described in the subsequent Section \ref{sec:FL procedure}.

\subsection{\gls{fft} Procedure}\label{sec:FL procedure}

To integrate the general knowledge in server-side public dataset with the domain-specific information in clients’ private datasets, while preserving privacy and reducing both computation and communication overhead on resource-limited clients, \gls{fft} adopts a two-stage framework: the server first pre-trains the model on the public dataset, followed by distributed fine-tuning on the clients’ private datasets.

\subsubsection{Server-Side Pre-Training}

The server first pre-trains a global model on its public dataset $\mathcal{D}_s$ using centralized learning.
In practice, the model can be trained from scratch solely on $\mathcal{D}_s$, or obtained by further fine-tuning an existing pre-trained model (e.g., adapting a ViT model pre-trained on ImageNet dataset to the CIFAR-100 classification task \cite{bafghi2024parameter}).
The resulting model, denoted ${\bar {\mathbf{w}}}_{\rm pre}$, provides a strong initialization for subsequent distributed fine-tuning, substantially reducing clients' computational and communication costs by avoiding training from scratch.

\subsubsection{Distributed Fine-Tuning}\label{sec:distributed training phase}

The distributed fine-tuning stage follows the standard \gls{fl} paradigm.
It further enhances the generalization ability of the pre-trained model ${\bar {\mathbf{w}}}_{\rm pre}$ by integrating knowledge from both public and private datasets.
During fine-tuning, the model parameters updated may include the full model (i.e., full-parameter fine-tuning) or only a subset (e.g., LoRA-based fine-tuning \cite{hu2022lora}).
As illustrated in Fig. \ref{fig:FL unreliable networks}, each communication round $r$ consists of the following steps:
\begin{enumerate}[(a)]
\item
\textbf{Broadcasting}:\label{sec:111}
The server selects a subset of clients ${\mathcal{K}}_{r} \subseteq [N]$ with $|{\mathcal{K}}_{r}| = K$ and broadcasts the current global model ${\bar {\mathbf{w}}}_{r-1}$ to them, where the initial model is set as ${\bar {\mathbf{w}}}_{0} = {\bar {\mathbf{w}}}_{\rm pre}$.

\item
\textbf{Local updating}:
Each selected client $i \in {\mathcal{K}}_{r}$ initializes its local model as $\mathbf{w}^{r,0}_{i} = {\bar {\mathbf{w}}}_{r-1}$ and performs $E$ steps of gradient descent\footnote{
When using mini-batch stochastic gradient descent (SGD), the full gradient $\nabla F_{i}(\mathbf{w}; \mathcal{D}_i)$ in (\ref{eq:local updating client}) is replaced by $\nabla F_{i}(\mathbf{w}; {\bm \xi}^{r}_{i})$, where ${\bm \xi}^{r}_{i}$ denotes a mini-batch sample and the corresponding loss is $F_{i}(\mathbf{w}; {\bm \xi}^{r}_{i}) = \mathbb{E}_{\xi \in {\bm \xi}^{r}_{i}}[\mathcal{L}(\mathbf{w}; \xi)]$.
For clarity, the theoretical analysis is based on full gradients, whereas the experiments employ mini-batch SGD.
} 
on its private dataset $\mathcal{D}_i$:
\begin{equation}\label{eq:local updating client}
\mathbf{w}^{r,t}_{i} = {\mathbf{w}}^{r,t-1}_{i}  -  \gamma \nabla F_{i}( {\mathbf{w}}^{r,t-1}_{i}; \mathcal{D}_i),
\;
t \in [E],
\end{equation}
where $\gamma>0$ is the learning rate.
In parallel, the server updates its own local model on the public dataset $\mathcal{D}_s$ to mitigate forgetting of public-domain knowledge:
\begin{equation}\label{eq:local updating server}
\mathbf{w}^{r,t}_{s} = {\mathbf{w}}^{r,t-1}_{s}  -  \gamma \nabla F_{s}( {\mathbf{w}}^{r,t-1}_{s}; \mathcal{D}^r_s),
\;
t \in [E],
\end{equation}
where the server-side local model is likewise initialized from the previous global model, i.e., $\mathbf{w}^{r,0}_{s} = {\bar {\mathbf{w}}}_{r-1}$. 

\item
\textbf{Global aggregation}:
The server updates the global model by aggregating its own local model together with those received from the selected clients:
\begin{equation}\label{eq:global aggregation}
{\bar {\mathbf{w}}}_{r} 
= 
\beta_s^r \mathbf{w}^{r,E}_{s} 
+ \sum_{i \in {\mathcal{K}}_{r}} \beta_i^r \mathbf{w}^{r,E}_{i},
\end{equation}
where $\beta_s^r$ and $\beta_i^r$ are the aggregation weights for the server and each selected client $i \in \mathcal{K}_r$.
These weights are nonnegative and satisfy the normalization condition $\beta_s^r + \sum_{i \in {\mathcal{K}}_{r}} \beta_i^r = 1$.
In standard \gls{fl} schemes such as \texttt{FedAvg}, these weights are assigned uniformly or in proportion to the coefficients $p_s$ and $p_i$ in \eqref{eq:distributed objective function}.
Importantly, during aggregation, only model parameters are transmitted from clients to the server, with no raw data exchanged, thereby preserving data privacy.
\end{enumerate}

\begin{remark}\label{remark: aggregation weights full partial}
\rm
Based on existing convergence analyses~\cite{liconvergence} and \cite{wang2025robust}, it is straightforward to show that under full client participation ($K=N$), the aggregation scheme in \eqref{eq:global aggregation} achieves optimality when $\beta_s^r = p_s$ and $\beta_i^r = p_i$ $\forall i \in [N]$, consistent with the distributed objective \eqref{eq:distributed objective function}.
Under partial participation ($K<N$), assigning uniform weights to selected clients, such as $\beta_s^r = p_s$ and $\beta_i^r = (1-p_s)/K$ $\forall i \in \mathcal{K}_r$, yields an unbiased estimate of the full-participation update:
$\mathbb{E}_{\mathcal{K}_r}[{\bar {\mathbf{w}}}_{r}] = p_s \mathbf{w}^{r,E}_{s} + \sum_{i=1}^N p_i \mathbf{w}^{r,E}_{i}$.
A detailed proof is provided in Appendix~\ref{appendix:simple averaging partial participation}.
\end{remark}

\subsection{Aggregation under Unreliable Connections}\label{sec:aggregation unreliable networks}

Although the \gls{fft} framework describe above effectively exploits both the server's public data and the clients' private data to improve global model generalization, it implicitly assumes reliable communication.
This assumption is often unrealistic, as real-world networks (Fig.~\ref{fig:FL unreliable networks}) frequently experience unstable connections that disrupt server-client communication and prevent timely model exchange.
To capture such connection failures, the aggregation scheme in \eqref{eq:global aggregation} is revised as
\begin{subequations}\label{global_model_failure}
\begin{align}
{\bar {\mathbf{w}}}_{r}
= &
\beta_s^r \mathbf{w}^{r,E}_{s} 
+
\sum_{{i} \in {\mathcal{K}}_{r}} \mathds{1}^{r}_{i} \beta^{r}_{i} \mathbf{w}^{r,E}_{i}
\label{global_model_failure aggregation scheme}
, 
\\
\text{s.t.} \quad
&
\beta_s^r 
+ 
\sum_{{i} \in {\mathcal{K}}_{r}} \mathds{1}^{r}_{i} \beta^{r}_{i} = 1,
\label{eq:aggregation weights sum}
\end{align}
\end{subequations}
where ${\mathds{1}^{r}_{i}}=1$ denotes a successful server-client transmission and ${\mathds{1}^{r}_{i}}=0$ denotes a failure.
By incorporating the revised aggregation rule \eqref{eq:global aggregation}, the complete \gls{fft} procedure under unreliable connections is summarized in Algorithm~\ref{algorithm:FL under unreliable networks}.

\begin{algorithm}[t]
\small
\caption{\small
\gls{fft} procedure under unreliable connections}
\begin{algorithmic}[1]
\Statex \textbf{\texttt{// Stage 1: Pre-training:}}
\State Server pre-trains the global model on the public dataset;
\Statex \textbf{\texttt{// Stage 2: Distributed fine-tuning:}}
\For{$r=1,2,\cdots,R$}
\Statex $\quad\,$ \textbf{\texttt{// (2-1) Broadcasting:}}
\State Server broadcasts global model ${\bar {\mathbf{w}}}_{r-1}$ to selected clients;
\Statex $\quad\,$ \textbf{\texttt{// (2-2) Local updating:}}
\For {each selected client ${i} \in \mathcal{K}_r$ and the server} \textbf{(in parallel)}
\State Update local model via (\ref{eq:local updating client}) or (\ref{eq:local updating server});
\EndFor
\Statex $\quad\,$ \textbf{\texttt{// (2-3) Global Aggregation:}}
\State {Server aggregates a new global model ${\bar {\mathbf{w}}}_{r}$ by (\ref{global_model_failure});}
\EndFor
\end{algorithmic}
\label{algorithm:FL under unreliable networks}
\end{algorithm}

\subsection{Network Unreliability May Lead to Significant Bias}

Similar to  Remark \ref{remark: aggregation weights full partial}, the aggregation weights for the global aggregation scheme \eqref{global_model_failure} under unreliable networks can be heuristically assigned either uniformly under partial participation or proportionally to the coefficients $p_s$ and $p_i$ in \eqref{eq:distributed objective function} under full participation\footnote{\label{footnote:aggregation weights FedAvg}
Based on Remark \ref{remark: aggregation weights full partial} and the normalization condition (\ref{eq:aggregation weights sum}), the aggregation weights in (\ref{global_model_failure aggregation scheme}) can be heuristically set as $\beta_s^r = \frac{p_s}{p_s + \sum_{{j} \in [N]} \mathds{1}^{r}_{j} p_{j}}$ for the server and $\beta_i^r = \frac{p_i}{p_s + \sum_{{j} \in [N]} \mathds{1}^{r}_{j} p_{j}}$ for each successfully connected client (with $\mathds{1}^{r}_{i}=1$) under full participation, or as $\beta_s^r = p_s$ and $\beta_i^r = \frac{1-p_s}{\sum_{{j} \in \mathcal{K}_r} \mathds{1}^{r}_{j}}$ under partial participation.
}.
However, these heuristic assignments can significantly degrade \gls{fft} performance, especially when both connection failures and non-i.i.d. data coexist.

\begin{figure}[t]
\begin{minipage}[h]{1\linewidth}
\centering
\includegraphics[width= 3.4 in ]{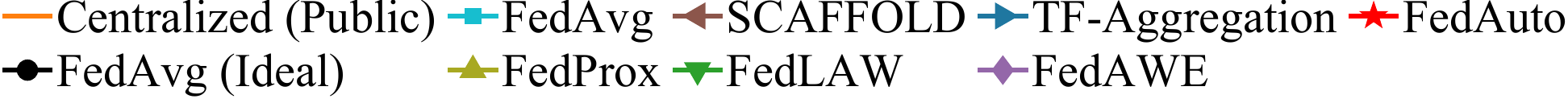}
\end{minipage}
\begin{minipage}[h]{1\linewidth}
\centering
\subfigure[i.i.d.\ data.]{
\includegraphics[width= 1.6 in ]{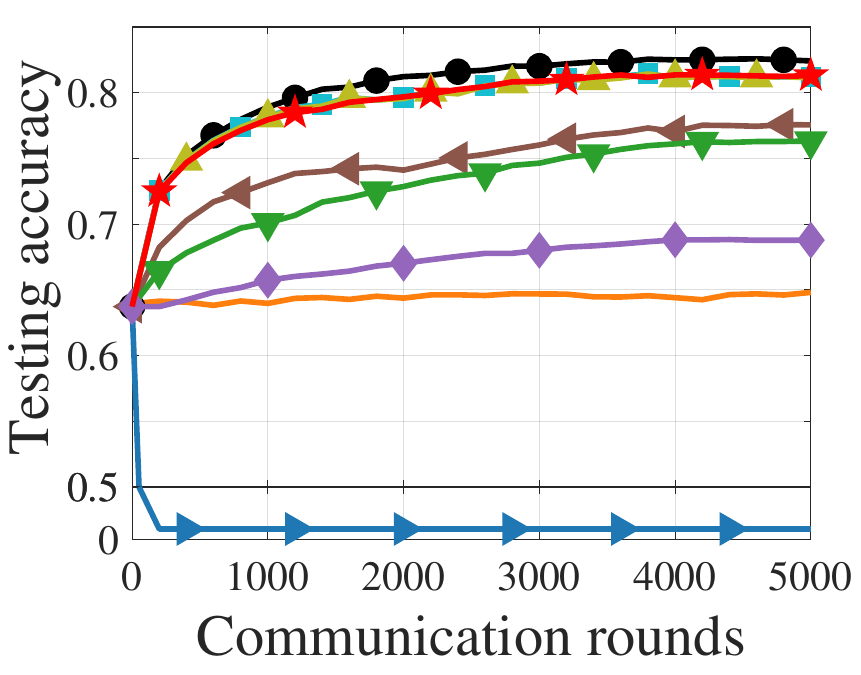} \label{subfig:FFT performance CIFAR10 mixed iid}}
\subfigure[Non-i.i.d.\ data.]{
\includegraphics[width= 1.6 in ]{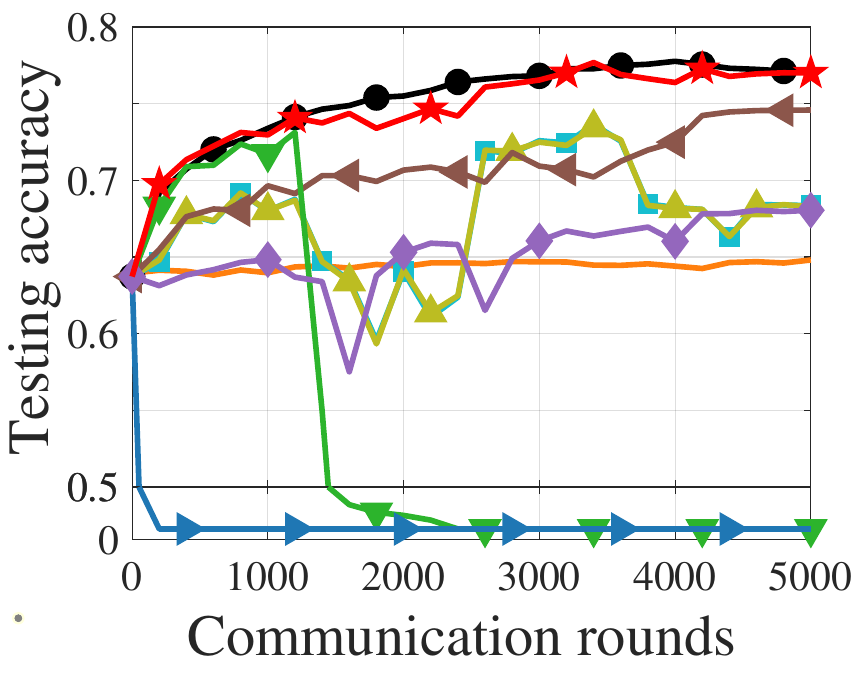} \label{subfig:FFT performance CIFAR10 mixed noniid}}
\caption{
\gls{fft} performance under heterogeneous networks with unreliable connections (CIFAR-10, mixed connection failures, $K=20$).}
\label{fig:FFT performance CIFAR10 mixed}
\end{minipage}
\end{figure}

Fig. \ref{fig:FFT performance CIFAR10 mixed} illustrates this degradation by comparing CIFAR-10 testing accuracy across various \gls{fl} strategies during distributed fine-tuning under unreliable networks, for both i.i.d. and non-i.i.d. data distributions.
The experiment considers a practical heterogeneous network environment, where clients connect to the server via diverse commercial network standards (wired, Wi-Fi, 4G, and 5G), with connection failures including both transient and intermittent disruptions;
refer to Section \ref{sec:Parameter Setting} for more detailed experimental settings.

As shown, most \gls{fl} strategies improve global generalization by utilizing private datasets, outperforming \texttt{Centralized(Pu-} \texttt{blic)} (orange curve), which relies solely on the server-side public data.
However, both the classical \texttt{FedAvg} and the advanced \texttt{FedProx}~\cite{li2020federated}, which rely on the aforementioned heuristic aggregation weights, degrade significantly under non-i.i.d.\ data.
Similarly, methods for mitigating data heterogeneity (e.g., \texttt{SCAFFOLD}~\cite{karimireddy2020scaffold}), specialized aggregation schemes such as \texttt{FedLAW}~\cite{li2023revisiting}, and approaches for handling connection failures (e.g., \texttt{TF-Aggregation}~\cite{salehi2021federated} and \texttt{FedAWE}~\cite{xiang2024efficient}), all exhibit unstable convergence or suboptimal performance, even under i.i.d.\ data.
These results highlight the need to address the combined effects of connection unreliability and data heterogeneity on \gls{fft} performance.

To address this issue, Section~\ref{Sec:Adaptive Aggregation} introduces \texttt{FedAuto}, a novel \gls{fft} framework with adaptive aggregation that mitigates both effects without requiring prior knowledge of connection failures or modifications to existing network infrastructure.
Section~\ref{sec:convergence analysis} further provides rigorous convergence guarantees, showing that \texttt{FedAuto} achieves robust convergence under heterogeneous and unreliable network conditions.
To the best of our knowledge, this is the first work to address unreliable connection issues in \gls{fft} within the context of practical heterogeneous networks, involving multiple network standards and diverse connection failure patterns.

\section{Adaptive Aggregation Strategy}\label{Sec:Adaptive Aggregation}

In global aggregation \eqref{global_model_failure}, regardless of network standards or connection failure types, the server integrates its own updated local model only with those from successfully connected clients.
However, stable \gls{fft} convergence requires balanced contributions across data classes \cite{wang2025robust, li2023revisiting}, which can be disrupted under unreliable connections.
In unreliable networks, clients with poor connections may suffer higher transmission failure probabilities, leading to an underrepresentation of their domain-specific knowledge in global aggregation.
This becomes particularly problematic when certain classes, primarily owned by clients with unstable connections, are lost, resulting in convergence bias and diminished \gls{fft} performance.
To address this issue, we propose \texttt{FedAuto}, a novel \gls{fft} scheme with adaptive aggregation that mitigates the impact of both connection unreliability and data heterogeneity.
The core idea is to use only the server's available resources (i.e, its public dataset and received local models) to balance class contributions and ensure robust convergence, without requiring modifications to existing network infrastructure.
This adaptive aggregation strategy consists of two core modules.

\subsection{Module 1: Server-Side Compensatory Training}

In unreliable networks, some class-specific local updates from clients may be missing due to connection failures.
Although the server's local model, $\mathbf{w}^{r,E}_{s}$, is trained via \eqref{eq:local updating server} on the public dataset $\mathcal{D}_s$, which covers a broad range of classes, its uniform contribution across classes is insufficient to specifically compensate for the missing ones in the received clients' updates.
Consequently, these classes remain underrepresented during global aggregation in \eqref{global_model_failure}.
To mitigate this, we propose a compensatory training mechanism that leverages the server’s public dataset to enhance the contributions of these missing classes in global aggregation.

Let $\mathcal{C}^{r}_{\text{miss}}$ denote the classes missing in received local models from clients in round $r$.
If $\mathcal{C}^{r}_{\text{miss}} \neq \emptyset$, the server trains a compensatory model ${\mathbf{w}}^{r,t}_{\text{miss}}$ using a data subset $\mathcal{D}^{r}_{\text{miss}} \subset \mathcal{D}_s$ that contains only samples from the missing classes.
Following the update rule in \eqref{eq:local updating server}, the compensatory model is initialized as ${\mathbf{w}}^{r,0}_{\text{miss}} = {\bar {\mathbf{w}}}_{r-1}$ and updated by $E$-step gradient descents:
\begin{equation}\label{eq:local updating server compensatory}
{\mathbf{w}}^{r,t}_{\text{miss}} = {\mathbf{w}}^{r,t-1}_{\text{miss}}  -  \gamma \nabla F_{s}( {\mathbf{w}}^{r,t-1}_{\text{miss}}; \mathcal{D}^{r}_{\text{miss}}),
\;
t \in [E].
\end{equation}
This procedure is triggered only when missing classes are detected, thus incurring negligible additional cost at the resource-rich server.

With this compensatory mechanism, the global aggregation scheme \eqref{global_model_failure} is refined as
\begin{equation}\label{global_model_failure_2}
{\bar {\mathbf{w}}}_{r}
=
\beta_s^r \mathbf{w}^{r,E}_{s} 
+ \beta^r_{\text{miss}} {\mathbf{w}}^{r,t}_{\text{miss}}
+ \sum_{{i} \in {\mathcal{K}}_{r}} \mathds{1}^{r}_{i} \beta^{r}_{i} \mathbf{w}^{r,E}_{i}
, 
\end{equation}
where the aggregation weight of compensatory model, $\beta^r_{\text{miss}}$, is set to zero if no classes are missing (i.e., $\mathcal{C}^{r}_{\text{miss}} = \emptyset$).

\subsection{Module 2: Aggregation Weight Optimization}\label{subsection:Module 2}

After compensating for missing classes, the next step is to ensure that each class's contribution to global aggregation aligns with its proportion in the global dataset $\mathcal{D}_g \triangleq \mathcal{D}_s \bigcup (\cup_{i = 1}^N \mathcal{D}_i)$, as defined in the learning objective \eqref{eq:distributed objective function}.
This will be formally justified by the convergence analysis of \gls{fft} under unreliable networks (Theorem \ref{theorem:convergence FL failure} and Corollary \ref{corollary:convergence FL failure}) presented in Section \ref{sec:convergence analysis}.
Based on the refined aggregation scheme \eqref{global_model_failure_2}, after collecting the updated local models ($\mathbf{w}^{r,E}_{s}$ and $\mathbf{w}^{r,E}_{i}$) and the compensatory model ${\mathbf{w}}^{r,t}_{\text{miss}}$ (if present), the server can adjust only the aggregation weights $\beta^r_s$, $\beta^r_{\text{miss}}$, and $\beta^r_i$.
Therefore, to balance each class's contribution to global aggregation, we propose optimizing these weights.

Let $\alpha_{g,c}$ denote the proportion of class-$c$ samples in the global dataset $\mathcal{D}_g$.
Similarly, let $\alpha_{s,c}$, $\alpha^{r}_{\text{miss},c}$ and $\alpha_{i,c}$ represent the class-$c$ proportions in the public dataset $\mathcal{D}_s$, the compensatory dataset $\mathcal{D}^{r}_{\text{miss}}$, and the private dataset $\mathcal{D}_i$, respectively\footnote{
From the definition of the global dataset $\mathcal{D}_g$ in the learning objective (\ref{eq:distributed objective function}), its class distribution is the weighted average of the local class distributions, i.e., $\alpha_{g,c} = p_s \alpha_{s,c} + \sum_{i=1}^N p_i \alpha_{i,c}$.
}.
We aim to ensure that each class’s effective contribution to the global aggregation aligns with its proportion in $\mathcal{D}_g$. This leads to the following minimization problem:
\begin{subequations}\label{eq:aggregation_optimization_2}
\begin{align}
\min_{\beta_s^r, \beta^r_{\text{miss}}, \atop \{ \beta^{r}_{i} \}_{ \mathds{1}^{r}_{i} = 1 } } 
&
{\small
\sum_{c=1}^C \frac{
\Big( 
\alpha_{g,c} 
 - 
\big(
\overbrace{
\beta_s^r \alpha_{s,c} 
+ \beta^r_{\text{miss}} \alpha^{r}_{\text{miss},c}
+ {\sum}_{{i} \in {\mathcal{K}}_{r}}  \mathds{1}^{r}_{i} \beta^{r}_{i} \alpha_{i,c}
}^{\triangleq {\tilde \alpha}^{r}_{c} }
\big)
\Big)^2}{\alpha_{g,c}}
}, 
\label{eq:aggregation_optimization_2 objective}
\\
\text{s.t.}
\quad
&
\beta_s^r + \beta^r_{\text{miss}} +
\sum_{{i} \in {\mathcal{K}}_{r}} \mathds{1}^{r}_{i} \beta^{r}_{i} = 1.
\label{eq:aggregation_optimization_2 constraint}
\end{align}
\end{subequations}
Here, ${\tilde \alpha}^{r}_{c}$ represents the effective proportion of class-$c$ samples contributing to the aggregated global model, and $C$ is the number of classes.
The denominator $\alpha_{g,c}$ prevents minority classes from being neglected.
The optimization formulation \eqref{eq:aggregation_optimization_2} is a convex weighted least squares problem, which can be easily solved using standard solvers such as CVX \cite{grant2009cvx} or Gurobi \cite{pedroso2011optimization}.

Despite its effectiveness, the optimization formulation \eqref{eq:aggregation_optimization_2} has a limitation:
when the public dataset $\mathcal{D}_s$ closely matches the global distribution (i.e., $\alpha_{s,c} \approx \alpha_{g,c}$ $\forall c \in [C]$), while private datasets remain heterogeneous, the optimization may assign an excessively large aggregation weight to the server, thereby underutilizing private updates.
To address this, we fix the server's aggregation weight as 
\begin{equation}\label{eq:server_aggregation_weight}
\beta_s^r = \frac{1}{1 + {\sum}_{{i} \in {\mathcal{K}}_{r}} \mathds{1}^{r}_{i}}, 
\end{equation}
where ${\sum}_{{i} \in {\mathcal{K}}_{r}} \mathds{1}^{r}_{i}$ is the number of clients successfully connected in round $r$.
This constraint ensures proportional contribution from the server’s public dataset while enabling the global model to continuously integrate knowledge from private datasets, thereby enhancing generalization.

The complete procedure of \texttt{FedAuto} is summarized in Algorithm \ref{algorithm:FedAuto}.
Since the adaptive aggregation strategy is fully executed on the server, \texttt{FedAuto} can be seamlessly integrated into existing systems without requiring modifications to the underlying network infrastructure.

\begin{algorithm}[t]
\small
\caption{\small
\texttt{FedAuto}: \gls{fft} with adaptive aggregation}
\begin{algorithmic}[1]
\Statex \textbf{\texttt{// Substitute Step 7 in Algorithm \ref{algorithm:FL under unreliable networks}:}}
\If {$\mathcal{C}^{r}_{\text{miss}} \neq \emptyset$}
\State Server performs compensatory training by (\ref{eq:local updating server compensatory});
\textbf{\texttt{(Module 1)}} 
\EndIf
\State Server optimizes aggregation weights via (\ref{eq:aggregation_optimization_2});
\textbf{\texttt{(Module 2)}} 
\State Server updates the global model ${\bar {\mathbf{w}}}_{r}$ using (\ref{global_model_failure_2}).
\end{algorithmic}
\label{algorithm:FedAuto}
\end{algorithm}

\begin{remark}
\rm
To solve (\ref{eq:aggregation_optimization_2}), clients only need to share their local class distributions $\{ \alpha_{i,c} \}$, which reveal minimal private information.
If stronger privacy guarantees are required, techniques such as secure multiparty computation can be applied \cite{ma2021client}.
As this work focuses on mitigating the effects of connection unreliability on \gls{fft} performance, privacy-preserving mechanisms are not further elaborated.
\end{remark}

\begin{remark}
\rm
This work assumes that the server has access to public data covering all client-owned classes, obtained from online resources \cite{gong2024ode} or generated using GANs~\cite{li2022federated} or diffusion models \cite{yang2024feddeo}.
Under this assumption, the server can fully compensate for missing classes during compensatory training (Module~1).
Future work may relax this assumption by considering scenarios in which some classes are unavailable at the server.
In such cases, class-incremental learning techniques \cite{dong2022federated, dong2023no}, combined with data generation techniques, could be employed to mitigate class unavailability.
\end{remark}

\section{Convergence analysis of \gls{fft}}\label{sec:convergence analysis}

In this section, we analyze the convergence of the proposed \texttt{FedAuto}.
We present a novel per-round aggregation view for \gls{fft} analysis, which removes the need for assumptions regarding the types and probabilities of connection failures, as well as the client selection strategies.
Unlike existing analyses that establish convergence only in expectation under unreliable networks, our approach ensures that the proposed \texttt{FedAuto} converges for each individual realization, thereby providing a stronger theoretical guarantee.
We first introduce two key foundations for deriving \gls{fft} convergence: the per-round view of global aggregation and the underlying causes of data heterogeneity.
Building on these, we derive the \gls{fft} convergence rate under unreliable connections, which establishes the theoretical convergence guarantee of \texttt{FedAuto}.

\subsection{Per-Round View of Global Aggregation}

In the global aggregation scheme \eqref{global_model_failure} for \gls{fft}, both connection reliability (i.e., $\mathds{1}^{r}_{i}$) and client selection (i.e., ${\mathcal{K}}_{r}$) directly influence the aggregated global model.
Existing convergence analyses of \gls{fl} under unreliable networks typically assume simplified models for connection failures (e.g., packet error rate \cite{chen2021joint} or transmission outage probability \cite{wang2022quantized}, etc) and fixed client selection strategies (e.g., full or partial participation, with or without replacement \cite{wang2025robust, salehi2021federated}).
These assumptions enable estimation of the expected global model, $\mathbb{E}_{{\mathcal{K}}_{r}, \mathds{1}^{r}_{i}} [{\bar {\mathbf{w}}}_{r} ]$, and provide convergence guarantees \textit{in expectation}.

However, such expectation-based approaches often fail to generalize well in real-world heterogeneous networks.
On one hand, connection unreliability arises from complex and mixed failure types \cite{li2023revisiting}.
On the other hand, client selection strategies vary across network standards and rarely follow uniform rules \cite{nguyen2021federated}.
Moreover, both connection reliability and client participation probabilities evolve dynamically due to client mobility and fluctuating communication conditions.
As a result, accurately modeling these probabilities throughout the training process is highly challenging.

To overcome these limitations, we adopt a \textit{per-round aggregation view} that embeds the combined effects of connection unreliability and client selection into the aggregation weights.
This reformulation eliminates the need for explicit modeling of connection failure probabilities or predefined client selection strategies in \gls{fft} convergence analysis. 
The key insight is formalized as follows:

\newpage

\begin{proposition}\label{proposition:global aggregation beta}
In each communication round, regardless of the connection failure type or client selection strategy, the aggregated global model in (\ref{global_model_failure}) can be reformulated based on the actual contribution of each client as
\begin{subequations}\label{eq:global_model_failure_proposition}
\begin{align}
{\bar {\mathbf{w}}}_{r}
=
&
\beta_s^r \mathbf{w}^{r,E}_{s} + \sum_{i = 1}^{N} \beta_i^r \mathbf{w}^{r,E}_{i},
\label{eq:global_model_failure_lemma obj}
\\
\text{s.t.}
\quad
&
\beta_s^r \geq 0,
\label{eq:global_model_failure_lemma constraint_a}
\\
&
\beta_i^r \geq 0 
\;\;
\text{if} \;\; i \in {\mathcal{K}}_{r} \;\; \text{and} \;\; \mathds{1}^{r}_{i}=1;
\;\;
\beta_i^r = 0 \;\; \text{otherwise},
\label{eq:global_model_failure_lemma constraint_b}
\\
&
\beta_s^r + \sum_{i = 1}^{N} \beta^{r}_{i} = 1.
\label{eq:global_model_failure_lemma constraint_c}
\end{align}
\end{subequations}
Here, the constraint (\ref{eq:global_model_failure_lemma constraint_b}) incorporates the effects of both network unreliability and client selection into the aggregation weight $\beta_i^r$, which reflects each client's actual contribution to the global model in round $r$.
Only the clients with successful connections contribute to the global model, while those unselected or disconnected have $\beta_i^r = 0$.
\end{proposition}

\subsection{Causes of Data Heterogeneity}

As indicated by \eqref{eq:global_model_failure_lemma obj}, the convergence direction of the aggregated global model depends not only on the aggregation weights $\beta_s^r$ and $\beta_i^r$, but also on the local models $\mathbf{w}^{r,E}_{s}$ and $\mathbf{w}^{r,E}_{i}$.
From the update rules in \eqref{eq:local updating client} and \eqref{eq:local updating server}, each local model is updated along its corresponding gradient, i.e., $\nabla F_{s}( {\mathbf{w}}; \mathcal{D}_s)$ for the server and $\nabla F_{i}( {\mathbf{w}}; \mathcal{D}_i)$ for each client $i \in [N]$.
These gradients are determined by the data distributions of public dataset $\mathcal{D}_s$ and private datasets $\mathcal{D}_i$, all of which serve as local datasets in \gls{fft} under the learning objective defined in \eqref{eq:distributed objective function}.

In contrast, the global gradient $\nabla F_g( {\mathbf{w}}; \mathcal{D}_g)$, derived from the global objective \eqref{eq:distributed objective function}, characterizes the ideal convergence direction of the fine-tuned global model.
It is determined by the data distribution of the global dataset $\mathcal{D}_g$, which integrates all local datasets.
Consequently, discrepancies between local and global data distributions inherently induce deviations between their gradients.
For this reason, the gradient gaps $\| \nabla F_s({\mathbf{w}}) - \nabla F ({\mathbf{w}}) \|^2$ and $\| \nabla F_i({\mathbf{w}}) - \nabla F ({\mathbf{w}}) \|^2$ are widely used to quantify data heterogeneity in \gls{fl} convergence analyses, especially in non-convex settings \cite{wang2022quantized, wang2023batch, karimireddy2020scaffold}. 
Prior studies consistently indicate that greater data heterogeneity significantly degrades convergence speed.

To more precisely characterize the impact of data heterogeneity on \gls{fft} performance, we adopt the decomposition framework from our prior work~\cite{wang2025robust}, which separates data heterogeneity into feature-related and label-related components.
Specifically, the global and local gradients can be reformulated as class-wise weighted averages:
\begin{subequations}\label{eq:nabla Fg Fsi alpha c}
\begin{equation}\label{eq:nabla Fg alpha c}
\nabla F_g({\mathbf{w}})  
= \sum_{c=1}^C \alpha_{g,c} \nabla F_{g,c} ({\mathbf{w}}),
\qquad\qquad\qquad
\end{equation}
\begin{equation}\label{eq:nabla Fsi alpha c}
\nabla F_j({\mathbf{w}}) 
= \sum_{c=1}^C \alpha_{j,c} \nabla F_{j,c} ({\mathbf{w}}),
\;
j \in \{s, [N]\}
,
\end{equation}
\end{subequations}
with $j=s$ denoting the server and $j \in [N]$ each client.
Here, $\nabla F_{g,c}$ and $\nabla F_{j,c}$ are the class-$c$ gradients computed on the global and local datasets, respectively;
$\alpha_{g,c}$ and $\alpha_{j,c}$ are the corresponding class proportions;
and $C$ is the total number of classes.
This reformulation yields the following upper bound on the discrepancy between local and global gradients.
\begin{proposition}\label{proposition:data heterogeneity}
\cite[Proposition 1]{wang2025robust}
The difference between a local gradient and the global gradient satisfies
\begin{equation}\label{ineq:nabla Fj - F}
\| \nabla F_j({\mathbf{w}}) - \nabla F_g ({\mathbf{w}}) \|^2
\leq 
2
\bigg(
\underbrace{
\sum_{c=1}^C  \alpha_{j,c}
\| \nabla F_{j,c} ({\mathbf{w}}) - \nabla F_{g,c} ({\mathbf{w}}) \|^2
}_{\text{\small (\ref{ineq:nabla Fj - F}a) related to data feature}}
+
\underbrace{
\chi^2_{\bm{\alpha}_j \| \bm{\alpha}_g}
\sum_{c=1}^C \alpha_{g,c} \| \nabla F_{g,c} ({\mathbf{w}}) \|^2
}_{\text{\small (\ref{ineq:nabla Fj - F}b) related to sample label}}
\bigg)
\end{equation}
for $j=s$ (server) and $j \in [N]$ (clients).
Term (\ref{ineq:nabla Fj - F}a) characterizes \textbf{feature-related heterogeneity}, where the within-class gradient deviation $\| \nabla F_{j,c} ({\mathbf{w}}) - \nabla F_{g,c} ({\mathbf{w}}) \|^2$ arises from factors such as limited local samples and cross-source feature shift \cite{wang2022hetvis, li2021fedbn}.
Term (\ref{ineq:nabla Fj - F}b) represents \textbf{label-related heterogeneity}, where the chi-square divergence $\chi^2_{\bm{\alpha}_i \| \bm{\alpha}_g} \triangleq \sum_{c=1}^C \frac{( \alpha_{i,c} - \alpha_{g,c} )^2}{\alpha_{g,c}}$ quantifies the mismatch between local and global class distributions.
\end{proposition}

\emph{Proof:}
See \cite[Appendix A]{wang2025robust}.
\hfill $\blacksquare$

\subsection{Convergence Rate of \gls{fft} under Unreliable Connections}\label{sec:theorem FFT failure}

\subsubsection{Assumptions}

Based on Proposition \ref{proposition:data heterogeneity}, to separately analyze the effects of feature-related heterogeneity and label-related heterogeneity on \gls{fft} convergence under unreliable connections, we introduce the following assumptions:
\begin{assumption}\label{assumption:class heterogeneity}
\textbf{(Bounded gradient deviation within a class)}
For each class $c$, the deviation between a local gradient and the corresponding global gradient is bounded as 
$\| \nabla F_{j,c}({\mathbf{w}}) - \nabla F_{g,c} ({\mathbf{w}})\|^2 \leq V_{j,c}^2$, $\forall j \in \{s,[N]\}$,
with $j=s$ denotes the server and $j \in [N]$ denotes each client.
\end{assumption}
\begin{assumption}\label{assumption:garident norm}
\textbf{(Bounded gradient norm)}
The global gradient norm is bounded as
$\| \nabla F_g ({\mathbf{w}}) \|^2 \leq G^2$.
\end{assumption}

Combining Proposition \ref{proposition:data heterogeneity} with Assumptions \ref{assumption:class heterogeneity} and \ref{assumption:garident norm} yields Corollary \ref{corollary:data heterogeneity}, which quantifies the degree of data heterogeneity.

\begin{corollary}\label{corollary:data heterogeneity}
Under Assumptions \ref{assumption:class heterogeneity} and \ref{assumption:garident norm}, the difference between the local and global gradients is bounded by
\begin{equation}\label{ineq:nabla Fi - F 2}
\| \nabla F_i({\mathbf{w}}) - \nabla F ({\mathbf{w}}) \|^2
\leq 
2 \sum_{c=1}^C
\alpha_{i,c} V_{i,c}^2
+ 2 \chi^2_{\bm{\alpha}_i \| \bm{\alpha}_g} G^2
,
\end{equation}
where the first term on the \gls{rhs} of (\ref{ineq:nabla Fi - F 2}) captures the impact of feature-related heterogeneity, and the second term reflects the effect of label-related heterogeneity.
\end{corollary}

In addition to Assumptions \ref{assumption:class heterogeneity} and \ref{assumption:garident norm}, we adopt the following standard smoothness assumption on local cost functions:
\begin{assumption}\label{assumption:L continuous}
\textbf{(L-smoothness)}
Each local cost function is differentiable with Lipschitz-continuous gradients, i.e., there exists a constant $L > 0$ such that $\forall {\mathbf{w}}$ and ${\mathbf{w}'}$, 
$\| \nabla F_j({\mathbf{w}}) - \nabla F_j({\mathbf{w}'}) \|_2 \leq L \| {\mathbf{w}} - {\mathbf{w}'} \|_2$, $\forall j \in \{s,[N]\}$.
\end{assumption}

\subsubsection{Theoretical Results}

The main convergence result for the \gls{fft} procedure under unreliable connections, as described in Algorithm \ref{algorithm:FL under unreliable networks}, is presented below.

\begin{theorem}\label{theorem:convergence FL failure}
Suppose Assumptions \ref{assumption:class heterogeneity} to \ref{assumption:L continuous} hold, and incorporate Propositions \ref{proposition:global aggregation beta} and \ref{proposition:data heterogeneity}.
Assume the total number of gradient descent updates $T = RE$ is sufficiently large with $T \geq N^{3}$, while the number of local update steps $E$ remains relatively small with $E \leq T^{\frac{1}{4}}/N^{\frac{3}{4}}$, where $R$ is the number of iterations in Algorithm \ref{algorithm:FL under unreliable networks}.
If the learning rate is chosen as $\gamma = N^{\frac{1}{2}} / (2L{T}^{\frac{1}{2}})$, then the convergence of \gls{fft} under unreliable connections in Algorithm \ref{algorithm:FL under unreliable networks} is upper bounded by
\begin{align}\label{eq:convergence bound of FL failure}
\frac{1}{R} \sum_{r=1}^R \left \| \nabla F_g(\bar{\mathbf{w}}_{r-1}) \right \|^2
\leq 
& 
\frac{20L}{\sqrt{TN}}
\big( F(\bar{\mathbf{w}}_{\rm pre}) - F(\bar{\mathbf{w}}_R) \big) 
+ 
\underbrace{
\frac{8}{\sqrt{TN}R} \sum_{r=1}^R \sum_{j \in \{s, [N]\}} \beta_{j}^r 
\sum_{c=1}^C \big( \alpha_{j,c} V_{j,c}^2 + {\chi}_{\bm{\alpha}_g || \bm{\alpha}_j}^2 G^2 \big)
}_{\text{\small (\ref{eq:convergence bound of FL failure}a) caused by non-i.i.d. data}}
\notag \\
&
+ 
\underbrace{
20 
\bigg (\chi^2_{\mathbf{p}\|\bm{\beta}} \sum_{c=1}^C \sum_{j \in \{s, [N]\}} p_i \alpha_{j,c} V_{j,c}^2
+ \frac{1}{R} \sum_{r=1}^R \chi^2_{\bm{\alpha}_g \| \bm{\tilde \alpha}^r} G^2 
\bigg ) 
}_{\text{\small (\ref{eq:convergence bound of FL failure}b) caused by connection unreliability and non-i.i.d. data}}
.
\end{align}
Here, the chi-square divergence $\chi^2_{\mathbf{p}\|\bm{\beta}} \triangleq \sum_{j \in \{s, [N]\}} \frac{({\beta}_j^r - p_j)^2}{p_j}$ measures the deviation between the aggregation weights $\{\beta_j^r \}$ and the weight coefficients $\{ p_j \}$ defined in (\ref{eq:distributed objective function}).
Similarly, $\chi^2_{\bm{\alpha}_g \| \bm{\tilde \alpha}^r} \triangleq \sum_{c=1}^C \frac{( \alpha_{g,c} - \sum_{j \in \{s, [N]\}} \beta_j^r \alpha_{j,c})^2}{\alpha_{g,c}}$ quantifies the divergence between the global class distribution $\{ \alpha_{g,c} \}$ and the effective class distribution $\{ {\tilde \alpha}_c^r ( \triangleq \sum_{j \in \{s, [N]\}} \beta_j^r \alpha_{j,c} ) \}$.
The aggregation weights $\{ \beta_j^r \}$ satisfy the constraints (\ref{eq:global_model_failure_lemma constraint_a})-(\ref{eq:global_model_failure_lemma constraint_c}). 
\end{theorem}

\emph{Proof:}
Unlike prior analyses of \gls{fl} convergence under unreliable networks~\cite{ribero2022federated, salehi2021federated, wang2025robust, chen2021joint}, we analyze \gls{fft} from a per-round aggregation perspective (Proposition \ref{proposition:global aggregation beta}).
This perspective offers two key advantages.
First, it eliminates the need to assume specific connection failure models or probabilities, thereby enhancing applicability to practical heterogeneous network environments.
Second, while previous works establish convergence only in expectation, we analyze convergence for each individual realization, providing a strictly stronger theoretical result and facilitating the design of more robust \gls{fft} strategies under unreliable connections.
The full proof is presented in Appendix \ref{appendix:Proof of convergence rate theorem}.
\hfill $\blacksquare$

From the \gls{rhs} of \eqref{eq:convergence bound of FL failure}, the convergence rate of Algorithm \ref{algorithm:FL under unreliable networks} is governed by several key factors, including:
(i) aggregation weights of the server and clients ($\beta_s^r$ and $\beta_i^r$);
(ii) data heterogeneity terms ($V_{i,c}^2$ and $\chi^2_{\bm{\alpha}_i \| \bm{\alpha}_g}$); 
and 
(iii) divergence terms induced by connection unreliability ($\chi^2_{\mathbf{p}\|\bm{\beta}}$ and $\chi^2_{\bm{\alpha}_g \| \bm{\tilde \alpha}^r}$).
Analyzing these terms leads to several important observations:
\begin{enumerate}[(a)]
\item \label{theorem observation:Fully reliable connections}
\textbf{Fully reliable connections}:
When all connections are reliable and full client participation holds (i.e., ${\mathds{1}^{r}_{i}}=1$ $\forall i,r$ and ${\mathcal{K}}_{r} = [N]$), choosing aggregation weights consistent with the distributed objective in~\eqref{eq:distributed objective function}, namely $\beta_j^r = p_j$ $\forall j \in \{s,[N]\}$, eliminates both divergence terms $\chi^2_{\mathbf{p}\|\bm{\beta}}$ and $\chi^2_{\bm{\alpha}_g \| \bm{\tilde \alpha}^r}$.
As a result, term (\ref{eq:convergence bound of FL failure}b) vanishes, and convergence is still affected by data heterogeneity through term~(\ref{eq:convergence bound of FL failure}a).

\item \label{theorem observation:iid data}
\textbf{i.i.d. datasets}:
If all local datasets (including both public and private) are i.i.d., their class distributions match the global distribution (i.e, $\alpha_{j,c} = \alpha_{g,c}$), and the within-class gradient deviations $\{ V_{j,c}^2 \}$ become negligible.
In this case, both terms (\ref{eq:convergence bound of FL failure}a) and (\ref{eq:convergence bound of FL failure}b) disappear, showing that i.i.d. data mitigates the negative effects of connection unreliability and ensures stable convergence.

\item \label{theorem observation:Joint effect of unreliability and heterogeneity}
\textbf{Joint effect of unreliability and heterogeneity}:
Term (\ref{eq:convergence bound of FL failure}b), which captures the joint impacts of connection unreliability and data heterogeneity, does not decay as $T$ increases.
This implies that network unreliability amplifies the adverse effects of data heterogeneity through the divergence factors $\chi^2_{\mathbf{p}\|\bm{\beta}}$ and $\chi^2_{\bm{\alpha}_g \| \bm{\tilde \alpha}^r}$, resulting in biased \gls{fft} convergence.

\item \label{theorem observation:Dominance of label-related heterogeneity}
\textbf{Dominance of label-related heterogeneity}:
Empirical evidence in \cite[Observation 1]{wang2025robust} shows that under non-i.i.d.\ data with heterogeneous class distributions, the global gradient norm $G^2$ is typically several orders of magnitude larger than the average within-class gradient deviation, $\sum_{j \in \{s, [N]\}} p_i \alpha_{j,c} V_{j,c}^2$.
Consequently, the label-related component $\frac{1}{R} \sum_{r=1}^R \chi^2_{\bm{\alpha}_g \| \bm{\tilde \alpha}^r} G^2$ in (\ref{eq:convergence bound of FL failure}b) dominates the convergence bias.
Therefore, enforcing $\chi^2_{\bm{\alpha}_g \| \bm{\tilde \alpha}^r} = 0$ in every communication round can eliminate this bias, rendering term~(\ref{eq:convergence bound of FL failure}b) negligible and enabling \gls{fft} to converge properly even under unreliable connections.

\end{enumerate}

\begin{remark}
\rm
To the best of our knowledge, Theorem \ref{theorem:convergence FL failure} provides the first theoretical characterization of how aggregation weights influence \gls{fft} convergence under unreliable connections.
Moreover, it is the first to establish convergence in such settings without requiring prior knowledge of connection failure types or probabilities, making it well suited for real-world \gls{fft} deployments in heterogeneous networks.
\end{remark}

\begin{remark}
\rm
If the connection failure probabilities and client selection are explicitly specified, Theorem \ref{theorem:convergence FL failure} reduces to expectation-based convergence analyses such as \cite{wang2025robust, salehi2021federated}.
In this case, \gls{lhs} of (\ref{eq:convergence bound of FL failure}) becomes $\frac{1}{R}\sum_{r=1}^R \mathbb{E}_{{\mathcal{K}}_{r}, \mathds{1}^{r}_{i}}[\left \| \nabla F_g(\bar{\mathbf{w}}_{r-1}) \right \| ^2]$,
indicating that the proposed analytical framework is more general and subsumes prior convergence results as special cases.
\end{remark}

\subsection{Convergence Guarantee of FedAuto}\label{sec:FedCote Guarantee}

From the observation~\eqref{theorem observation:Dominance of label-related heterogeneity} of Theorem \ref{theorem:convergence FL failure}, proper \gls{fft} convergence critically depends on minimizing the divergence between the global and effective class distributions, $\chi^2_{\bm{\alpha}_g \| \bm{\tilde \alpha}^r}$, in each round.
The proposed \texttt{FedAuto} in Algorithm \ref{algorithm:FedAuto} achieves this through server-side compensatory training for underrepresented classes and  adaptive optimization of aggregation weights to balance class contributions in the aggregated global model.
As a result, the effective class proportion ${\tilde \alpha}_c^r$ closely matches the global proportion $\alpha_{g,c}$ for each class $c$, driving $\chi^2_{\bm{\alpha}_g \| \bm{\tilde \alpha}^r}$ towards zero.
This leads to the following result:

\begin{corollary}\label{corollary:convergence FL failure}
Under the conditions of Theorem \ref{theorem:convergence FL failure} and incorporating \cite[Observation 1]{wang2025robust}, if the adaptive aggregation strategy in Algorithm \ref{algorithm:FedAuto} is employed such that the effective class distribution $\{ {\tilde \alpha}_c^r \}$ closely approximates the global class distribution $\{ \alpha_{g,c} \}$, i.e., $\chi^2_{\bm{\alpha}_g \| \bm{\tilde \alpha}^r} \approx 0$, then the convergence rate in (\ref{eq:convergence bound of FL failure}) approximately reduces to
\begin{align}\label{eq:convergence bound of FL failure corollary}
\frac{1}{R} \sum_{r=1}^R \left \| \nabla F_g(\bar{\mathbf{w}}_{r-1}) \right \|^2
\lesssim
\frac{20L}{\sqrt{TN}}
\big( F(\bar{\mathbf{w}}_{\rm pre}) - F(\bar{\mathbf{w}}_R) \big) 
+ 
\frac{8}{\sqrt{TN}R} \sum_{r=1}^R \sum_{j \in \{s, [N]\}} \beta_{j}^r 
\sum_{c=1}^C \big( \alpha_{j,c} V_{j,c}^2 + {\chi}_{\bm{\alpha}_g || \bm{\alpha}_j}^2 G^2 \big)
.
\end{align}
\end{corollary}

From the \gls{rhs} of \eqref{eq:convergence bound of FL failure corollary}, the global gradient satisfies $\nabla F_g(\bar{\mathbf{w}}_{r-1}) \to 0$ as $T \to \infty$, indicating that \gls{fft} converges to a proper solution even in the presence of both unreliable connections and data heterogeneity.
This establishes a rigorous theoretical convergence guarantee for \texttt{FedAuto}.
Moreover, since convergence is ensured for each individual realization rather than only in expectation, \texttt{FedAuto} exhibits strong robustness in unreliable network environments.

\begin{remark}
\rm
An alternative to adjusting aggregation weights to minimize $\chi^2_{\bm{\bar \alpha} \| \bm{\alpha}_g}$ is to modify the local label distributions $\{ {\alpha}_{i,c} \}$, for example, through augmenting local datasets with server-side public or synthetic data \cite{gong2024delta,li2024feature}.
Such approaches implicitly render all datasets approximately i.i.d., thereby mitigating the impact of connection unreliability and enabling stable convergence, as noted in the observation \ref{theorem observation:iid data} of Theorem \eqref{theorem:convergence FL failure}.
However, they incur additional computational and storage overhead on clients, making them less suitable for resource-constrained edge devices.
Regardless of the strategy, the core objective remains to balance the effective class distribution $\{ {\tilde \alpha}_c^r \}$ during global aggregation, preventing dominance by overrepresented classes while preserving the contributions of underrepresented ones.
\end{remark}

\section{Experimental Results}\label{sec:experimental results}

\subsection{Parameter Settings}\label{sec:Parameter Setting}

\subsubsection{Heterogeneous Network Environment}

We simulate a practical heterogeneous network scenario in which clients connect to the server via diverse communication standards, including wired Ethernet and wireless links over 4G, 5G, and Wi-Fi (2.4\,GHz and 5\,GHz).
As illustrated in Fig. \ref{fig:FL unreliable networks}, the 4G and 5G base stations are deployed outdoors at the center of a 200\,m-radius cell with an antenna height of 20\,m, while the Wi-Fi \gls{ap} is installed indoors at the center of a $20 \times 20$\,m area with a height of 3\,m.
The central server coordinates $N = 20$ clients for FFT of \glspl{dnn}. 
Among them, 8 clients are uniformly distributed indoors, and the remaining 12 clients are uniformly distributed in the outdoor region. 
Detailed network standards and communication resource allocations for each client are provided in Appendix~\ref{subsec:Communication Resource Allocation}.

\subsubsection{Modes of Connection Failures}

We consider three patterns of connection failures:
\begin{itemize}
\item
\textbf{Transient}:
Short, random disruptions modeled as instantaneous events with stochastic occurrence.
In our experiments, each client’s transient failure is induced by transmission outages and its probability is determined by a classical path-loss model with shadowing effects \cite{goldsmith2005wireless}.

\item
\textbf{Intermittent}:
Longer disruptions (e.g., power depletion) that occur randomly and persist for a duration.
Intermittent failure occurrences follow an exponential distribution, while the disconnection durations follow a uniform distribution once triggered \cite{correcher2012intermittent}.

\item
\textbf{Mixed}: 
A composite mode in which clients experience both transient and intermittent failures.
\end{itemize}

The full mathematical formulations of these failure models are provided in Appendix~\ref{subsec:Failure Probability Formulation}.

\subsubsection{Datasets, Data Distributions, and Models}\label{section:datasets_distributions_models}

We evaluate \texttt{FedAuto} on three widely used image-classification benchmarks under both i.i.d.\ and non-i.i.d.\ data settings.
For each dataset, we consider two representative \glspl{dnn}, one small-scale and one large-scale.

\begin{itemize}
\item
\textbf{MNIST} \cite{lecun1998gradient}:  
60,000 training and 10,000 test images from ten classes.  
The i.i.d.\ setting shuffles and uniformly partitions samples across clients.  
For the non-i.i.d.\ setting, each client receives data from two designated classes (e.g., clients 1--4: \{1,2\}; clients 5--8: \{3,4\}, etc).  
For classification, we adopt a small-scale CNN architecture with two convolutional layers followed by fully connected layers 
(0.22\,M parameters), 
and a large-scale ViT model 
(86\,M parameters) 
. 

\item
\textbf{CIFAR-10} \cite{krizhevsky2009learning}:  
50,000 training and 10,000 test images from ten classes.  
Both i.i.d.\ and non-i.i.d.\ partitioning follow the MNIST setup.  
ResNet with \gls{gn} \cite{hsieh2020non} 
(0.27\,M parameters) 
is used as the small-scale model, and ViT 
(86\,M parameters) 
as the large one.

\item
\textbf{CIFAR-100} \cite{krizhevsky2009learning}:
50,000 training and 10,000 test images from 100 classes.  
The i.i.d.\ partitioning mirrors MNIST, whereas in the non-i.i.d.\ setting, each client receives data from 20 classes.  
We use the ResNet-18 model with \gls{gn} from \cite{hsieh2020non} 
(11\,M parameters) 
as the small model and again adopt ViT 
(86\,M parameters) 
as the large one.

\end{itemize}

The architectures of all employed \glspl{dnn} are detailed in Appendix~\ref{subsec:DNN architectures}.  
In all experiments, the global model produced by FFT is evaluated on the full test dataset to assess convergence and generalization performance.

\subsubsection{Fine-Tuning Strategies}\label{section:fine_tuning_ways}

We compare two strategies:
\begin{itemize}
\item
\textbf{Full-parameter fine-tuning}:
All model parameters are updated during training and used for small-scale models.

\item
\textbf{Partial-parameter fine-tuning}:
Only a subset of parameters is updated using \gls{lora} \cite{hu2022lora}.
This strategy is employed to fine-tune large-scale ViT models initialized from an ImageNet-1k-pretrained ViT model \cite{rw2019timm}.
\end{itemize}

The training hyperparameters for both strategies are summarized in Appendix~\ref{section:FFT training_paras}.

\subsubsection{Baselines}\label{section:Baselines}

To comprehensively assess the effectiveness of \texttt{FedAuto}, we compare it against a broad set of baselines spanning centralized learning, classical \gls{fl} algorithms, aggregation-enhanced methods, and communication resource-aware schemes.
We begin with the centralized learning setting:
\begin{itemize}
\item
\textbf{\texttt{Central(Public)}}:
A centralized training scheme in which the server learns the model solely from the public dataset, without leveraging any distributed private data.
\end{itemize}

We next include classical and the advanced \gls{fl} algorithms designed to address data heterogeneity:
\begin{itemize}
\item
\textbf{\texttt{FedAvg(Ideal)}} \cite{mcmahan2017communication}:
The classical \texttt{FedAvg} executed without connection failures. 
It follows Algorithm~\ref{algorithm:FL under unreliable networks}, adopts the aggregation rule in~\eqref{eq:global aggregation}, and uses aggregation weights specified in Remark~\ref{remark: aggregation weights full partial}. 
This configuration serves as the performance upper bound in our experiments.

\item
\textbf{\texttt{FedAvg}}:
The practical \texttt{FedAvg} deployed under realistic connection failures, implemented via Algorithm~\ref{algorithm:FL under unreliable networks} using the heuristic aggregation weights defined in footnote~\ref{footnote:aggregation weights FedAvg}.

\item
\textbf{\texttt{FedProx}} \cite{li2020federated}:
An extension of \texttt{FedAvg} that introduces a proximal term into each client's local objective to alleviate the effects of data heterogeneity.

\item
\textbf{\texttt{SCAFFOLD}} \cite{karimireddy2020scaffold}:
A variance-reduction method that employs control variates on both the server and client sides to correct the client drift induced by non-i.i.d.\ data.
\end{itemize}

We further consider aggregation-enhanced methods:

\begin{itemize}
\item
\textbf{\texttt{FedLAW}} \cite{li2023revisiting}:
A refinement of \texttt{FedAvg} that leverages a server-side proxy dataset to jointly optimize a global shrinking factor and client-specific aggregation weights, thus enhancing robustness under data heterogeneity.

\item
\textbf{\texttt{TF-Aggregation}} \cite{salehi2021federated}:
A method designed to mitigate transient connection failures via incorporating clients’ estimated failure probabilities into the global aggregation. 

\item
\textbf{\texttt{FedAWE}} \cite{xiang2024efficient}:
An adaptive weighting scheme that adjusts each client's local stepsize based on its number of failed rounds, thereby balancing global aggregation.

\item
\textbf{\texttt{FedEx-LoRA}} \cite{singhal2024fedex}:
An enhancement of \texttt{FedAvg} that incorporates the error residual induced by global aggregation of \gls{lora} modules into local model updates.

\end{itemize}

Finally, we consider \gls{fft}-based baselines that explicitly optimize communication resources:
\begin{itemize}
\item
\textbf{\texttt{ResourceOpt-1}}: 
Motivated by \cite[Corollary 1]{wang2022quantized}, which indicates that equalizing clients’ failure probabilities can reduce convergence bias, this baseline jointly optimizes transmit power and bandwidth to balance transient connection failure probabilities across clients.

\item
\textbf{\texttt{ResourceOpt-2}}: 
Recognizing that joint optimization of communication resources across multiple standards is often infeasible in heterogeneous commercial networks, this baseline provides a practical variant of \texttt{ResourceOpt-1} by independently optimizing transmit power and bandwidth allocation within each standard.

\end{itemize}

Detailed mathematical formulations for the above baselines are provided in Appendix~\ref{section:Baselines formulations}.
All reported results are averaged over five independent runs to ensure statistical reliability.

\subsection{Performance under Full-Parameter Fine-Tuning}\label{section:experiment Full-Parameter Fine-Tuning}

In this part, we evaluate the robustness of \texttt{FedAuto} under full-parameter fine-tuning across various combinations of data distributions and connection failure modes.

\subsubsection{Full Client Participation}

\begin{figure}[t]
\begin{minipage}[h]{1\linewidth}
\centering
\includegraphics[width= 3.4 in ]{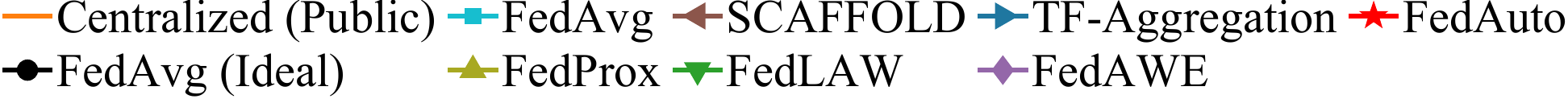}
\end{minipage}
\begin{minipage}[h]{1\linewidth}
\centering
\subfigure[MNIST, i.i.d.]{
\includegraphics[width= 1.6 in ]{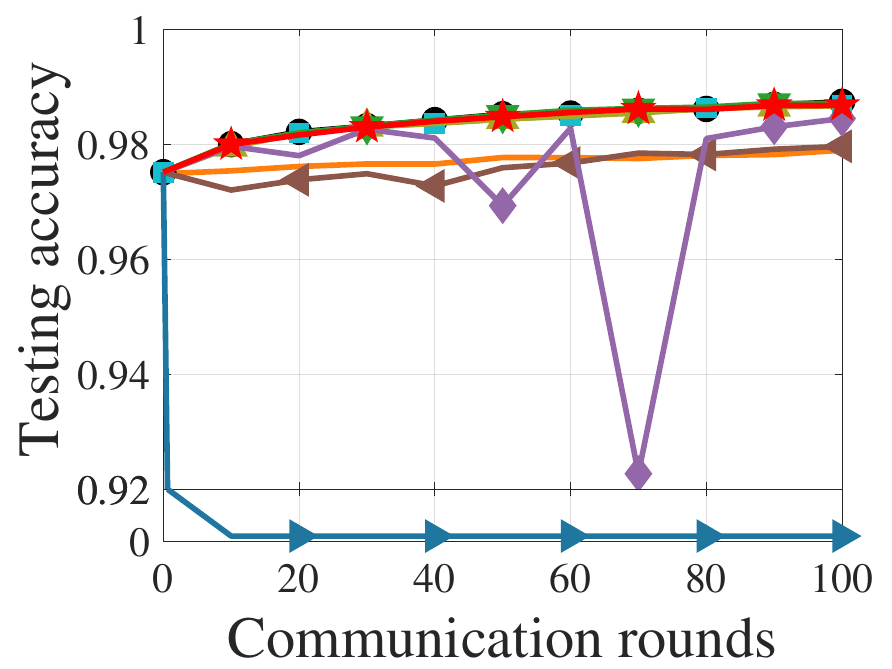}\label{fig:mnist mixed iid}}
\subfigure[MNIST, non-i.i.d.]{
\includegraphics[width= 1.6 in ]{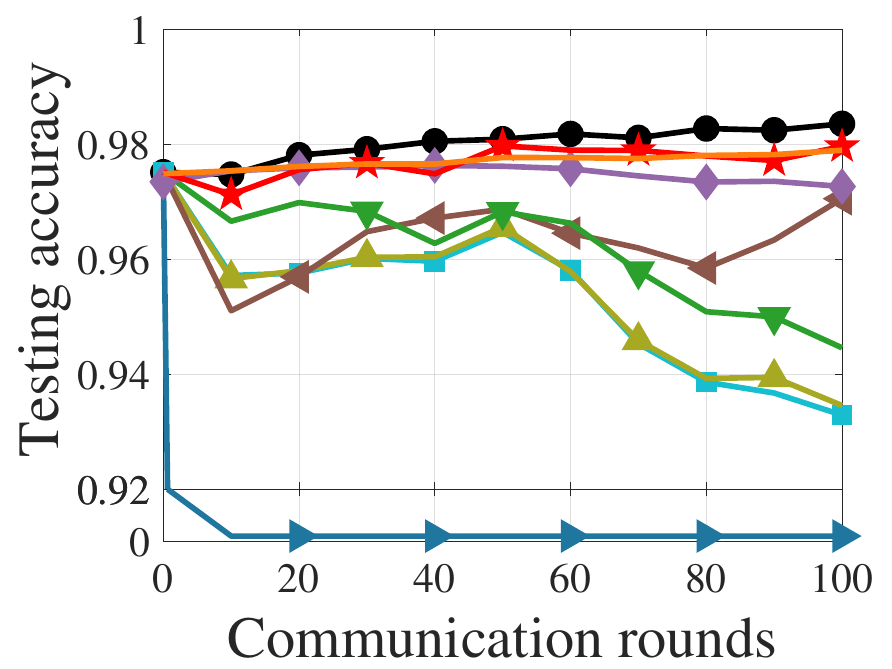}\label{fig:mnist mixed noniid}}
\subfigure[CIFAR-10, i.i.d.]{
\includegraphics[width= 1.6 in ]{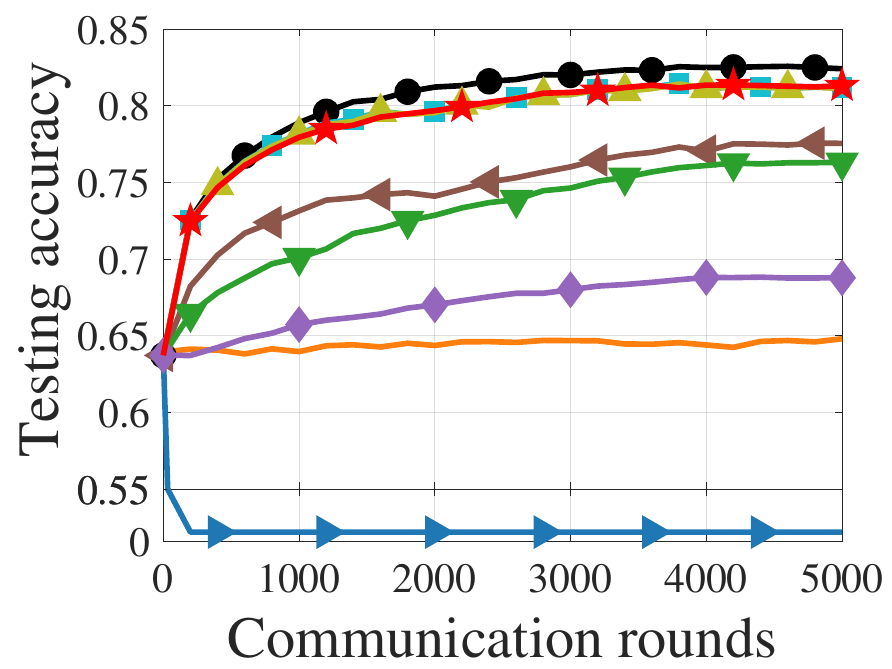}\label{fig:cifar10 mixed iid}}
\subfigure[CIFAR-10, non-i.i.d.]{
\includegraphics[width= 1.6 in ]{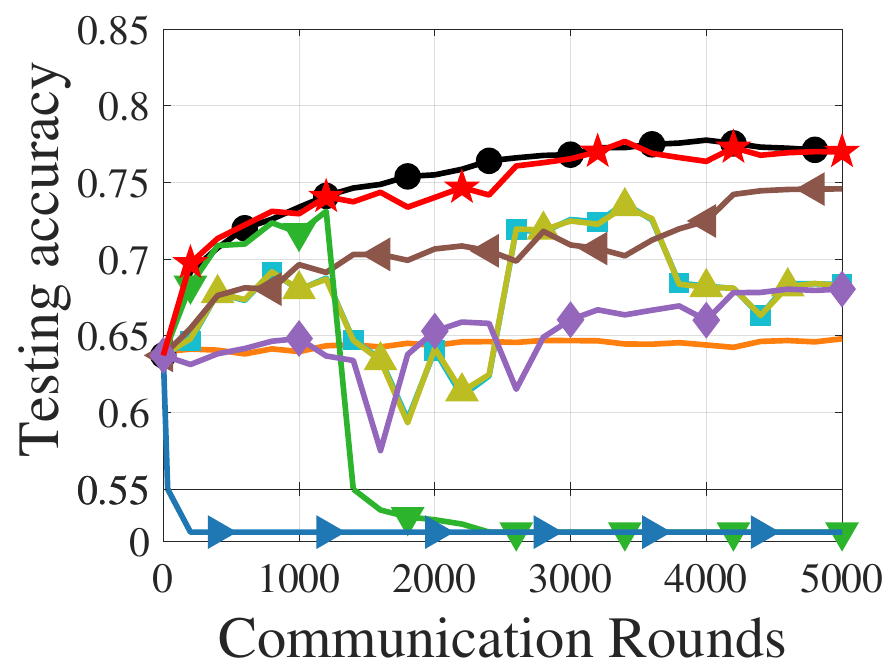}\label{fig:cifar10 mixed noniid}}
\subfigure[CIFAR-100, i.i.d.]{
\includegraphics[width= 1.6 in ]{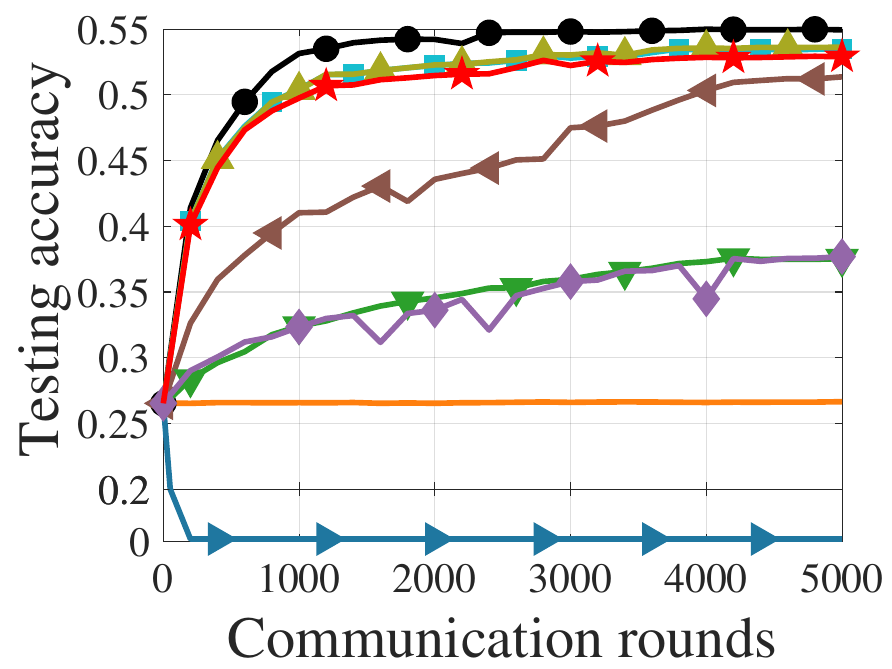}\label{fig:cifar100 mixed iid}}
\subfigure[CIFAR-100, non-i.i.d.]{
\includegraphics[width= 1.6 in ]{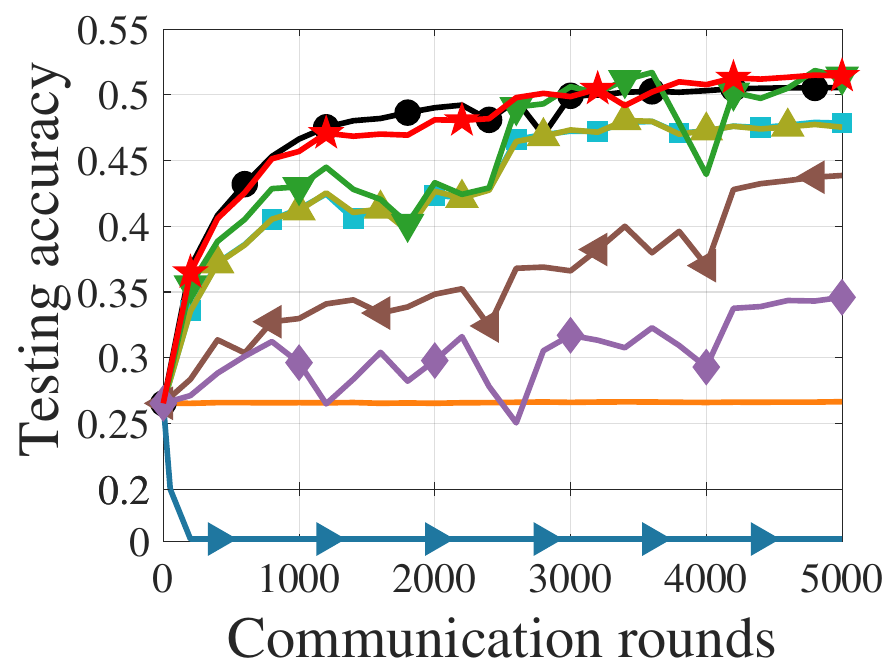}\label{fig:cifar100 mixed noniid}}
\caption{
Convergence of different \gls{fft} strategies under full-parameter fine-tuning with full participation ($K=20$, mixed failures).}
\label{fig:performance diff dataset mixed iid full}
\end{minipage}
\end{figure}

We first examine various \gls{fft} strategies under full client participation with $K = N = 20$. 
The convergence behaviors on multiple datasets are shown in Fig.~\ref{fig:performance diff dataset mixed iid full}, and the corresponding testing accuracies are summarized in Tables~\ref{table:FFT performance iid full} and~\ref{table:FFT performance non-iid full}. 
Since the convergence trends under transient and intermittent failures closely resemble those under the mixed failure mode in Fig.~\ref{fig:performance diff dataset mixed iid full}, we present them in Appendix~\ref{section:supplementary transient intermittent}.

\begin{table*}[t]
\centering
\caption{ 
Testing accuracy(\%) of different \gls{fft} strategies under full-parameter fine-tuning with full participation ($K=20$, i.i.d.\ data).
}\label{table:FFT performance iid full}
{\scriptsize
\begin{tabular}{c|ccc|ccc|ccc}
\hline
\multirow{2}{*}{\textbf{\gls{fft} Strategy}} & \multicolumn{3}{c|}{\textbf{Transient}} & \multicolumn{3}{c|}{\textbf{Intermittent}} & \multicolumn{3}{c}{\textbf{Mixed}} \\
& MNIST & CIFAR-10 &  CIFAR-100 & MNIST & CIFAR-10 & CIFAR-100 & MNIST & CIFAR-10 & CIFAR-100 \\
\hline
\texttt{Centralized(Public)}
& 97.89\tiny{$\pm$0.1337} & 64.92\tiny{$\pm$0.9380} & 26.68\tiny{$\pm$0.8543}
& 97.89 \tiny{$\pm$0.1337} & 64.92\tiny{$\pm$0.9380} & 26.68\tiny{$\pm$0.8543}
& 97.89\tiny{$\pm$0.1337} & 64.92\tiny{$\pm$0.9380} & 26.68\tiny{$\pm$0.8543} \\
\texttt{FedAvg}~\cite{mcmahan2017communication}
& 98.70\tiny{$\pm$0.1330} & 81.13\tiny{$\pm$0.3437} & 51.86\tiny{$\pm$1.9350} & 98.65\tiny{$\pm$0.1236} & 81.39\tiny{$\pm$0.4807} & 53.36\tiny{$\pm$1.5590}& 98.68\tiny{$\pm$0.1251} & 81.18\tiny{$\pm$0.4720} & 53.49\tiny{$\pm$1.9090}\\
\texttt{FedProx}~\cite{li2020federated}
& 98.70\tiny{$\pm$0.1330} & 81.0l\tiny{$\pm$0.3503} & 52.29\tiny{$\pm$1.7081} & 98.65\tiny{$\pm$0.1210} & 81.35\tiny{$\pm$0.5428} & 53.53\tiny{$\pm$1.7631}& 98.67\tiny{$\pm$0.1221} & 81.12\tiny{$\pm$0.5482} & 53.62\tiny{$\pm$1.7831}\\
\texttt{SCAFFOLD}~\cite{karimireddy2020scaffold}
& 97.83\tiny{$\pm$0.2483} & 75.49\tiny{$\pm$0.6683} & 46.37\tiny{$\pm$2.2444} & 98.01\tiny{$\pm$0.2716} & 77.09\tiny{$\pm$0.6215} &50.46\tiny{$\pm$1.3387}& 97.96\tiny{$\pm$0.2022} & 77.55\tiny{$\pm$0.5254} & 51.37\tiny{$\pm$1.5439}\\
\texttt{FedLAW}~\cite{li2023revisiting}
& 98.74\tiny{$\pm$0.0936} & 76.66\tiny{$\pm$1.2494} & 37.70\tiny{$\pm$2.8919} & 98.70 \tiny{$\pm$0.0850} & 76.65\tiny{$\pm$1.4020} & 37.93\tiny{$\pm$2.4467}& 98.71\tiny{$\pm$0.1041} & 76.32\tiny{$\pm$1.5242} & 37.54\tiny{$\pm$2.7535}\\
\texttt{TF-Aggregation}~\cite{salehi2021federated}
& 9.80\tiny{$\pm$0.0000} & 10.00\tiny{$\pm$0.0000} & 1.00\tiny{$\pm$0.0000}  & 9.80\tiny{$\pm$0.0000} & 10.00\tiny{$\pm$0.0000} & 1.00\tiny{$\pm$0.0000}& 9.80\tiny{$\pm$0.0000} & 10.00\tiny{$\pm$0.0000} & 1.00\tiny{$\pm$0.0000}\\
\texttt{FedAWE}~\cite{xiang2024efficient}
& 91.17\tiny{$\pm$16.2217} & 68.84\tiny{$\pm$0.8884} & 34.29\tiny{$\pm$4.1678}  & 98.07\tiny{$\pm$0.7321} & 68.84\tiny{$\pm$0.9319} & 37.98\tiny{$\pm$1.3004}& 98.45\tiny{$\pm$0.0885} & 68.78\tiny{$\pm$1.0224} & 37.69\tiny{$\pm$1.2323}\\
\textbf{\texttt{FedAuto} (Ours)}
& 98.69\tiny{$\pm$0.0940} & 81.08\tiny{$\pm$0.4163} & 51.48\tiny{$\pm$1.8997} & 98.67\tiny{$\pm$0.1244} & 81.28\tiny{$\pm$0.3567} & 52.67\tiny{$\pm$1.6424} & 98.67\tiny{$\pm$0.1282} & 81.31\tiny{$\pm$0.5199} & 52.90\tiny{$\pm$1.8059}\\
\hline
\texttt{FedAvg(Ideal)}
& 98.74\tiny{$\pm$0.1182} & 82.42\tiny{$\pm$0.2982} & 54.95\tiny{$\pm$1.7194}& 98.74\tiny{$\pm$0.1182} & 82.42\tiny{$\pm$0.2982} & 54.95\tiny{$\pm$1.7194}& 98.74\tiny{$\pm$0.1182} & 82.42\tiny{$\pm$0.2982} & 54.95\tiny{$\pm$1.7194}\\
\hline
\end{tabular}
}
\end{table*}

\begin{table*}[t]
\centering
\caption{
Testing accuracy(\%) of different \gls{fft} strategies under full-parameter fine-tuning with full participation ($K=20$, non-i.i.d.\ data).
}\label{table:FFT performance non-iid full}
{\scriptsize
\begin{tabular}{c|ccc|ccc|ccc}
\hline
\multirow{2}{*}{\textbf{\gls{fft} Strategy}} & \multicolumn{3}{c|}{\textbf{Transient}} & \multicolumn{3}{c|}{\textbf{Intermittent}} & \multicolumn{3}{c}{\textbf{Mixed}} \\
& MNIST & CIFAR-10 & CIFAR-100 & MNIST & CIFAR-10 & CIFAR-100 & MNIST & CIFAR-10 & CIFAR-100 \\
\hline
\texttt{Centralized(Public)}
& 97.89\tiny{$\pm$0.1337} & 64.92\tiny{$\pm$0.9380} & 26.68\tiny{$\pm$0.8543}
& \textbf{97.89}\tiny{$\pm$0.1337} & 64.92\tiny{$\pm$0.9380} & 26.68\tiny{$\pm$0.8543}
& 97.89\tiny{$\pm$0.1337} & 64.92\tiny{$\pm$0.9380} & 26.68\tiny{$\pm$0.8543} \\
\texttt{FedAvg}~\cite{mcmahan2017communication}
& 96.56\tiny{$\pm$0.8938} & 72.82\tiny{$\pm$1.1859} & 47.30\tiny{$\pm$1.3502} & 94.98\tiny{$\pm$1.7455} & 70.33\tiny{$\pm$1.3170} & 47.77\tiny{$\pm$0.9096} & 93.28\tiny{$\pm$1.1806} & 68.35\tiny{$\pm$0.3418} & 47.85\tiny{$\pm$1.0632}\\
\texttt{FedProx}~\cite{li2020federated}
& 96.60\tiny{$\pm$0.8895} & 72.73\tiny{$\pm$1.1765} & 47.36\tiny{$\pm$1.5320} & 95.00\tiny{$\pm$1.7403} & 70.05\tiny{$\pm$1.4086} & 47.76\tiny{$\pm$0.9871} & 93.45\tiny{$\pm$1.3919} & 68.27\tiny{$\pm$0.3757} & 47.54\tiny{$\pm$1.0059}\\
\texttt{SCAFFOLD}~\cite{karimireddy2020scaffold}
& 97.04\tiny{$\pm$0.9013} & 72.26\tiny{$\pm$0.7260} & 39.80\tiny{$\pm$1.1494} & 96.96\tiny{$\pm$0.5209} & 74.50\tiny{$\pm$0.5466} & 42.96\tiny{$\pm$1.5065} & 97.05\tiny{$\pm$0.4798} & 74.58\tiny{$\pm$0.4641} & 43.87\tiny{$\pm$0.8793}\\
\texttt{FedLAW}~\cite{li2023revisiting}
& 97.24\tiny{$\pm$0.7967} & 76.92\tiny{$\pm$0.9772} & 32.70\tiny{$\pm$29.0020} & 95.80\tiny{$\pm$1.6932} & 36.80\tiny{$\pm$36.7075} & \textbf{54.74}\tiny{$\pm$1.0620}& 94.46\tiny{$\pm$1.4583} & 10.00\tiny{$\pm$0.0000} & 51.40\tiny{$\pm$1.4886}\\
\texttt{TF-Aggregation}~\cite{salehi2021federated}
& 9.80\tiny{$\pm$0.0000} & 10.00\tiny{$\pm$0.0000} & 1.00\tiny{$\pm$0.0000} & 9.80\tiny{$\pm$0.0000} & 10.00\tiny{$\pm$0.0000} & 1.00\tiny{$\pm$0.0000} & 9.80\tiny{$\pm$0.0000} & 10.00\tiny{$\pm$0.0000} & 1.00\tiny{$\pm$0.0000}\\
\texttt{FedAWE}~\cite{xiang2024efficient}
& 73.30\tiny{$\pm$34.0446} & 67.16\tiny{$\pm$1.4748} & 35.08\tiny{$\pm$0.8634} & 73.06\tiny{$\pm$7.1127} & 68.34\tiny{$\pm$0.8812} & 36.44\tiny{$\pm$1.0254} & 67.97\tiny{$\pm$8.9996} & 68.05\tiny{$\pm$1.5370} & 34.60\tiny{$\pm$0.5924}\\
\textbf{\texttt{FedAuto} (Ours)}
& \textbf{98.22}\tiny{$\pm$0.2499} & \textbf{77.37}\tiny{$\pm$0.4824} & \textbf{50.76}\tiny{$\pm$1.5694} & 97.77\tiny{$\pm$0.2156} & \textbf{76.91}\tiny{$\pm$0.6739} & 51.38\tiny{$\pm$1.6592}& \textbf{97.96}\tiny{$\pm$0.2489} & \textbf{77.00}\tiny{$\pm$0.5494} & \textbf{51.43}\tiny{$\pm$1.4919}\\
\hline
\texttt{FedAvg(Ideal)}
& 98.35\tiny{$\pm$0.1599} & 77.19\tiny{$\pm$0.7742} & 50.57\tiny{$\pm$1.3264}& 98.35\tiny{$\pm$0.1599} & 77.19\tiny{$\pm$0.7742} & 50.57\tiny{$\pm$1.3264}& 98.35\tiny{$\pm$0.1599} & 77.19\tiny{$\pm$0.7742} & 50.57\tiny{$\pm$1.3264}\\
\hline
\end{tabular}
}
\end{table*}

\textbf{The i.i.d.\ data case.}
From Fig.~\ref{fig:mnist mixed iid}, \ref{fig:cifar10 mixed iid} and~\ref{fig:cifar100 mixed iid}, together with Table \ref{table:FFT performance iid full}, we observe that across all scenarios, \texttt{FedAuto} consistently achieves performance close to the idealized \texttt{FedAvg(Ideal)}, which assumes a failure-free network.
Most \gls{fft} strategies outperform \texttt{Centralized(Public)} (orange curve), which relies exclusively on server-side public data, and this advantage becomes more pronounced on more complex datasets such as CIFAR-10 and CIFAR-100.
Moreover, classical \texttt{FedAvg} and advanced \texttt{FedProx}, both adopting aggregation weights aligned with the distributed objective function~\eqref{eq:distributed objective function}, converge steadily to the optimal solution, in agreement with the observation \eqref{theorem observation:iid data} of Theorem \ref{theorem:convergence FL failure}. 

In contrast, several advanced \gls{fft} strategies exhibit unstable convergence or suboptimal performance.
\texttt{SCAFFOLD} suffers from performance degradation due to its reliance on dual transmissions of model parameters and control variables, both of which are vulnerable to connection failures.
In the specialized aggregation scheme \texttt{FedLAW}, joint optimization of the introduced shrinking factor and the aggregations weights becomes ineffective when connection failures distort the estimation of client importance, thereby degrading its adaptive aggregation performance.
Strategies explicitly designed to handle connection unreliability also face limitations:
\texttt{TF-Aggregation} incorporates failure probabilities into the aggregation denominator, which leads to instability under high failure rates.
\texttt{FedAWE} increases local stepsizes to compensate for accumulated failure rounds, but becomes ineffective under prolonged connection failures.

\textbf{The non-i.i.d. data case.}
As shown in Fig.~\ref{fig:mnist mixed noniid}, \ref{fig:cifar10 mixed noniid}, and~\ref{fig:cifar100 mixed noniid}, as well as Table~\ref{table:FFT performance non-iid full}, \texttt{FedAuto} consistently maintains stable convergence and high accuracy under non-i.i.d.\ data, demonstrating strong robustness to both data heterogeneity and connection failures.
In contrast, other baselines suffer varying degrees of degradation.
Both \texttt{FedAvg} and \texttt{FedProx}, which perform well under i.i.d.\ data, experience significant drops under non-i.i.d.\ settings, as prolonged client absence induces biased aggregation toward active clients.
These results aligns with the observation~\eqref{theorem observation:Joint effect of unreliability and heterogeneity} of Theorem \ref{theorem:convergence FL failure}.
Besides, \texttt{TF-Aggregation}, \texttt{SCAFFOLD}, \texttt{FedLAW}, and \texttt{FedAWE} all exhibit unstable convergence or suboptimal performance.
A closer comparison of Fig.~\ref{fig:cifar10 mixed noniid} and Fig.~\ref{fig:cifar100 mixed noniid} reveals that \texttt{FedLAW} fails to converge on the simpler CIFAR-10 dataset but performs comparatively better on the more complex CIFAR-100 dataset. 
This behavior stems from the instability of \texttt{FedLAW}’s joint optimization of the shrinking factor and aggregation weights, which relies heavily on client-importance estimation and becomes unreliable under sustained intermittent failures, resulting in task-dependent performance variability.

\subsubsection{Partial Client Participation}\label{section:FedAuto Partial}

We next evaluate \gls{fft} strategies under partial client participation with $K=10$, focusing on the most challenging setting that combines mixed connection failures and non-i.i.d.\ data. 
Testing accuracies are reported Table~\ref{table:FFT performance partial participation}.
Since most \gls{fft} strategies exhibit convergence trends under partial participation that are qualitatively similar to those under full participation (Fig. \ref{fig:performance diff dataset mixed iid full}), the corresponding convergence curves are provided in Appendix~\ref{fig:supp performance partial participation}.
As shown in Table~\ref{table:FFT performance partial participation}, \texttt{FedAuto} continues to converge properly and achieves the best testing accuracy across all datasets, with only modest reductions compared to the full-participation results in Table~\ref{table:FFT performance non-iid full}.
In contrast, several baselines degrade substantially under partial participation.
\texttt{SCAFFOLD} suffers severe performance loss, as reduced participation destabilizes the control variate updates in the presence of connection failures.
\texttt{FedLAW} also deteriorates markedly, since fluctuating client participation disrupts the optimization of shrinking factors and aggregation weights.

\begin{table}[t]
\centering
\caption{
Testing accuracy(\%) of different \gls{fft} strategies under full-parameter fine-tuning with partial participation ($K=10$, mixed failures, non-i.i.d.\ data).
}\label{table:FFT performance partial participation}
{\scriptsize
\begin{tabular}{c|ccc}
\hline
\multirow{2}{*}{\textbf{\gls{fft} Strategy}} & \multicolumn{3}{c}{\textbf{Dataset}} \\
& MNIST & CIFAR-10 & CIFAR-100 \\
\hline
\texttt{Centralized(Public)}
& 97.89\tiny{ $\pm$0.1337} & 64.92\tiny{ $\pm$0.9380} & 26.68\tiny{ $\pm$0.8543} \\
\texttt{FedAvg}~\cite{mcmahan2017communication}
& 93.15\tiny{ $\pm$2.5508} & 67.90\tiny{ $\pm$2.5059} & 49.28\tiny{ $\pm$1.6928}\\
\texttt{FedProx}~\cite{li2020federated}
& 93.10\tiny{ $\pm$2.5112} & 67.53\tiny{ $\pm$2.5610} & 48.99\tiny{ $\pm$1.7383}\\
\texttt{SCAFFOLD}~\cite{karimireddy2020scaffold}
& 14.64\tiny{ $\pm$8.6962} & 10.00\tiny{ $\pm$0.0000} & 1.00\tiny{ $\pm$0.0000}\\
\texttt{FedLAW}~\cite{li2023revisiting}
& 92.39\tiny{ $\pm$3.4460} & 10.00\tiny{ $\pm$0.0000} & 1.00\tiny{ $\pm$0.0000}\\
\texttt{TF-Aggregation}~\cite{salehi2021federated}
& 9.80\tiny{ $\pm$0.0000} & 10.00\tiny{ $\pm$0.0000} & 1.00\tiny{ $\pm$0.0000}\\
\texttt{FedAWE}~\cite{xiang2024efficient}
& 97.16\tiny{ $\pm$0.2792} & 64.15\tiny{ $\pm$6.8564} & 33.24\tiny{ $\pm$0.9452}\\
\textbf{\texttt{FedAuto} (Ours)}
& \textbf{98.12}\tiny{ $\pm$0.1456} & \textbf{75.86}\tiny{ $\pm$0.8343} & \textbf{50.04}\tiny{ $\pm$1.8081}\\
\hline
\texttt{FedAvg(Ideal)}
& 97.84\tiny{ $\pm$0.4446} & 76.54\tiny{ $\pm$1.4070} & 51.95\tiny{ $\pm$1.6991}\\
\hline
\end{tabular}
}
\end{table}

\subsection{Performance under Partial-Parameter Fine-Tuning (LoRA)}\label{section:experiment Partial-Parameter Fine-Tuning}

We further evaluate the robustness of \texttt{FedAuto} under the widely adopted partial-parameter fine-tuning strategy, \gls{lora}, on both CIFAR-10 and CIFAR-100 datasets, focusing on the most challenging setting that combines mixed connection failures with non-i.i.d.\ data.
The resulting convergence curves and testing accuracies are reported in Fig.~\ref{fig:performance diff dataset mixed noniid partial} and Table~\ref{table:FFT performance non-iid LoRA}.

The results show that \texttt{FedAuto} maintains stable convergence and achieves superior testing accuracy on both datasets.
It consistently performs closest to the idealized \texttt{FedAvg(Ideal)}, with more pronounced accuracy gains than \gls{fft} baselines on the more complex CIFAR-100 dataset.
In contrast, on the simpler CIFAR-10 dataset, all \gls{fft} baselines that leverage both public and private data underperform \texttt{Centralized(Public)}, which trains solely on server-side public data.
This indicates that connection failures can disrupt the generalization benefits of pre-trained models, highlighting the fragility of existing \gls{fft} baselines under complex and unreliable network conditions.

\begin{figure}[t]
\begin{minipage}[h]{1\linewidth}
\centering
\includegraphics[width= 3.4 in ]{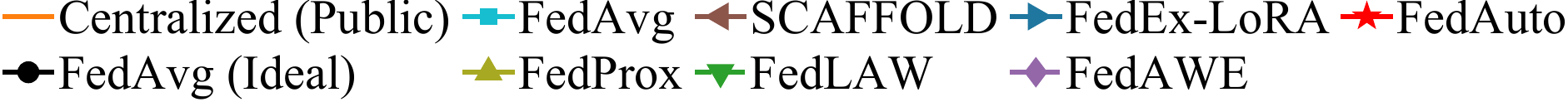}
\end{minipage}
\begin{minipage}[h]{1\linewidth}
\centering
\subfigure[CIFAR-10, non-i.i.d.]{
\includegraphics[width= 1.6 in ]{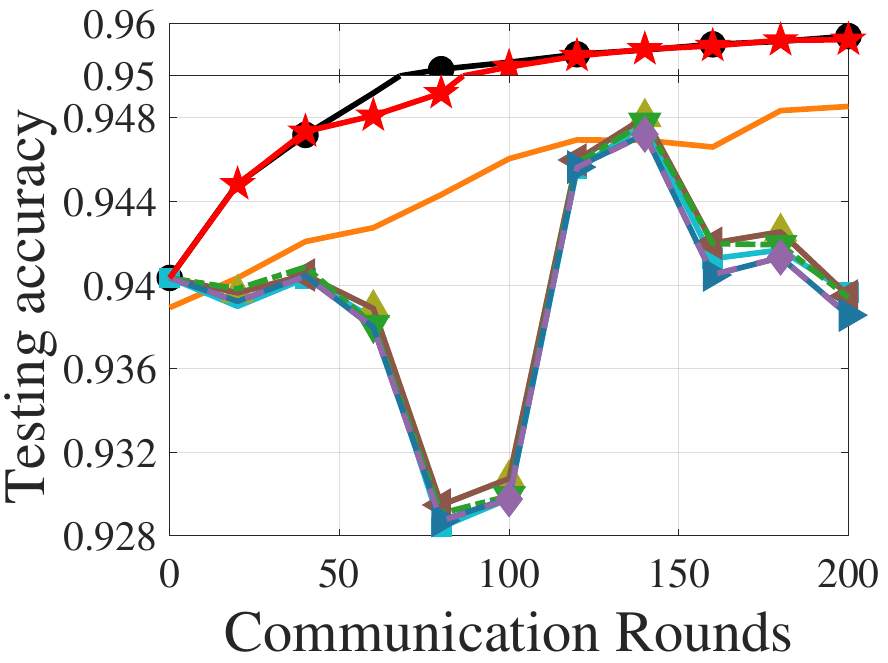}\label{fig:cifar10 mixed noniid LoRA}}
\subfigure[CIFAR-100, non-i.i.d.]{
\includegraphics[width= 1.6 in ]{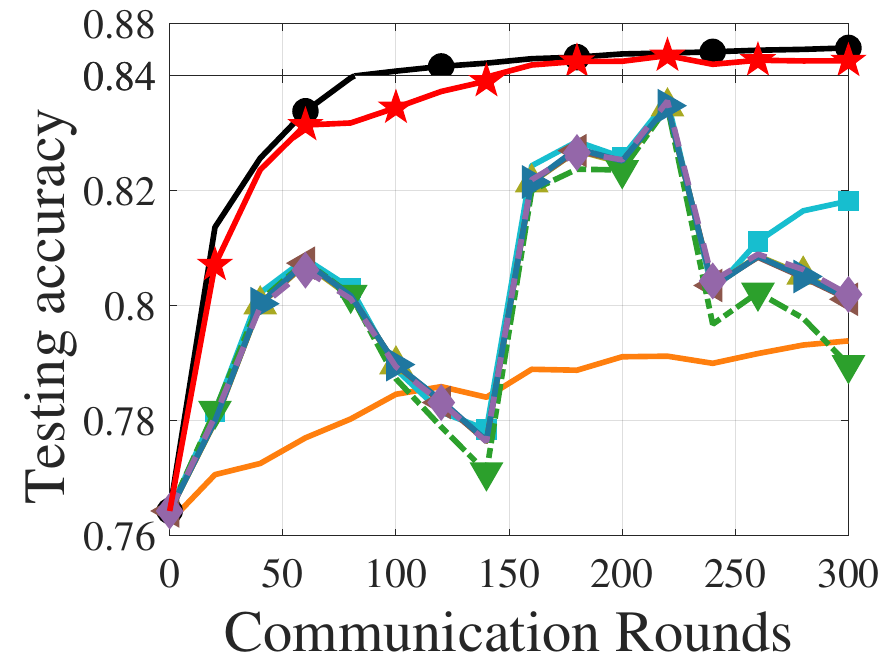}\label{fig:cifar100 mixed noniid LoRA}}
\caption{
Convergence trends of different \gls{fft} strategies under \gls{lora} ($K=20$, mixed failures).
}
\label{fig:performance diff dataset mixed noniid LoRA}
\end{minipage}
\end{figure}

\begin{table}[t]
\centering
\caption{
Testing accuracy(\%) of different \gls{fft} strategies under \gls{lora} ($K=20$, mixed failures, non-i.i.d.\ data).
}\label{table:FFT performance non-iid LoRA}
{\scriptsize 
\begin{tabular}{c|cc}
\hline
\multirow{2}{*}{\textbf{\gls{fft} Strategy}} & \multicolumn{2}{c}{\textbf{Dataset}} \\
& CIFAR-10 & CIFAR-100 \\
\hline
\texttt{Centralized(Public)}
& 94.93 \tiny{$\pm$0.1629} & 79.48 \tiny{$\pm$0.6347} \\
\texttt{FedAvg}~\cite{mcmahan2017communication}
& 93.96 \tiny{$\pm$0.4465} & 81.82 \tiny{$\pm$0.6588} \\
\texttt{FedProx}~\cite{li2020federated}
& 93.95 \tiny{$\pm$0.4426} & 80.14 \tiny{$\pm$1.7574} \\
\texttt{SCAFFOLD}~\cite{karimireddy2020scaffold}
& 93.95 \tiny{$\pm$0.4426} & 80.12 \tiny{$\pm$1.7358} \\
\texttt{FedLAW}~\cite{li2023revisiting}
& 93.94 \tiny{$\pm$0.5050} & 78.97 \tiny{$\pm$2.4693} \\
\texttt{FedAWE}~\cite{xiang2024efficient}
& 93.85 \tiny{$\pm$0.5453} & 80.20 \tiny{$\pm$1.7281} \\
\texttt{FedEx-LoRA}~\cite{singhal2024fedex}
& 93.86 \tiny{$\pm$0.5462} & 80.13 \tiny{$\pm$1.7555}  \\
\textbf{\texttt{FedAuto} (Ours)}
& \textbf{95.69} \tiny{$\pm$0.1321} & \textbf{85.15} \tiny{$\pm$0.2928} \\
\hline
\texttt{FedAvg(Ideal)}
& 95.76 \tiny{$\pm$0.1310} & 86.14 \tiny{$\pm$0.2066} \\
\hline
\end{tabular}
}
\end{table}

\subsection{FedAuto v.s. Resource Allocation Optimization}\label{section:FedAuto resource}

\begin{figure}[t]
\begin{minipage}[h]{1\linewidth}
\centering
\includegraphics[width= 3.4 in ]{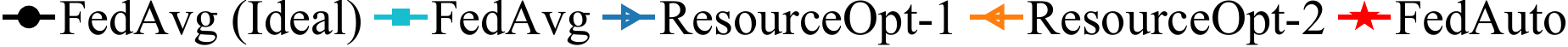}
\end{minipage}

\begin{minipage}[h]{1\linewidth}
\centering
\subfigure[MNIST, transient.]{
\includegraphics[width= 1.6 in ]{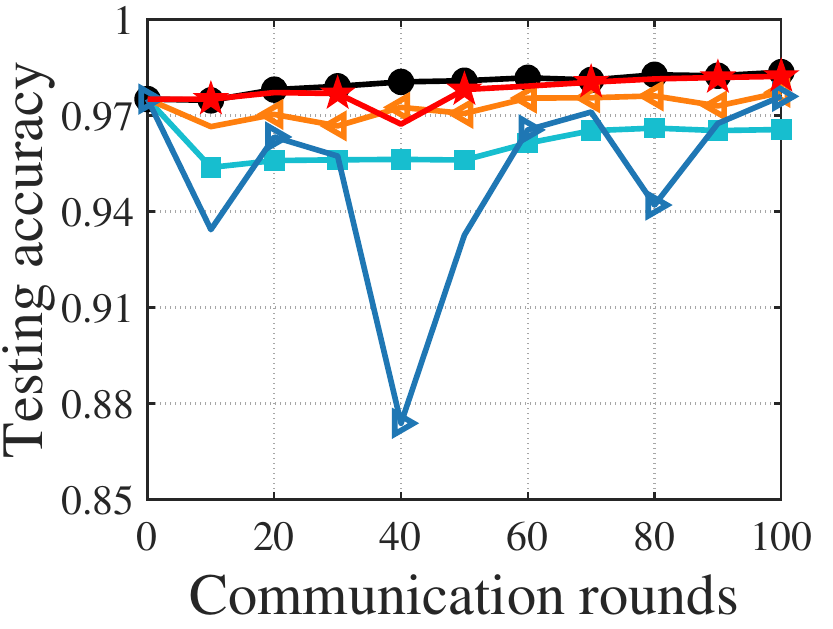}\label{fig:mnist ResourceOpt transient}}
\subfigure[MNIST, mixed.]{
\includegraphics[width= 1.6 in ]{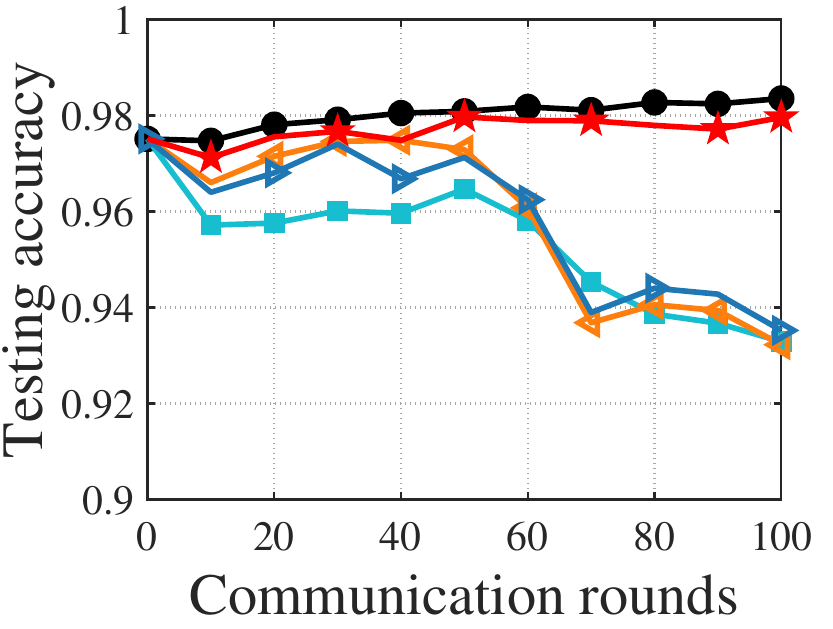}\label{fig:mnist ResourceOpt mixed}}
\subfigure[CIFAR-10, transient.]{
\includegraphics[width= 1.6 in ]{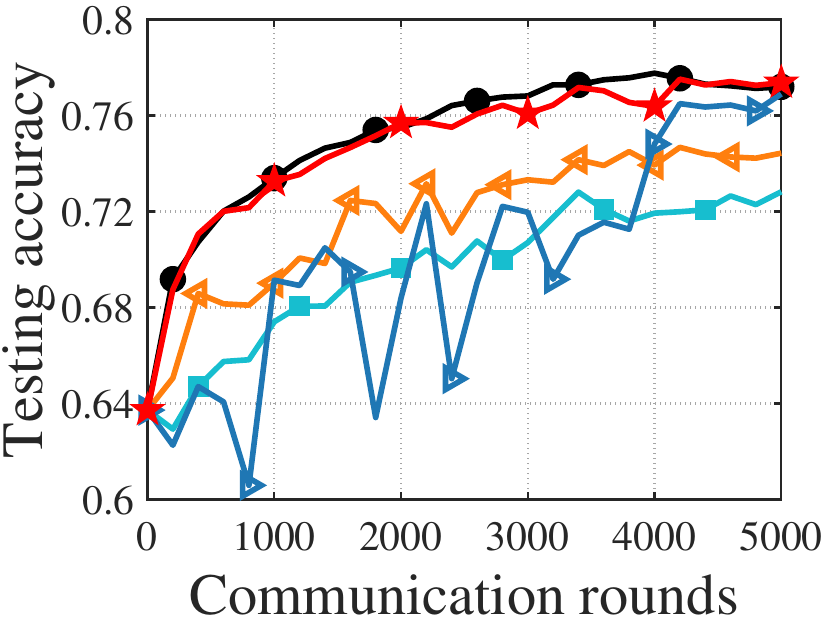}\label{fig:cifar10 ResourceOpt transient}}
\subfigure[CIFAR-10, mixed.]{
\includegraphics[width= 1.6 in ]{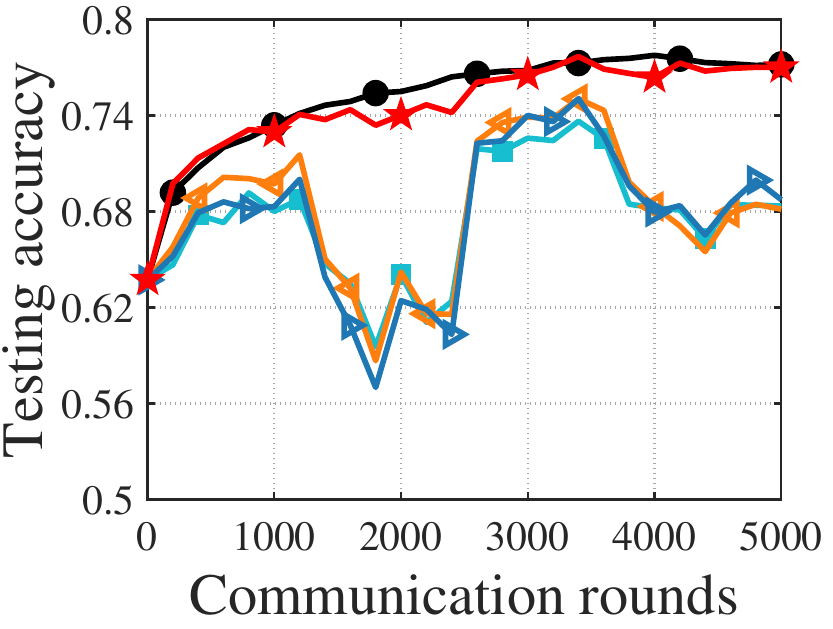}\label{fig:cifar10 ResourceOpt mixed}}

\subfigure[CIFAR-100, transient.]{
\includegraphics[width= 1.6 in ]{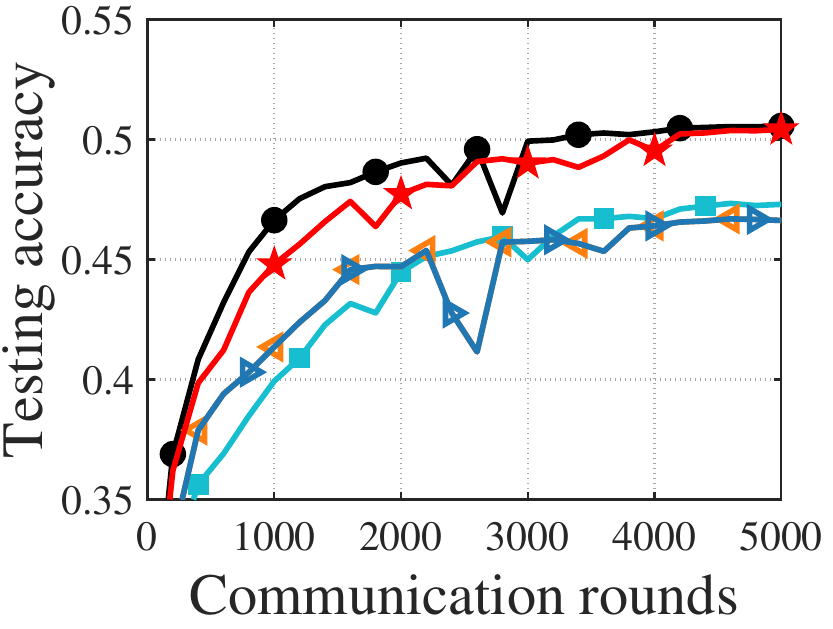}\label{fig:cifar100 ResourceOpt transient}}
\subfigure[CIFAR-100, mixed.]{
\includegraphics[width= 1.6 in ]{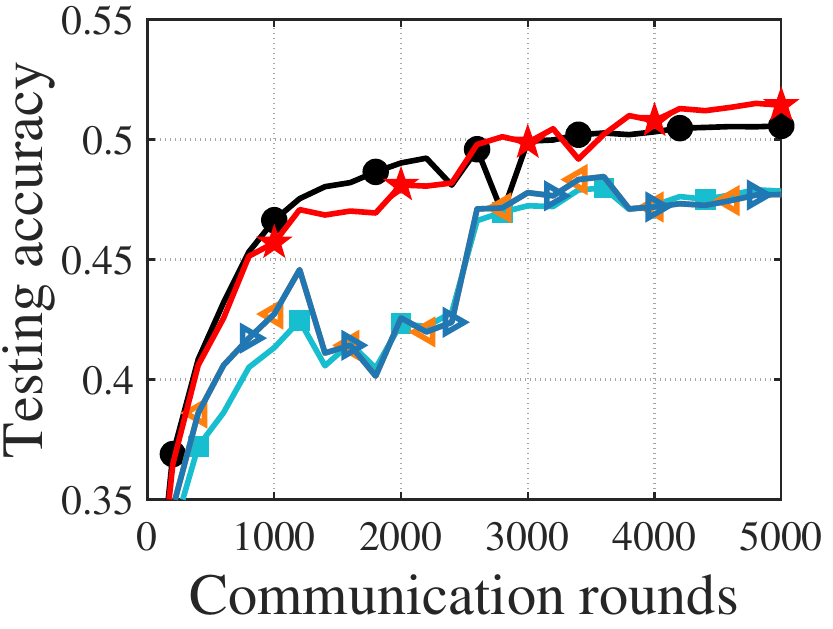}\label{fig:cifar100 ResourceOpt mixed}}

\caption{
Convergence trends of \texttt{FedAuto} and resource allocation-based baselines (full-parameter fine-tuning, $K=20$, non-i.i.d.).}
\label{fig:performance resource allocation}
\end{minipage}
\end{figure}

In this subsection, we compare \texttt{FedAuto} with resource allocation-based baselines to further validate its superiority.
Since communication resource optimization can only mitigate transient failures caused by short-term channel fluctuations (e.g., transmission outages in our simulations), but cannot address intermittent failures due to device faults or battery depletion, we consider only transient and mixed failure scenarios involving transient failures. 
The corresponding results are shown in Fig.~\ref{fig:performance resource allocation}.

As illustrated in Fig.~\ref{fig:performance resource allocation}, \texttt{ResourceOpt-1} exhibits large performance fluctuations.
To equalize connection probabilities, it allocates more communication resources to the clients with poor channel conditions and less to those with good conditions, constraining all clients by the worst connections and reducing the number of successfully connected clients per communication round. 
\texttt{ResourceOpt-2} achieves suboptimal convergence due to cross-standard differences in connection failure probabilities, leading to imbalanced class contributions during global aggregation.
In contrast, \texttt{FedAuto} achieves stable and accurate convergence under different failure modes without requiring network-level resource interventions.

\subsection{Ablation Studies}\label{section:Ablation Studies}

We finally conduct ablation studies to validate the necessity of each module in the adaptive aggregation strategy of \texttt{FedAuto} (Algorithm \ref{algorithm:FedAuto}).
The experiments are performed under challenging training conditions with mixed failures and non-i.i.d. data, and the results are summarized in Table \ref{table:Ablation Study}.

\begin{table*}[t]
\centering
\caption{
Testing accuracy (\%) of ablation studies ($K=20$, mixed failures, non-i.i.d. data).
}\label{table:Ablation Study}
{\scriptsize
\begin{tabular}{cc|ccc|ccc}
\hline
\multicolumn{2}{c|}{\textbf{Settings}} & \multicolumn{3}{c|}{\textbf{Full-Parameter Fine-Tuning}} & \multicolumn{3}{c}{\textbf{Partial-Parameter Fine-Tuning (LoRA)}} \\
Server-Side Compensatory Training & Aggregation Weight Optimization & \multirow{2}{*}{MNIST} & \multirow{2}{*}{CIFAR-10} & \multirow{2}{*}{CIFAR-100} & \multirow{2}{*}{MNIST} & \multirow{2}{*}{CIFAR-10} & \multirow{2}{*}{CIFAR-100} \\
(Module 1) & (Module 2) & & & \\
\hline
\ding{55} & \ding{55} & 93.28\tiny{$\pm$1.1806} & 68.35\tiny{$\pm$0.3418} & 47.85\tiny{$\pm$1.0632} & 93.87\tiny{$\pm$0.4356} & 93.96\tiny{$\pm$0.4465} & 81.82\tiny{$\pm$0.6588} \\
\ding{51} & \ding{55} & 96.68\tiny{$\pm$0.2654} & 73.83\tiny{$\pm$0.5587} & 49.75\tiny{$\pm$1.3102} & 96.15\tiny{$\pm$0.2168} & 95.13\tiny{$\pm$0.2480} & 83.61\tiny{$\pm$0.7697} \\
\ding{55} & \ding{51} & 94.06\tiny{$\pm$1.5616} & 72.21\tiny{$\pm$0.7936} & 47.88\tiny{$\pm$1.0179} & 95.10\tiny{$\pm$0.5654} & 94.96\tiny{$\pm$0.3224} & 82.28\tiny{$\pm$1.9158} \\
\ding{51} & \ding{51} & \textbf{97.96}\tiny{$\pm$0.2489} & \textbf{77.00}\tiny{$\pm$0.5494} & \textbf{51.43}\tiny{$\pm$1.4919} & \textbf{97.06}\tiny{$\pm$0.0750} & \textbf{95.69}\tiny{$\pm$0.1321} & \textbf{85.15}\tiny{$\pm$0.2928} \\
\hline
\end{tabular}
}
\end{table*}

\subsubsection{Impact of Server-Side Compensation Training}

When server-side compensatory training is disabled, the aggregation weight $\beta^r_{\text{miss}}$ in \eqref{global_model_failure_2} is set to zero, and the associated parameters $\beta^r_{\text{miss}}$ and $\alpha^{r}_{\text{miss},c}$ are removed from the optimization problem in \eqref{eq:aggregation_optimization_2}.
Detailed formulations are provided in Appendix~\ref{appendix:Formulations Ablation Studies}.
Comparing the first and second rows of Table~\ref{table:Ablation Study}, it is evident that server-side compensatory training markedly improves testing accuracy by mitigating the absence of class-specific local updates caused by unreliable connections.
Furthermore, contrasting with third and fourth rows shows that omitting compensatory training from the adaptive aggregation strategy results in significant performance degradation.
Although aggregation weight optimization can partially balance the contributions of different data classes, it remains insufficient to compensate for the missing classes, leaving them underrepresented in the global aggregation.

\subsubsection{Impact of Aggregation Weight Optimization}

In the absence of aggregation weight optimization, the aggregation weights in \eqref{global_model_failure_2} reduce to simple averaging.
The comparison between the first and third rows of Table~\ref{table:Ablation Study} demonstrates that aggregation weight optimization improves testing accuracy by better balancing class contributions in the global aggregation.
Moreover, comparing the second and fourth rows reveals that while compensatory training alone achieves testing accuracy close to the full strategy on the simpler MNIST dataset, a notable gap remains for the more complex CIFAR-100 dataset.
This is because MNIST’s simpler feature space allows compensatory training to effectively enhance data diversity and class balance, whereas datasets with richer feature representations require aggregation weight optimization to further balance class contributions and achieve optimal performance.

Overall, these results highlight that both server-side compensatory training and aggregation weight optimization are crucial for improving \gls{fft} accuracy under unreliable network conditions.
They play complementary roles: compensatory training offsets missing classes, while weight optimization ensures balanced contributions across data classes.

\section{Conclusion}\label{section:Conclusion} 

This paper investigated the fundamental challenges of \gls{fft} in heterogeneous and unreliable network environments, a common yet underexplored setting in practical communication systems. 
Through theoretical analysis (Theorem~\ref{theorem:convergence FL failure}) and empirical evaluations (e.g., Fig.~\ref{fig:FFT performance CIFAR10 mixed}), we showed that the combined effects of connection unreliability and data heterogeneity can severely impair \gls{fft} convergence. 
To address these challenges, we proposed \texttt{FedAuto}, a robust \gls{fft} framework that mitigates performance degradation through adaptive aggregation alone. 
\texttt{FedAuto} requires neither prior knowledge of connection failure statistics nor no modification to existing network infrastructure, enabling seamless plug-and-play deployment in heterogeneous and unreliable networks. 
We further established a rigorous convergence analysis that provides per-realization convergence guarantees for \texttt{FedAuto}. 
Extensive experiments on fine-tuning pre-trained models demonstrate that \texttt{FedAuto} consistently outperforms state-of-the-art \gls{fft} baselines, underscoring the importance of aggregation-aware design for enhancing practical deployability of \gls{fft}. 
Future work will explore extensions to personalized \gls{fl} with heterogeneous model architectures, as well as semi-supervised learning under time-varying and class-incremental data distributions.

\bibliographystyle{IEEEtran}
\bibliography{refs_journal}

%
%

\vfill

\clearpage

\onecolumn

\appendices

\section{Proof of Remark \ref{remark: aggregation weights full partial}}\label{appendix:simple averaging partial participation}

Under reliable network conditions (${\mathds{1}^{r}_{i}}=1$ $\forall i,r$), global aggregation follows \eqref{eq:global aggregation}.
In the case of full participation, where all clients and the server contribute to the aggregation, the aggregation weights are assigned as $\beta_s^r = p_s$ and $\beta_i^r = p_i$ $\forall i \in [N]$.
The resulting global model is expressed as

\vspace{-0.3cm}
\begin{small}
\begin{align}\label{eq:global aggregation full p}
{\bar {\mathbf{w}}}_{r} 
= 
p_s \mathbf{w}^{r,E}_{s} 
+ \sum_{i=1}^N p_i \mathbf{w}^{r,E}_{i},
\end{align}
\end{small}

\noindent
which aligns exactly with the distributed learning objective in \eqref{eq:distributed objective function}.

Differently, under partial participation, a subset of $K$ clients is randomly sampled with replacement from the total $N$ clients at each round $r$.
The probability of selecting client $i \in [N]$ is $\frac{p_i}{1-p_s}$.
In this setting, the aggregation weight of the server remains $\beta_s^r = p_s$, while each selected client $i \in \mathcal{K}_r$ is assigned a uniform weight $\beta_i^r = \frac{1-p_s}{K}$.
The aggregated global model is then given by

\vspace{-0.3cm}
\begin{small}
\begin{align}\label{eq:global aggregation partial p}
{\bar {\mathbf{w}}}_{r} 
= 
p_s \mathbf{w}^{r,E}_{s} 
+ \sum_{i \in {\mathcal{K}}_{r}} \frac{1-p_s}{K} \mathbf{w}^{r,E}_{i}.
\end{align}
\end{small}

Because clients are sampled independently according to the defined probability distribution, the expectation of the aggregated global model over the randomness of $\mathcal{K}_r$ is

\vspace{-0.3cm}
\begin{small}
\begin{align}\label{eq:global aggregation unbiased estimate}
\mathbb{E}_{\mathcal{K}_r}[{\bar {\mathbf{w}}}_{r}]
=
p_s \mathbf{w}^{r,E}_{s} 
+ 
\frac{1-p_s}{K} 
\sum_{i \in {\mathcal{K}}_{r}} 
\mathbb{E}_{i \in [N]} [ \mathbf{w}^{r,E}_{i} ]
= 
p_s \mathbf{w}^{r,E}_{s} 
+ 
\frac{1-p_s}{K} 
\sum_{i \in {\mathcal{K}}_{r}}
\bigg(
\sum_{i=1}^N \frac{p_i}{1-p_s} \mathbf{w}^{r,E}_{i}
\bigg)
= 
p_s \mathbf{w}^{r,E}_{s} 
+ \sum_{i=1}^N p_i \mathbf{w}^{r,E}_{i}.
\end{align}
\end{small}

\noindent
Thus, the aggregation strategy in \eqref{eq:global aggregation partial p} yields an unbiased estimate of the global model obtained under full participation, as defined in \eqref{eq:global aggregation full p}.
\hfill $\blacksquare$

\section{Proof of Theorem \ref{theorem:convergence FL failure}}\label{appendix:Proof of convergence rate theorem}

For notational clarity in the subsequent derivations, we unify unify the indices of the server and clients by introducing $j \in \{s,[N]\}$, where $j=s$ denotes the server and $j \in [N]$ refers to the clients.

\subsection{Proof of Convergence Rate}

Under Assumption \ref{assumption:L continuous}, we have

\vspace{-0.3cm}
\begin{small}
\begin{align}\label{ineq:F_w_r_bound}
F({\bar {\mathbf{w}}}_{r})
\leq &
F({\bar {\mathbf{w}}}_{r-1})
+ \langle\nabla F_g({\bar {\mathbf{w}}}_{r-1}), {\bar {\mathbf{w}}}_{r} - {\bar {\mathbf{w}}}_{r-1}\rangle
+ \frac{L}{2} \| {\bar {\mathbf{w}}}_{r} - {\bar {\mathbf{w}}}_{r-1}  \|^2
.
\end{align}
\end{small}

We also need the following three key lemmas, which are proved in subsequent subsections.

\begin{lemma}\label{lemma:bound nabla F wr wr-1}
Under Assumptions \ref{assumption:class heterogeneity}, \ref{assumption:garident norm} and \ref{assumption:L continuous}, the following bound holds:

\vspace{-0.3cm}
\begin{small}
\begin{align}\label{ineq:bound nabla F wr wr-1}
&
\left\langle \nabla F_g({\bar {\mathbf{w}}}_{r-1}), {\bar {\mathbf{w}}}_{r} - {\bar {\mathbf{w}}}_{r-1} \right\rangle 
\notag \\
\leq 
&
- \frac{\gamma E}{2} {\sum}_{t=1}^E \left \| \nabla F_g(\bar{\mathbf{w}}_{r-1}) \right \| ^2 
- \frac{\gamma}{2} {\sum}_{t = 1}^{E} 
\Big\|  
{\sum}_{j \in \{s,[N]\}} \beta_{j}^{r} \nabla F_{j} (\mathbf{w}^{r,t-1}_{j}) 
\Big\|^2 
\notag \\
&
+ 2 \gamma E \chi^2_{\mathbf{p}\|\bm{\beta}} {\sum}_{c=1}^C {\sum}_{j \in \{s, [N]\}} p_j \alpha_{j,c} V_{j,c}^2
+ 2 \gamma E \chi^2_{\bm{\alpha}_g \| \bm{\tilde \alpha}^r} G^2 
+ 
\gamma L^2 {\sum}_{j \in \{s, [N]\}} \beta_{j}^r {\sum}_{t=1}^E \left \| {\mathbf{w}}_{j}^{r,t-1} - \bar{\mathbf{w}}_{r-1} \right \|^2
,
\end{align}
\end{small}

\noindent
where $\chi^2_{\mathbf{p}\|\bm{\beta}}$ and $\chi^2_{\bm{\alpha}_g \| \bm{\tilde \alpha}^r}$ denote the chi-square divergences defined in (\ref{eq:convergence bound of FL failure}).
\end{lemma}

\begin{lemma}\label{lemma:Proof_Thm1_lemma_2}
The deviation between the local model at round $r$ and the global model from the previous round is bounded by

\vspace{-0.3cm}
\begin{small}
\begin{align}\label{Proof_Thm1_lemma_2_formulation}
{\sum}_{t=1}^E \big\|{\mathbf{w}}_{j}^{r,t-1} - \bar{\mathbf{w}}_{r-1} \big\|^2
\leq 
\frac{4\gamma^2 E^3}{2-3\gamma^2 E^2 L^2} 
{\sum}_{c=1}^C \big( \alpha_{j,c} V_{j,c}^2 + {\chi}_{\bm{\alpha}_g || \bm{\alpha}_j}^2 G^2 \big)
+ \frac{2\gamma^2 E^3}{2-3\gamma^2 E^2 L^2} \big\| \nabla F_g(\bar{\mathbf{w}}_{r-1}) \big\|^2
.
\end{align}
\end{small}
\end{lemma}

\begin{lemma}\label{lemma:Proof_Thm1_lemma_3}
Based on the local updating rules in (\ref{eq:local updating client}) and (\ref{eq:local updating server}) and the global aggregation rule (\ref{eq:global aggregation}), the difference between the global models at two consecutive rounds satisfies

\vspace{-0.3cm}
\begin{small}
\begin{align}\label{eq:wr-wr-1}
{\bar {\mathbf{w}}}_{r} - {\bar {\mathbf{w}}}_{r-1}
=
- \gamma {\sum}_{j \in \{s,[N]\}} \beta_{j}^{r} {\sum}_{t=1}^{E} \nabla F_j (\mathbf{w}_j^{r, t-1})
,
\end{align}
\end{small}

\noindent
which implies

\vspace{-0.3cm}
\begin{small}
\begin{align}\label{eq:wr-wr-1 bound}
\big\| {\bar {\mathbf{w}}}_{r} - {\bar {\mathbf{w}}}_{r-1} \big\|^2
=
\gamma^2 \Big\| {\sum}_{t=1}^{E} {\sum}_{j \in \{s,[N]\}} \beta_{j}^{r} \nabla F_j (\mathbf{w}_j^{r, t-1}) \Big\|^2
\overset{(a)}{\leq}
\gamma^2 E {\sum}_{t=1}^{E} \Big\| {\sum}_{j \in \{s,[N]\}} \beta_{j}^{r} \nabla F_j (\mathbf{w}_j^{r, t-1}) \Big\|^2
,
\end{align}
\end{small}

\noindent
where inequality (a) follows from the Cauchy–Schwarz inequality.
\end{lemma}

Substituting \eqref{ineq:bound nabla F wr wr-1} into the second term on the \gls{rhs} of \eqref{ineq:F_w_r_bound}, \eqref{eq:wr-wr-1 bound} into the third term, and incorporating \eqref{Proof_Thm1_lemma_2_formulation}, we obtain

\vspace{-0.3cm}
\begin{small}
\begin{align}\label{ineq:Proof_Fwr_less_Fwr1}
F(\bar{\mathbf{w}}_r) 
\leq & 
F(\bar{\mathbf{w}}_{r-1}) - \frac{\gamma E }{2} \big \| \nabla F_g(\bar{\mathbf{w}}_{r-1}) \big \| ^2 
- \underbrace{\frac{\gamma(1-\gamma EL)}{2}}_{\rm (\ref{ineq:Proof_Fwr_less_Fwr1}b)} {\sum}_{t = 1}^{E} 
\Big\|  
{\sum}_{j \in \{s,[N]\}} \beta_{j}^{r} \nabla F_{j} (\mathbf{w}^{r,t-1}_{j}) 
\Big\|^2  
\notag \\
& 
+ 2 \gamma E \chi^2_{\mathbf{p}\|\bm{\beta}} {\sum}_{c=1}^C {\sum}_{j \in \{s, [N]\}} p_j \alpha_{j,c} V_{j,c}^2
+ 2 \gamma E  \chi^2_{\bm{\alpha}_g \| \bm{\tilde \alpha}^r} G^2
\notag \\
& 
+ \frac{4\gamma^3 E^3 L^2}{2-3\gamma^2 E^2 L^2} {\sum}_{j \in \{s, [N]\}} \beta_{j}^r 
{\sum}_{c=1}^C \big( \alpha_{j,c} V_{j,c}^2 + {\chi}_{\bm{\alpha}_g || \bm{\alpha}_j}^2 G^2 \big)
+ \frac{2\gamma^3 E^3 L^2}{2-3\gamma^2 E^2 L^2} \left\| \nabla F_g(\bar{\mathbf{w}}_{r-1}) \right\|^2
\notag \\
\overset{(a)}{\leq} & 
F(\bar{\mathbf{w}}_{r-1}) 
-
\Big ( \frac{\gamma E }{2} - \frac{2\gamma^3 E^3 L^2}{2-3\gamma^2 E^2 L^2} \Big ) \big \| \nabla F_g(\bar{\mathbf{w}}_{r-1}) \big \| ^2 
+ 2 \gamma E \chi^2_{\mathbf{p}\|\bm{\beta}} {\sum}_{c=1}^C {\sum}_{j \in \{s, [N]\}} p_j \alpha_{j,c} V_{j,c}^2
+ 2 \gamma E \chi^2_{\bm{\alpha}_g \| \bm{\tilde \alpha}^r} G^2 
\notag \\
& 
+ 
\frac{4\gamma^3 E^3 L^2}{2-3\gamma^2 E^2 L^2} {\sum}_{j \in \{s, [N]\}} \beta_{j}^r 
{\sum}_{c=1}^C \big( \alpha_{j,c} V_{j,c}^2 + {\chi}_{\bm{\alpha}_g || \bm{\alpha}_j}^2 G^2 \big)
,
\end{align}
\end{small}

\noindent
where inequality (a) holds by setting $\gamma EL \leq 1$, ensuring that (\ref{ineq:Proof_Fwr_less_Fwr1}b) $\geq 0$.

Next, summing the above inequality from $r = 1$ to $R$ and dividing both sides by $\gamma T$ (where $T = RE$ is the total number of local gradient descent steps) yields

\vspace{-0.3cm}
\begin{small}
\begin{align}\label{Proof_Thm1_formulation}
&
\underbrace{\Big( \frac{1}{2} - \frac{2\gamma^2E^2L^2}{2-3\gamma^2E^2L^2} \Big)}_{\rm \triangleq (\ref{Proof_Thm1_formulation}a)}\frac{1}{R}{\sum}_{r=1}^R \left \| \nabla F_g(\bar{\mathbf{w}}_{r-1}) \right \|^2
\notag \\
\leq & 
\underbrace{\frac{1}{\gamma T}}_{\rm \triangleq (\ref{Proof_Thm1_formulation}b)} \big( F(\bar{\mathbf{w}}_{0}) - F(\bar{\mathbf{w}}_R) \big) 
+ \underbrace{\frac{4\gamma^2 E^2 L^2}{2-3\gamma^2E^2L^2}}_{\rm \triangleq (\ref{Proof_Thm1_formulation}c)}\frac{1}{R} {\sum}_{r=1}^R {\sum}_{j \in \{s, [N]\}} \beta_{j}^r 
{\sum}_{c=1}^C \big( \alpha_{j,c} V_{j,c}^2 + {\chi}_{\bm{\alpha}_g || \bm{\alpha}_j}^2 G^2 \big)
\notag \\
&
+ 2 \chi^2_{\mathbf{p}\|\bm{\beta}} {\sum}_{c=1}^C {\sum}_{j \in \{s, [N]\}} p_j \alpha_{j,c} V_{j,c}^2
+ \frac{2}{R} {\sum}_{r=1}^R \chi^2_{\bm{\alpha}_g \| \bm{\tilde \alpha}^r} G^2 
.
\end{align}
\end{small}

Let the learning rate be $\gamma = N^{\frac{1}{2}}/(2LT^{\frac{1}{2}}) $ and the number of local updating steps satisfy $E \leq T^{\frac{1}{4}}/N^{\frac{3}{4}}$, where $T \geq N^{3}$ ensures $E \geq 1$.
Since $T \geq 1$ and $N \geq 1$, these parameter settings guarantee 

\vspace{-0.3cm}
\begin{small}
\begin{align}
\gamma E L 
\leq \frac{1}{2(TN)^{\frac{1}{4}}}
\leq \frac{1}{2} < 1
,
\end{align}
\end{small}

\noindent
satisfying the condition required for inequality (a) in \eqref{ineq:Proof_Fwr_less_Fwr1}.
Moreover, we obtain ${\rm (\ref{Proof_Thm1_formulation}b)} = 2L/\sqrt{TN}$ and

\vspace{-0.3cm}
\begin{small}
\begin{subequations}
\begin{align}
{\rm (\ref{Proof_Thm1_formulation}c)}
\leq &
\frac{\frac{1}{\sqrt{TN}}}{2 - \frac{3}{4\sqrt{TN}}}
\overset{(a)}{\leq}
\frac{\frac{1}{\sqrt{TN}}}{2 - \frac{3}{4}}
=
\frac{4}{5\sqrt{TN}}
,
\label{eq:Proof_Thm1_formulation_c_value}
\\
{\rm (\ref{Proof_Thm1_formulation}a)}
= &
\frac{1}{2} - \frac{\rm (\ref{Proof_Thm1_formulation}c)}{2}
\geq 
\frac{1}{2}
\Big(
1 -  \frac{4}{5\sqrt{TN}}
\Big)
\geq 
\frac{1}{2}
\Big(
1 -  \frac{4}{5}
\Big)
=
\frac{1}{10}
,
\end{align}
\end{subequations}
\end{small}

\noindent
where inequality (a) in \eqref{eq:Proof_Thm1_formulation_c_value} is due to $TN \geq 1$.

Finally, substituting the above coefficients and $\bar{\mathbf{w}}_{0} = \bar{\mathbf{w}}_{\rm pre}$ into \eqref{Proof_Thm1_formulation} completes the proof of Theorem \ref{theorem:convergence FL failure}.
\hfill $\blacksquare$

\subsection{Proof of Lemma \ref{lemma:bound nabla F wr wr-1}}

From \eqref{eq:wr-wr-1}, we have

\vspace{-0.3cm}
\begin{small}
\begin{align}\label{eq:bound nabla F wr wr-1}
& \langle \nabla F_g({\bar {\mathbf{w}}}_{r-1}), {\bar {\mathbf{w}}}_{r} - {\bar {\mathbf{w}}}_{r-1} \rangle
\notag \\
=
& \bigg\langle \nabla F_g({\bar {\mathbf{w}}}_{r-1}),
- \gamma {\sum}_{j \in \{s,[N]\}} \beta_{j}^{r} {\sum}_{t=1}^{E} \nabla F_j (\mathbf{w}_j^{r, t-1})
\Big\rangle
\notag \\
=
&
- \gamma {\sum}_{t = 1}^{E} \Big\langle\nabla F_g({\bar {\mathbf{w}}}_{r-1}),
{\sum}_{j \in \{s,[N]\}} \beta_{j}^{r} \nabla F_{j} (\mathbf{w}^{r,t-1}_{j}) \Big\rangle
\notag \\
\overset{(a)}{=}
&
- \frac{\gamma}{2} {\sum}_{t = 1}^{E} \| \nabla F_g({\bar {\mathbf{w}}}_{r-1}) \|^2
- \frac{\gamma}{2} {\sum}_{t = 1}^{E} 
\Big\|  
{\sum}_{j \in \{s,[N]\}} \beta_{j}^{r} \nabla F_{j} (\mathbf{w}^{r,t-1}_{j}) 
\Big\|^2 
+ \frac{\gamma}{2} {\sum}_{t = 1}^{E}
\Big\| 
\nabla F_g({\bar {\mathbf{w}}}_{r-1}) 
-  
{\sum}_{j \in \{s,[N]\}} \beta_{j}^{r} \nabla F_{j} (\mathbf{w}^{r,t-1}_{j}) 
\Big\|^2
\notag \\
\overset{(b)}{\leq}
&
- \frac{\gamma E}{2} \| \nabla F_g({\bar {\mathbf{w}}}_{r-1}) \|^2
- \frac{\gamma}{2} {\sum}_{t = 1}^{E} 
\Big\|  
{\sum}_{j \in \{s,[N]\}} \beta_{j}^{r} \nabla F_{j} (\mathbf{w}^{r,t-1}_{j}) 
\Big\|^2 
\notag \\
&
+ 
\gamma E
\Big\| 
\nabla F_g(\bar{\mathbf{w}}_{r-1}) 
- {\sum}_{j \in \{s, [N]\}} \beta_{j}^{r} \nabla F_{j} (\bar{\mathbf{w}}_{r-1}) 
\Big \|^2
+
\gamma {\sum}_{t = 1}^{E}
\Big\| 
{\sum}_{j \in \{s, [N]\}} \beta_{j}^{r} \nabla F_{j} (\bar{\mathbf{w}}_{r-1}) 
- {\sum}_{j \in \{s, [N]\}} \beta_{j}^{r} \nabla F_{j} (\mathbf{w}^{r,t-1}_{j}) 
\Big \|^2
\notag \\
=
&
- \frac{\gamma E}{2} \| \nabla F_g({\bar {\mathbf{w}}}_{r-1}) \|^2
- \frac{\gamma}{2} {\sum}_{t = 1}^{E} 
\Big\|  
{\sum}_{j \in \{s,[N]\}} \beta_{j}^{r} \nabla F_{j} (\mathbf{w}^{r,t-1}_{j}) 
\Big\|^2 
\notag \\
&
+ 
\gamma E
\underbrace{
\Big\| 
{\sum}_{j \in \{s, [N]\}} ( p_{j} - \beta_{j}^{r} ) \nabla F_{j}(\bar{\mathbf{w}}_{r-1})
\Big\| ^2
}_{\rm \triangleq (\ref{eq:bound nabla F wr wr-1}c)}
+
\gamma {\sum}_{t = 1}^{E}
\underbrace{
\Big\| 
{\sum}_{j \in \{s, [N]\}} \beta_{j}^{r} \big( \nabla F_{j} (\bar{\mathbf{w}}_{r-1}) - \nabla F_{j} (\mathbf{w}^{r,t-1}_{j}) \big) 
\Big\|^2
}_{\rm \triangleq (\ref{eq:bound nabla F wr wr-1}d)}
,
\end{align}
\end{small}

\noindent
where equality (a) follows from the identity $\langle {\mathbf{x}}_1,{\mathbf{x}}_2 \rangle = \frac{1}{2}( \| {\mathbf{x}}_1 \|^2 + \| {\mathbf{x}}_2 \|^2 - \| {\mathbf{x}}_1 - {\mathbf{x}}_2 \|^2 )$, and inequality (b) is derived from the property $\| {\mathbf{x}}_1 + {\mathbf{x}}_2 \|^2 \leq 2\| {\mathbf{x}}_1 \|^2 + 2\| {\mathbf{x}}_2 \|^2$.

In \eqref{eq:bound nabla F wr wr-1}, term (\ref{eq:bound nabla F wr wr-1}c) can be further bounded as

\vspace{-0.3cm}
\begin{small}
\begin{align}\label{eq:bound A11}
{\rm (\ref{eq:bound nabla F wr wr-1}c)}
\overset{(a)}{=} 
& 
\Big \| 
{\sum}_{c=1}^C {\sum}_{j \in \{s, [N]\}} (p_j \alpha_{j,c} - \beta_{j}^r \alpha_{j,c})\nabla F_{j,c} (\bar{\mathbf{w}}_{r-1}) 
\Big \|^2 
\notag \\
=
& 
\Big \| 
{\sum}_{c=1}^C {\sum}_{j \in \{s, [N]\}} (p_j \alpha_{j,c} - \beta_{j}^r \alpha_{j,c})\nabla F_{j,c} (\bar{\mathbf{w}}_{r-1}) 
- {\sum}_{c=1}^C {\sum}_{j \in \{s, [N]\}} (p_j \alpha_{j,c} - \beta_{j}^r \alpha_{j,c}) \nabla F_{g,c}(\bar{\mathbf{w}}_{r-1})
\notag \\
& 
+ {\sum}_{c=1}^C {\sum}_{j \in \{s, [N]\}} (p_j \alpha_{j,c} - \beta_{j}^r \alpha_{j,c}) \nabla F_{g,c}(\bar{\mathbf{w}}_{r-1})
\Big\|^2 
\notag \\
\overset{(b)}{\leq} 
& 
2 
\Big \| 
{\sum}_{c=1}^C {\sum}_{j \in \{s, [N]\}} (p_j \alpha_{j,c} -
\beta_{j}^r \alpha_{j,c})\big( \nabla F_{j,c}(\bar{\mathbf{w}}_{r-1}) - \nabla F_{g,c}(\bar{\mathbf{w}}_{r-1}) \big) 
\Big \|^2
\notag \\
& 
+ 2 
\Big \| 
{\sum}_{c=1}^C \Big(\alpha_{g,c} - {\sum}_{j \in \{s, [N]\}} \beta_{j}^r \alpha_{j,c} \Big) \nabla F_{g,c}(\bar{\mathbf{w}}_{r-1}) 
\Big \|^2 
\notag \\
= 
&
2 
\Big \| 
{\sum}_{c=1}^C {\sum}_{j \in \{s, [N]\}} \frac{p_j \alpha_{j,c} - \beta_{j}^r \alpha_{j,c}}{\sqrt{p_j \alpha_{j,c}}} \sqrt{p_j \alpha_{j,c}} 
\big(\nabla F_{j,c} (\bar{\mathbf{w}}_{r-1}) - \nabla F_{g,c}(\bar{\mathbf{w}}_{r-1}) \big) 
\Big \|^2
\notag \\
&
+ 2 
\Big \| 
{\sum}_{c=1}^C \frac{\alpha_{g,c} - {\sum}_{j \in \{s, [N]\}} \beta_{j}^r \alpha_{j,c}}{\sqrt{\alpha_{g,c}}}\sqrt{\alpha_{g,c}}\nabla F_{g,c}(\bar{\mathbf{w}}_{r-1}) 
\Big \| ^2 
\notag \\
\overset{(c)}{\leq} 
& 
2 
\underbrace{{\sum}_{c=1}^C {\sum}_{j \in \{s, [N]\}} \frac{(p_j \alpha_{j,c} - \beta_{j}^r \alpha_{j,c})^2}{p_j \alpha_{j,c}}}_{\rm \triangleq (\ref{eq:bound A11}e)} {\sum}_{c=1}^C {\sum}_{j \in \{s, [N]\}} p_j \alpha_{j,c}
\left \| \nabla F_{j,c}(\bar{\mathbf{w}}_{r-1}) - F_{g,c}(\bar{\mathbf{w}}_{r-1}) \right \|^2 \notag \\
& 
+ 
2 
{\sum}_{c=1}^C \frac{(\alpha_{g,c} - {\sum}_{j \in \{s, [N]\}} \beta_{j}^r \alpha_{j,c})^2}{\alpha_{g,c}} {\sum}_{c=1}^C \alpha_{g,c}
\| \nabla F_{g,c}(\bar{\mathbf{w}}_{r-1}) \|^2 \notag \\
\overset{(d)}{\leq}
& 
2 \chi^2_{\mathbf{p}\|\bm{\beta}} {\sum}_{c=1}^C {\sum}_{j \in \{s, [N]\}} p_j \alpha_{j,c} V_{j,c}^2
+ 
2 \chi^2_{\bm{\alpha}_g \| \bm{\tilde \alpha}^r} G^2
,
\end{align}
\end{small}

\noindent
where equality (a) follows from \eqref{eq:nabla Fsi alpha c}, inequality (b) is due to the global class distribution $\alpha_{g,c} = p_s \alpha_{s,c} + \sum_{i=1}^N p_i \alpha_{i,c}$ together with the inequality $\| {\mathbf{x}}_1 + {\mathbf{x}}_2 \|^2 \leq 2\| {\mathbf{x}}_1 \|^2 + 2\| {\mathbf{x}}_2 \|^2$,
and inequality (c) is obtained via the Cauchy–Schwarz inequality.
In inequality (c), term (\ref{eq:bound A11}e) is further bounded as

\vspace{-0.3cm}
\begin{small}
\begin{align}\label{ref:lemma proof term chi beta p}
{\rm (\ref{eq:bound A11}e)}
=
{\sum}_{c=1}^C {\sum}_{j \in \{s, [N]\}} \frac{(p_j - \beta_j^r)^2 \alpha_{j,c}}{p_j}
=
 \underbrace{{\sum}_{j \in \{s, [N]\}} \frac{(p_j - \beta_j^r)^2}{p_j}}_{\triangleq \chi^2_{\mathbf{p}\|\bm{\beta}}}
\underbrace{{\sum}_{c=1}^C \alpha_{j,c}}_{= 1}
=
\chi^2_{\mathbf{p}\|\bm{\beta}}
.
\end{align}
\end{small}

\noindent
Combining this with Assumptions \ref{assumption:class heterogeneity} and \ref{assumption:garident norm}, as well as the definition of the chi-square divergence $\chi^2_{\bm{\alpha}_g \| \bm{\tilde \alpha}^r}$ in \eqref{eq:convergence bound of FL failure}, yields inequality (d) in \eqref{eq:bound A11}.

Similarly, term (\ref{eq:bound nabla F wr wr-1}d) is bounded by

\vspace{-0.3cm}
\begin{small}
\begin{align}\label{eq:bound A12}
{\rm (\ref{eq:bound nabla F wr wr-1}d)}
\overset{(a)}{\leq} 
{\sum}_{j \in \{s, [N]\}} \beta_{j}^r \left \| \nabla F_{j}(\bar{\mathbf{w}}_{r-1}) - F_{j}({\mathbf{w}}_{j}^{r,t-1}) \right \|^2
\overset{(b)}{\leq} L^2 {\sum}_{j \in \{s, [N]\}} \beta_{j}^r \left \| {\mathbf{w}}_{j}^{r,t-1} - \bar{\mathbf{w}}_{r-1} \right \|^2
,
\end{align}
\end{small}

\noindent
where inequality (a) follows from Jenson's inequality, and inequality (b) is derived under Assumption \ref{assumption:L continuous}.

Finally, substituting the bounds obtained in (\ref{eq:bound A11}) and (\ref{eq:bound A12}) into (\ref{eq:bound nabla F wr wr-1}) establishes Lemma \ref{lemma:bound nabla F wr wr-1}.
\hfill $\blacksquare$

\subsection{Proof of Lemma \ref{lemma:Proof_Thm1_lemma_2}}

According to the local update rules in \eqref{eq:local updating client} and \eqref{eq:local updating server}, the local model at iteration $r$ is updated as
\begin{align}
{\mathbf{w}}_{j}^{r,t-1} = \bar{\mathbf{w}}_{r-1} - \gamma {\sum}_{e=1}^{t-1} \nabla F_j({\mathbf{w}}_{j}^{r,e-1}),
\;
j \in \{s,[N]\}.
\end{align}
Consequently, the deviation between the local and global models is bounded by

\vspace{-0.3cm}
\begin{small}
\begin{align}\label{the third bound}
&
{\sum}_{t=1}^E \big\|{\mathbf{w}}_{j}^{r,t-1} - \bar{\mathbf{w}}_{r-1} \big\|^2
\notag \\
= & 
{\sum}_{t=1}^E \Big\| \gamma {\sum}_{e=1}^{t-1} \nabla F_j({\mathbf{w}}_{j}^{r,e-1}) \Big\|^2 
\notag \\
\overset{(a)} \leq & 
\gamma^2 {\sum}_{t=1}^E (t-1) {\sum}_{e=1}^{t-1} \big \| \nabla F_j({\mathbf{w}}_{j}^{r,e-1}) \big \|^2 
\notag \\
= &
\gamma^2 {\sum}_{t=1}^E (t-1) {\sum}_{e=1}^{t-1} \big \| \nabla F_j({\mathbf{w}}_j^{r,e-1}) - \nabla F_j(\bar{\mathbf{w}}_{r-1}) + \nabla F_j(\bar{\mathbf{w}}_{r-1}) - \nabla F_g(\bar{\mathbf{w}}_{r-1}) + \nabla F_g(\bar{\mathbf{w}}_{r-1}) \big \|^2 
\notag \\
\overset{(b)} \leq &
3 \gamma^2 {\sum}_{t=1}^E (t-1) {\sum}_{e=1}^{t-1} 
\Big( 
\big \| \nabla F_j({\mathbf{w}}_j^{r,e-1}) - \nabla F_j(\bar{\mathbf{w}}_{r-1}) \big \|^2
+ \big \| \nabla F_j(\bar{\mathbf{w}}_{r-1}) - \nabla F_g(\bar{\mathbf{w}}_{r-1}) \big \|^2
+ \big \|\nabla F_g(\bar{\mathbf{w}}_{r-1}) \big \|^2 
\Big) 
\notag \\
\overset{(c)} \leq & 
3 \gamma^2 L^2 \underbrace{{\sum}_{t=1}^E (t-1) {\sum}_{e=1}^{t-1} \big \| {\mathbf{w}}_j^{r,e-1} - \bar{\mathbf{w}}_{r-1} \big \|^2}_{\rm (\ref{the third bound}d)}
+ 6 \gamma^2 \underbrace{{\sum}_{t=1}^E (t-1)^2}_{\rm (\ref{the third bound}e)} {\sum}_{c=1}^C \big( \alpha_{j,c} V_{j,c}^2 + {\chi}_{\bm{\alpha}_g || \bm{\alpha}_j}^2 G^2 \big) \notag \\
& 
+ 3 \gamma^2 \underbrace{{\sum}_{t=1}^E (t-1)^2}_{\rm (\ref{the third bound}e)} \big \| \nabla F_g(\bar{\mathbf{w}}_{r-1}) \big \|^2,
\end{align}
\end{small}

\noindent
where inequality (a) follows from the Cauchy-Schwarz Inequality, inequality (b) results from
$\| {\mathbf{x}}_1 + {\mathbf{x}}_2 + {\mathbf{x}}_3 \|^2 \leq 3\| {\mathbf{x}}_1 \|^2 + 3\| {\mathbf{x}}_2 \|^2 + 3\| {\mathbf{x}}_3 \|^2$, and inequality (c) holds under Assumptions \ref{assumption:L continuous} and Corollary \ref{corollary:data heterogeneity}.

In \eqref{the third bound}, the term (\ref{the third bound}d) is further bounded by

\vspace{-0.3cm}
\begin{small}
\begin{align}\label{the third subbound 1}
{\rm (\ref{the third bound}d)}
\overset{(a)}{=} & 
{\sum}_{e=1}^{E-1} \Big( {\sum}_{t=e+1}^E t-1 \Big) \big \| {\mathbf{w}}_{j}^{r,e-1} - \bar{\mathbf{w}}_{r-1} \big \|^2 
= 
{\sum}_{e=1}^{E-1} \Big( {\sum}_{t=e}^{E-1} t \Big) \big \| {\mathbf{w}}_{j}^{r,e-1} - \bar{\mathbf{w}}_{r-1} \big \|^2 
\notag \\
\overset{(b)}{=} & 
{\sum}_{t=1}^{E-1} \Big( {\sum}_{e=t}^{E-1} e \Big) \big \| {\mathbf{w}}_{j}^{r,t-1} - \bar{\mathbf{w}}_{r-1} \big \|^2 
\notag \\
\overset{(c)}{\leq} & 
{\sum}_{t=1}^{E-1}\frac{E^2}{2} \big \| {\mathbf{w}}_{j}^{r,t-1} - \bar{\mathbf{w}}_{r-1} \big \|^2
,
\end{align}
\end{small}

\noindent
where equality (a) is obtained by interchanging the order of summation, 
equality (b) follows from re-indexing variables $e$ and $t$,
and inequality (c) is derived from ${\sum}_{e=t}^{E-1} e = \frac{(t+E-1)(E-1-t+1)}{2} < \frac{(E+t)(E-t)}{2} = \frac{E^2-t^2}{2} < \frac{E^2}{2}$.
Similarly, term (\ref{the third bound}e) is bounded as

\vspace{-0.3cm}
\begin{small}
\begin{align}\label{the third subbound 2}
{\rm (\ref{the third bound}e)}
= 
{\sum}_{t=1}^E (t-1)^2 = {\sum}_{t=1}^{E-1} t^2 
\overset{(a)}{=}
\frac{(E-1)E(2E-1)}{6} \leq \frac{E^3}{3}
,
\end{align}
\end{small}

\noindent
where equality (a) uses the identity $\sum_{x=1}^{m} x^2 = m(m+1)(2m+1)/6$.

Substituting the bounds from \eqref{the third subbound 1} and \eqref{the third subbound 2} into \eqref{the third bound} gives

\vspace{-0.3cm}
\begin{small}
\begin{align}\label{the third exbound}
&
{\sum}_{t=1}^E \left\|{\mathbf{w}}_{j}^{r,t-1} - \bar{\mathbf{w}}_{r-1} \right\|^2
\notag \\
\leq &
\frac{3}{2}\gamma^2 E^2 L^2 {\sum}_{t=1}^E \left\|{\mathbf{w}}_{j}^{r,t-1} - \bar{\mathbf{w}}_{r-1} \right\|^2
+ 2 \gamma^2 E^3 {\sum}_{c=1}^C \big( \alpha_{j,c} V_{j,c}^2 + {\chi}_{\bm{\alpha}_g || \bm{\alpha}_j}^2 G^2 \big)
+ \gamma^2 E^3 \left\| \nabla F_g(\bar{\mathbf{w}}_{r-1}) \right\|^2
.
\end{align}
\end{small}

Finally, rearranging terms in \eqref{the third exbound} completes the proof of Lemma \ref{lemma:Proof_Thm1_lemma_2}.
\hfill $\blacksquare$

\section{Detailed settings of experiments}

\subsection{Network Standards and Communication Resource Allocation}\label{subsec:Communication Resource Allocation}

Table~\ref{table:network_resource_allocation} summarizes the network standard assigned to each client, along with the corresponding communication resources, including transmit power, bandwidth, and operating frequency band. \cite{ETSI_EN301893_V221_2024, 3GPP_TS_38_101_1_V17_8_0, ieee8023_2018}

\begin{table}[h]
\centering
\caption{Network standard and communication resources allocated to each client.}
\begin{tabular}{c|c|cccc}
\hline
& & \multicolumn{4}{c}{\textbf{Wireless}} \\ \cline{3-6}
\textbf{Standard} & \textbf{Wired} & \textbf{Wi-Fi (2.4\,GHz)} & \textbf{Wi-Fi (5\,GHz)} & \textbf{4G} & \textbf{5G} \rule{0pt}{1.1em} \\
(Index)  & (0) & (1) & (2) & (3) & (4) \\
\hline
\textbf{Client Index} & 1, 2, 3, 4 & 5, 9, 13, 17 & 6, 10, 14, 18 & 7, 11, 15, 19 & 8, 12, 16, 20 \\
\textbf{Transmit Power} & -20\,dBm (0.01\,mW) & 20\,dBm  (100\,mW) & 23\,dBm (200\,mW) & 23\,dBm (200\,mW) & 23\,dBm (200\,mW) \\
\textbf{Bandwidth} & 10\,MHz & 10\,MHz & 10\,MHz & 1.8\,MHz & 2.88\,MHz \\
\textbf{Frequency Band} & Baseband (0) & 2.4\,GHz & 5\,GHz & 1.8\,GHz & 3.5\,GHz \\
\hline
\end{tabular}\label{table:network_resource_allocation}
\end{table}

\subsection{Connection Failure Modeling}\label{subsec:Failure Probability Formulation}

\subsubsection{Transient Failures}\label{subsec:Transient Failure Probability Formulation}

In our setting, wired transmissions are assumed to be stable with negligible loss, whereas wireless transmissions may experience transient failures due to channel fluctuations. 
For each wireless client, the transient failure probability is derived from a classical path-loss model with shadowing effects \cite{wang2025robust, goldsmith2005wireless}.
Assuming \gls{fdma} for uplink transmission, the channel capacity of each client $i \in [N]$ in communication round $r$ is
\begin{equation}\label{channel_capacity}
{C}^{r}_{i} = W_{i} \log_2 \left( 1 + \frac{P_{i} | h^{r}_{i} |^2 }{{W_{i}N_0}} \right) \; \text{bps},
\end{equation}
where $P_i$ and $W_i$ denote the transmit power and allocated bandwidth (from Table~\ref{table:network_resource_allocation}), and $N_0 = -174$\,dBm/Hz is the \gls{psd} of additive noise.
The channel gain $h^{r}_{i}$ follows the log-distance path-loss model with shadowing \cite{goldsmith2005wireless}:
\begin{equation}
[ | h^{r}_{i} |^2 ]_{\rm dB}
= - [{\rm PL_0}(d_0)]_{\rm dB} - \lambda [d_{i}]_{\rm dB} 
+ [\psi_{\rm shadow}]_{\rm dB}
+ [\psi_{\rm wall}]_{\rm dB},
\end{equation}
where $\lambda=3$ is the path-loss exponent, $d_i$ is the distance to the server, and $[x]_{\rm dB}$ denotes the dB representation of $x$.
Shadowing is modeled as $[\psi_{\rm shadow}]_{\rm dB} \sim \mathcal{N}(0, \sigma^2_{\rm dB})$, with $\sigma_{\rm dB}=4$ for \gls{los} and $8$ for \gls{nlos}.
The wall-loss term is set to 12\,dB, 18\,dB, 10\,dB, and 15\,dB per wall under Wi-Fi (2.4\,GHz), Wi-Fi (5\,GHz), 4G, and 5G, respectively. \cite{3GPP_TS_38_101_1_V17_8_0, ieee8023_2018}
The free-space reference loss is
\begin{equation}
[{\rm PL_0}(d_0)]_{\rm dB} = 20\log_{10}(d_i) + 20\log_{10}(f) + 32.44,
\end{equation}
where $d_i$ is in kilometers and $f$ (MHz) is the carrier frequency from Table~\ref{table:network_resource_allocation}. \cite{balanis2016antenna}

According to the channel coding theorem \cite{goldsmith2005wireless}, a transmission outage occurs if the transmission rate $R_i$ exceeds the instantaneous channel capacity $C^{r}_{i}$. Thus, the transient failure probability is
\begin{equation}\label{eq:failure_probability TF}
{\epsilon}^{\rm TF}_{i, r} = {\rm Pr} ( C^{r}_{i} \leq R_{i} ),
\end{equation}
where the transmission rate is given by 
\begin{equation}
R_{i} = \frac{{L}_{i}}{\tau_i},
\end{equation}
with ${L}_{i}$ denoting the uploaded model size and $\tau_i$ the transmission delay, specified in Table \ref{table:transmission_delay} for each dataset and fine-tuning configuration.

\begin{table}[h]
\centering
\caption{Transmission delay for different datasets and fine-tuning configurations.}\label{table:transmission_delay}
\begin{tabular}{c|ccc|ccc}
\hline
\multirow{2}{*}{\textbf{Parameter}} & \multicolumn{3}{c|}{\textbf{Full-Parameter Fine-Tuning}} & \multicolumn{3}{c}{\textbf{Partial-Parameter Fine-Tuning}} \\ 
& MNIST & CIFAR-10 & CIFAR-100 & MNIST & CIFAR-10 & CIFAR-100 \\
\hline
\textbf{Number of Fine-Tuned Parameters} & 215,466 & 269,722 & 11,220,132 & 302,602 & 302,602 & 371,812 \\
\textbf{Model Size} ($L_i$) & 0.86\,M & 1.08\,M & 44.88\,M & 1.21\,M & 1.21\,M & 1.49\,M \\ 
\textbf{Transmission Delay per Round} ($\tau_i$) & 0.8\,s & 1\,s & 45\,s & 1.2\,s & 1.2\,s & 1.4\,s \\
\hline
\end{tabular}
\end{table}

\subsubsection{Intermittent Failures}
Intermittent failures occur randomly and persist for multiple communication rounds.
The occurrence of such failures is modeled by an exponential distribution \cite{correcher2012intermittent}, thus the probability that client $i$ experiences an intermittent failure at round $r$ is given by
\begin{equation}\label{eq:failure_probability IF}
{\epsilon}^{\rm IF}_{i, r} = 1 - e^{-\lambda_i(r-r_0)}, 
\end{equation}
where $r_0$ denotes the most recent recovery round, and $\lambda_i$ is the client-specific failure-rate parameter, as specified in Table \ref{table:lambda_i}.
Once triggered, the disconnection duration is modeled as a uniform random variable over $[1, {100}/{\alpha}]$, where smaller values of $\alpha$ corresponds to longer interruptions.
\begin{table}[h]
\centering
\caption{Intermittent-failure rate of each client.}
\begin{tabular}{c|ccccc}
\hline
\textbf{Client Index} & 1 -- 4 & 5 -- 8 & 9 -- 12 & 13 -- 16 & 17 -- 20 \rule{0pt}{1.1em} \\
\hline
\textbf{Failure Rate} ($\lambda_i$) & $10^{-5}$ & $10^{-4}$ & $10^{-3}$ & $10^{-2}$ & $10^{-1}$ \rule{0pt}{1.1em} \\ 
\hline
\end{tabular}\label{table:lambda_i}
\end{table}

\subsection{\gls{dnn} Architectures for Different Datasets}\label{subsec:DNN architectures}

\subsubsection{MNIST}

The small-scale model follows the CNN architecture summarized in Table~\ref{tab:cnn_gn}, whereas the large-scale model adopts the ViT architecture specified in Table~\ref{tab:vit MNIST}.

\begin{table}[H]
\centering
\caption{CNN architecture for MNIST dateset}
\begin{tabular}{lccccc}
\hline
\textbf{Layer Name} & \textbf{Output Size} & \textbf{Kernel / Stride} & \textbf{Channels} & \textbf{Normalization} & \textbf{Activation} \\
\hline
Conv2d + GroupNorm & 28$\times$28 & 5$\times$5 / 1 & 16 & \gls{gn} (4 groups) & ReLU \\
MaxPool2d & 14$\times$14 & 2$\times$2 / 2 & 16 & -- & -- \\
Conv2d + GroupNorm & 14$\times$14 & 5$\times$5 / 1 & 32 & \gls{gn} (4 groups) & ReLU \\
MaxPool2d & 7$\times$7 & 2$\times$2 / 2 & 32 & -- & -- \\
Flatten & 1568 & -- & -- & -- & -- \\
Linear & 128 & -- & -- & -- & ReLU \\
Linear & 10 & -- & -- & -- & Softmax \\
\hline
\end{tabular}
\label{tab:cnn_gn}
\end{table}

\begin{table}[H]
\centering
\caption{ViT architecture for MNIST dataset}
\begin{tabular}{lccccc}
\hline
\textbf{Layer Name} & \textbf{Output Size} & \textbf{Kernel / Stride} & \textbf{Channels} & \textbf{Normalization} & \textbf{Activation} \\
\hline
Input Image & $224 \times 224$ & -- & 3 & -- & -- \\
Patch Embedding (Conv2d) & $14 \times 14$ (196 tokens) & $16 \times 16$ / 16 & 768 & -- & -- \\
Positional Dropout & 196 tokens & -- & 768 & -- & -- \\
$12 \times$ Transformer Encoder Block (with LoRA) & 196 tokens & -- & 768 & LayerNorm & GELU \\
\quad LayerNorm & 196 tokens & -- & 768 & LayerNorm & -- \\
\quad Multi-Head Self-Attention & 196 tokens & -- & 768 & -- & -- \\
\quad\quad QKV Projection (LoRA, $r{=}8$) & 196 tokens & -- & 2304 & -- & -- \\
\quad Linear Projection & 196 tokens & -- & 768 & -- & -- \\
\quad LayerNorm & 196 tokens & -- & 768 & LayerNorm & -- \\
\quad MLP (FC $\rightarrow$ GELU $\rightarrow$ FC) & 196 tokens & -- & $3072 \rightarrow 768$ & -- & GELU \\
Final LayerNorm & 196 tokens & -- & 768 & LayerNorm & -- \\
Classification Head (Linear) & 10 & -- & -- & -- & Softmax \\
\hline
\end{tabular}
\label{tab:vit MNIST}
\end{table}

\subsubsection{CIFAR-10}

The small-scale model uses a ResNet architecture, as shown in Table~\ref{tab:network-architecture}, whereas the large-scale model adopts the same ViT configuration as Table \ref{tab:vit MNIST}.

\begin{table}[H]
\centering
\caption{ResNet architecture for CIFAR-10 dateset}
\begin{tabular}{lcccccc}
\hline
\textbf{Layer Name} & \textbf{Output Size} & \textbf{Kernel / Stride} & \textbf{Channels} & \textbf{Normalization} & \textbf{Activation} \\
\hline
Conv2d + GroupNorm & 32$\times$32 & 3$\times$3 / 1 & 16 & \gls{gn} (4 groups) & ReLU \\
3 $\times$ BasicBlock & 32$\times$32 & 3$\times$3 / 1 & 16 & \gls{gn} (4 groups) & ReLU \\
3 $\times$ BasicBlock & 16$\times$16 & 3$\times$3 / 2 & 32 & \gls{gn} (8 groups) & ReLU \\
3 $\times$ BasicBlock & 8$\times$8 & 3$\times$3 / 2 & 64 & \gls{gn} (16 groups) & ReLU \\
Global AvgPool & 1$\times$1 & - & 64 & - & - \\
Linear & 10 & - & - & - & Softmax \\
\hline
\end{tabular}
\label{tab:network-architecture}
\end{table}

\subsubsection{CIFAR-100}

The small-scale model employs the ResNet-18 architecture in Table~\ref{tab:resnet18_gn}, while the large-scale model adopts a ViT architecture similar to that in Table~\ref{tab:vit MNIST}, with the output dimension of the final classification head adjusted to 100 to support the 100-class classification task.

\begin{table}[H]
\centering
\caption{ResNet-18 architecture for CIFAR-100 dateset}
\begin{tabular}{lccccc}
\hline
\textbf{Layer Name} & \textbf{Output Size} & \textbf{Kernel / Stride} & \textbf{Channels} & \textbf{Normalization} & \textbf{Activation} \\
\hline
Conv2d + GroupNorm & 32$\times$32 & 3$\times$3 / 1 & 64 & \gls{gn} (32 groups) & ReLU \\
2 $\times$ BasicBlock & 32$\times$32 & 3$\times$3 / 1 & 64 & \gls{gn} (32 groups) & ReLU \\
2 $\times$ BasicBlock & 16$\times$16 & 3$\times$3 / 2 & 128 & \gls{gn} (32 groups) & ReLU \\
2 $\times$ BasicBlock & 8$\times$8 & 3$\times$3 / 2 & 256 & \gls{gn} (32 groups) & ReLU \\
2 $\times$ BasicBlock & 4$\times$4 & 3$\times$3 / 2 & 512 & \gls{gn} (32 groups) & ReLU \\
Global AvgPool & 1$\times$1 & -- & 512 & -- & -- \\
Linear & 100 & -- & -- & -- & Softmax \\
\hline
\end{tabular}
\label{tab:resnet18_gn}
\end{table}

\subsection{Training Hyperparameters for Fine-Tuning Strategies}\label{section:FFT training_paras}

The training hyperparameters used for different fine-tuning strategies across all datasets are summarized in Table~\ref{table:Training Hyperparameters}.

\begin{table}[h]
\centering
\caption{Training hyperparameters for different datasets and fine-tuning configurations.}\label{table:Training Hyperparameters}
\begin{tabular}{c|c|ccc}
\hline
\multirow{2}{*}{\textbf{Fine-Tuning Strategy}} & \multirow{2}{*}{\textbf{Hyperparameter}} & \multicolumn{3}{c}{\textbf{Dataset}} \\
& & \textbf{MNIST} & \textbf{CIFAR-10} & \textbf{CIFAR-100} \\ 
\hline
\multirow{3}{*}{\textbf{Full-Parameter}}
& \textbf{Batch Size} & 128  & 128  & 128 \\
& \textbf{Learning Rate} & 0.05 &  0.1 ($r \leq 4000$); 0.01 ($r > 4000$) &  0.1 ($r \leq 4000$); 0.01 ($r > 4000$)  \\
\hline
\multirow{3}{*}{\textbf{Partial-Parameter (LoRA)}}
& \textbf{Batch Size} & 32 & 32 & 32   \\
& \textbf{Learning Rate} & 0.01 & 0.01 & 0.1  \\
& \textbf{Rank} & 8 & 8 & 8  \\
\hline
\end{tabular}
\end{table}

\subsection{Formulations of Baselines}\label{section:Baselines formulations}

This section summarizes the mathematical formulations of the baselines used in our comparisons, including \gls{fl} schemes with enhanced aggregation or local training, and \gls{fl} schemes incorporating communication-resource optimization.

\begin{itemize}
\item
\textbf{\texttt{FedProx}} \cite{li2020federated}:
This baseline augments each client's local objective with a proximal regularization term.
The local objective becomes
\begin{equation}
F_i({\mathbf{w}}) + \frac{\mu}{2}\| \mathbf{w} - {\bar {\mathbf{w}}}_{r-1}\|^2,
\end{equation}
where $\mu$ is the proximal coefficient.
For fairness, we report the best performance over $\mu  \in \{ 0.1, 0.01, 0.001\}$.

\item
\textbf{\texttt{SCAFFOLD}} \cite{karimireddy2020scaffold}:
This baseline introduces server- and client-side control variates, $\mathbf{c}$ and $\mathbf{c}_i$, to mitigate data heterogeneity.
Each client $i$ updates its local model and control variate by \eqref{eq:SCAFFOLD local model} and \eqref{eq:SCAFFOLD control variate}, respectively.
\begin{subequations}
\begin{equation}\label{eq:SCAFFOLD local model}
\mathbf{w}^{r,t}_{i} 
= 
{\mathbf{w}}^{r,t-1}_{i}  -  \gamma_l ( \nabla F_{i}( {\mathbf{w}}^{r,t-1}_{i} ) - \mathbf{c}_i + \mathbf{c} ), \forall t \in [E], 
\end{equation}
\begin{equation}\label{eq:SCAFFOLD control variate}
\mathbf{c}_i^{+} 
= 
\mathbf{c}_i - \mathbf{c} + \tfrac{1}{K\gamma_l} ({\bar {\mathbf{w}}}_{r-1} - \mathbf{w}^{r,E}_{i}).    
\end{equation}
\end{subequations}
where $\gamma_l$ is the local learning rate.
The server then updates the global model and the server-side control variates via
\begin{subequations}\label{eq:scaffold_global}
\begin{equation}
{\bar {\mathbf{w}}}_{r}
=
{\bar {\mathbf{w}}}_{r-1}
+ \frac{\gamma_g}{{\sum}_{{j} \in {\mathcal{K}}_{r}} \mathds{1}^{r}_{j}} 
{\sum}_{{i} \in {\mathcal{K}}_{r}} \mathds{1}^{r}_{i} ( \mathbf{w}^{r,E}_{i} - {\bar {\mathbf{w}}}_{r-1} )
,
\end{equation}
\begin{equation}\label{eq:scaffold_c_r}
\mathbf{c}^{r}
=
\mathbf{c}^{r-1}
+ \frac{1}{N} {\sum}_{{i} \in {\mathcal{K}}_{r}} \mathds{1}^{r}_{i} ( \mathbf{c}_i^{+} - \mathbf{c}_i )
,
\end{equation}
\end{subequations}

\noindent
where $\gamma_g$ is the global learning rate.
In our experiments, we set the local learning rate $\gamma_l$ = $\gamma$, consistent with the learning rate specified in Table~\ref{table:Training Hyperparameters}, and fix the global learning rate at $\gamma_g = 1$.

\item
\textbf{\texttt{FedLAW}} \cite{li2023revisiting}:
This baseline leverages a proxy dataset at the server to learn adaptive aggregation weights.
Instead of directly averaging local models, the server jointly optimizes a global shrinking factor $\rho$ and client-specific aggregation weights $\beta_i$ by solving

\begin{equation}\label{eq:fedlaw optimization}
\rho^*, \beta_i^* = 
\mathop{\mathrm{arg \, min}}_{\rho > 0, \, \beta_i \ge 0}
\; \mathcal{L}_{\text{proxy}}
\Big( \rho {\sum}_{{i} \in {\mathcal{\hat K}}_{r}} \beta_i \mathbf{w}^{r,E}_{i} \Big),
\end{equation}
where $\mathcal{L}_{\text{proxy}}$ denotes the empirical loss evaluated on the proxy dataset.
The optimized parameters are then used to update the global model as
\begin{equation}
{\bar {\mathbf{w}}}_{r} = \rho^* {\sum}_{{i} \in {\mathcal{\hat K}}_{r}} \beta_i^* \mathbf{w}^{r,E}_{i}.
\end{equation}

\item
\textbf{\texttt{TF-Aggregation}} \cite{salehi2021federated}:
This baseline incorporates the transient connection failure probability into the denominator of the global aggregation.
Specifically, the global aggregation in Step~9 of Algorithm~\ref{algorithm:FL under unreliable networks} is modified as
\begin{equation}\label{eq:aggregation TF}
{\bar {\mathbf{w}}}_{r} = \frac{1}{K} {\sum}_{{i} \in {\mathcal{K}}_{r}} \mathds{1}^{r}_{i} \frac{p_i}{s_i \left(1- \epsilon_{i} \right)} \mathbf{w}^{r,E}_{i}
,
\end{equation}
where $\epsilon_{i}$ denotes the connection failure probability of client $i$, and client selection probabilities $\{s_i\}$ are optimized by
\begin{subequations}\label{eq:optimization problem TF}
\begin{align}
\min_{s_i} \quad & {\sum}_{i=1}^N \;\; \frac{p_i}{s_i (1- \epsilon_{i})}, \\
{\rm s.t.} \quad 
& 
s_i \geq 0, \;\; \forall i \in [N]; \;\; {\sum}_{i=1}^N s_i = 1.
\end{align}
\end{subequations}
To avoid inefficient selection when $\epsilon_i \to 1$, clients with excessively high failure probabilities are excluded via a reliability threshold $\epsilon_{\mathrm{th}}$.
The optimization problem in \eqref{eq:optimization problem TF} is then modified as 
\begin{subequations}
\begin{align}
\min_{s_i} \quad &
{\sum}_{i=1, \epsilon^0_i \leq \epsilon_{\mathrm{th}}}^N \; \frac{p_i}{s_i \left(1- \epsilon^0_i \right)},
\\
{\rm s.t.} \quad  & 
s_i \geq 0 \;\; {\rm if} \;\; \epsilon^0_i \leq \epsilon_{\mathrm{th}};
\;\;
s_i = 0 \;\; {\rm if} \;\; \epsilon^0_i > \epsilon_{\mathrm{th}}, \\
& 
{\sum}_{ i=1, \epsilon^0_i \leq \epsilon_{\mathrm{th}} }^N s_i = 1.
\end{align}
\end{subequations}
where the initial transient failure probability $\epsilon^{0}_{i}$ is computed according to the formulations in Appendix~\ref{subsec:Failure Probability Formulation}, and the threshold is set to $\epsilon_{\mathrm{th}} = 0.9$ in our experiments.

\item
\textbf{\texttt{FedAWE}} \cite{xiang2024efficient}:
This baseline adapts each client’s local update according to its history of successful connections.
After performing the $E$-step local update in \eqref{eq:local updating client}, client $i$ further adjusts its local model $\mathbf{w}^{r,E}_{i}$ as
\begin{equation}\label{eq:fedawe_local}
\mathbf{w}^{r,E}_{i} = \mathbf{w}^{r,E}_{i} - \gamma_g (r - \tau_i) 
({\bar {\mathbf{w}}}_{r-1} - \mathbf{w}^{r,E}_{i})
,
\end{equation}
where $\gamma_g$ denotes the global learning rate (set to 0.001), and $\tau_i$ is the most recent round in which client $i$ successfully connected to the server.
If client $i$ connects in every round, then $\tau_i = r-1$.

\item
\textbf{\texttt{FedEx-LoRA}} \cite{singhal2024fedex}:
This baseline incorporates an error-residual correction induced by the global aggregation of \gls{lora} modules into local updates.
Given a pretrained model $\mathbf{w}_{\rm pre}$ fine-tuned via low-rank adapters $\mathbf{A}$ and $\mathbf{B}$, the server aggregates client adapters at round $r$ following \texttt{FedAvg}:
\begin{equation}\label{eq:fedex_server_lora}
\mathbf{\bar B}^{r} = \frac{1}{{\sum}_{{j} \in {\mathcal{K}}_{r}} \mathds{1}^{r}_{j} } {\sum}_{{i} \in {\mathcal{K}}_{r}} \mathds{1}^{r}_{i} \mathbf{B}_{i}^{r,E}
, \;\; 
\mathbf{\bar A}^{r} = \frac{1}{{\sum}_{{j} \in {\mathcal{K}}_{r}} \mathds{1}^{r}_{j} } {\sum}_{{i} \in {\mathcal{K}}_{r}} \mathds{1}^{r}_{i} \mathbf{A}_{i}^{r,E}
.
\end{equation}
The induced error-residual is computed as
\begin{equation}\label{eq:fedex_server_w}
\Delta \mathbf{w}_{\rm res}^{r}
=
\Big(
\frac{1}{{\sum}_{{j} \in {\mathcal{K}}_{r}} \mathds{1}^{r}_{j} } {\sum}_{{i} \in {\mathcal{K}}_{r}} \mathds{1}^{r}_{i} \mathbf{B}_{i}^{r,E} \mathbf{A}_{i}^{r,E}
\Big)
- \mathbf{\bar B}^{r} \times \mathbf{\bar A}^{r}
.
\end{equation}
Clients update the pretrained model as $\mathbf{w}_{\rm pre}^{r} = \mathbf{w}_{\rm pre}^{r} + \Delta \mathbf{w}_{\rm res}^{r}$, and initialize the next-round adapters as $\mathbf{B}_{i}^{r+1,0} = \mathbf{\bar B}^{r}$ and $\mathbf{A}_{i}^{r+1,0} = \mathbf{\bar A}^{r}$.

\item
\textbf{\texttt{ResourceOpt-1}}: 
Motivated by \cite[Corollary~1]{wang2022quantized}, this baseline seeks to reduce convergence bias by equalizing clients’ transient connection failure probabilities.  
To this end, the optimization problem in \eqref{eq:objective ResourceOpt 1} jointly allocates each client’s transmit power $P_i$ and bandwidth $W_i$ to minimize the variance of their failure probabilities.

\vspace{-0.3cm}
\begin{subequations}\label{eq:objective ResourceOpt 1}
\begin{small}
\begin{align}
\min_{P_i, W_i, i \in [N]}  
\;\;\; & 
\frac{1}{2} \sum_{i \in [N], \; \epsilon^{0}_{i} \leq \epsilon_{\rm th}} \bigg\| \epsilon_i - 
\underbrace{\frac{{\sum}_{i \in [N], \; \epsilon^{0}_{i} \leq \epsilon_{\rm th}} \epsilon_i}{\sum_{i \in [N], \; \epsilon^{0}_{i} \leq \epsilon_{\rm th}} 1}}_{\triangleq \bar{\epsilon}} \bigg\|^2, \label{eq:ResourceOpt 1 obj func}
\\
{\mathrm{s.t.}} \quad
& 
P_i \leq P_{s}^{\max}, \; \forall i \in [N] \;\; {\rm and} \;\; \epsilon^{0}_{i} \leq \epsilon_{\rm th}, \label{eq:ResourceOpt 1 constraint power} \\
& 
{\sum}_{i \in \mathcal{S}_s} W_i \leq W_{s}^{\rm total}, \; \forall i \in [N] \;\; {\rm and} \;\; \epsilon^{0}_{i} \leq \epsilon_{\rm th}. \label{eq:ResourceOpt 1 constraint band} 
\end{align}
\end{small}
\end{subequations}

\noindent
In \eqref{eq:ResourceOpt 1 obj func}, $\epsilon_i$ denotes the transient failure probability of client $i$, and $\bar{\epsilon}$ is the corresponding average across all eligible clients.
We impose $\epsilon^{0}_{i} \leq \epsilon_{\rm th} = 0.9$ to avoid selecting highly unreliable clients, consistent with the \texttt{TF-Aggregation} baseline.
Constraints \eqref{eq:ResourceOpt 1 constraint power}-\eqref{eq:ResourceOpt 1 constraint band} limit each client's transmit power and bandwidth based on the maximum allowable values $P_{s}^{\max}$ and $W_{s}^{\mathrm{total}}$, respectively, where $s \in \{0,1,2,3,4\}$ corresponds to the communication standards in Table~\ref{table:network_resource_allocation} and $\mathcal{S}_s$ denotes the set of clients operating under standard~$s$.
Since wired clients ($s=0$) exhibit negligible failure probability, we refine \eqref{eq:objective ResourceOpt 1} into the modified formulation \eqref{eq:objective ResourceOpt 1-modified}.
This formulation first optimizes wireless clients ($s=1$--$4$) under \eqref{eq:ResourceOpt 1 obj func-modified}-\eqref{eq:ResourceOpt 1 constraint band-modified}, then enforces \eqref{eq:failure prob wired-modified} by aligning wired clients’ failure probabilities to $\bar{\epsilon}$ via random dropping of received models at the server:

\vspace{-0.3cm}
\begin{subequations}\label{eq:objective ResourceOpt 1-modified}
\begin{small}
\begin{align}
\min_{P_i, W_i, \{\mathcal{S}_s\}_{s=1}^{4}}  
\;\;\; & 
\frac{1}{2} \sum_{i \in \{\mathcal{S}_s\}_{s=1}^{4}, \; \epsilon^{0}_{i} \leq \epsilon_{\rm th}} \bigg\| \epsilon_i - 
\underbrace{\frac{{\sum}_{i \in [N], \epsilon^{0}_{i} \leq \epsilon_{\rm th}} \epsilon_i}{\sum_{i \in [N], \epsilon^{0}_{i} \leq \epsilon_{\rm th}} 1}}_{\triangleq \bar{\epsilon}} \bigg\|^2, \label{eq:ResourceOpt 1 obj func-modified}
\\
{\rm s.t.} \quad
& 
P_i \leq P_{s}^{\max}, \; \forall i \in \{ \mathcal{S}_s \}_{s=1}^{4} \;\; {\rm and} \;\; \epsilon^{0}_{i} \leq \epsilon_{\rm th}, \label{eq:ResourceOpt 1 constraint power-modified} \\
& 
{\sum}_{i \in \mathcal{S}_s} W_i \leq W_{s}^{\rm total}, \; \forall i \in \{ \mathcal{S}_s \}_{s=1}^{4} \;\; {\rm and} \;\; \epsilon^{0}_{i} \leq \epsilon_{\rm th}, \label{eq:ResourceOpt 1 constraint band-modified} \\
& 
\epsilon_i = \bar{\epsilon}, \; \forall i \in \mathcal{S}_0. 
\label{eq:failure prob wired-modified}
\end{align}
\end{small}
\end{subequations}

\noindent
Given the smooth and continuous gradients of \eqref{eq:ResourceOpt 1 obj func-modified} due to the squared terms, we apply gradient descent to optimize the transmit power $P_i$ and bandwidth $W_i$ for clients satisfying $\epsilon^{0}_{i} \leq \epsilon_{\rm th}$.

\item
\textbf{\texttt{ResourceOpt-2}}: 
Because joint resource optimization across heterogeneous communication standards is often impractical in real-world deployments, this baseline performs standard-wise optimization.
For each standard $s \in \{0,1,2,3,4\}$, the following problem is solved independently:

\vspace{-0.3cm}
\begin{subequations}\label{eq:objective ResourceOpt 2}
\begin{small}
\begin{align}
\min_{P_i, W_i, \atop i \in \mathcal{S}_s} \quad & \frac{1}{2} \sum_{i \in \mathcal{S}_s, \, \epsilon^{0}_{i} \leq \epsilon_{\rm th}} \bigg\| \epsilon_i - \frac{{\sum}_{i \in \mathcal{S}_s, \epsilon^{0}_{i} \leq \epsilon_{\rm th}} \epsilon_i}{\sum_{i \in \mathcal{S}_s, \epsilon^{0}_{i} \leq \epsilon_{\rm th}} 1} \bigg\|^2, \label{eq:ResourceOpt 2 obj func}
\\
{\rm s.t.} \quad
& P_i \leq P_{s}^{\max}, \; \forall i \in \mathcal{S}_s \;\; {\rm and} \;\; \epsilon^{0}_{i} \leq \epsilon_{\rm th}, \label{eq:ResourceOpt 2 constraint power} \\
& {\sum}_{i \in \mathcal{S}_s} W_i \leq W_{s}^{\rm total}, \; \forall i \in \mathcal{S}_s \;\; {\rm and} \;\; \epsilon^{0}_{i} \leq \epsilon_{\rm th}. \label{eq:ResourceOpt 2 constraint band}
\end{align}
\end{small}
\end{subequations}

\end{itemize}

\subsection{Formulations for Ablation Studies}\label{appendix:Formulations Ablation Studies}

\subsubsection{Without Server-Side Compensation Training}

When the server-side compensatory training module is disabled, the global aggregation rule in \eqref{global_model_failure_2} reduces to \eqref{global_model_failure}.
Consequently, the aggregation weight optimization problem \eqref{eq:aggregation_optimization_2} simplifies to

\vspace{-0.3cm}
\begin{small}
\begin{subequations}\label{eq:aggregation_optimization_2_appendix}
\begin{align}
\min_{\beta_s^r, \{ \beta^{r}_{i} \}_{ \mathds{1}^{r}_{i} = 1 } } 
&
{\small
\sum_{c=1}^C \frac{
\Big( 
\alpha_{g,c}  - 
\big(
\beta_s^r \alpha_{s,c}  +  {\sum}_{{i} \in {\mathcal{K}}_{r}}  \mathds{1}^{r}_{i} \beta^{r}_{i} \alpha_{i,c}
\big)
\Big)^2}{\alpha_{g,c}}
}, 
\label{eq:aggregation_optimization_2 objective appendix}
\\
\text{s.t.}
\quad
&
\beta_s^r + {\sum}_{{i} \in {\mathcal{K}}_{r}} \mathds{1}^{r}_{i} \beta^{r}_{i} = 1.
\label{eq:aggregation_optimization_2 constraint appendix}
\end{align}
\end{subequations}
\end{small}

\noindent
Analogous to \eqref{eq:aggregation_optimization_2}, the formulation in \eqref{eq:aggregation_optimization_2_appendix} remains a convex weighted least squares problem and can be easily solved by standard quadratic optimization solvers.
Following the procedure in Section~\ref{subsection:Module 2}, the server-side aggregation weight continues to be fixed as \eqref{eq:server_aggregation_weight}.

\subsubsection{Without Aggregation Weight Optimization}

When aggregation weight optimization is removed, the global aggregation rule in \eqref{global_model_failure_2} degenerates into a simple averaging scheme.
In this case, while the server weight remains fixed as \eqref{eq:server_aggregation_weight}, all remaining aggregation weights, including $\beta^r_{\text{miss}}$ and $\beta^{r}_{i}$, are assigned by simple averaging:
\begin{align}
\begin{cases}
\beta^r_{\text{miss}} = 0, 
\quad
\beta^{r}_{i} = \frac{1}{1 + {\sum}_{{j} \in {\mathcal{K}}_{r}} \mathds{1}^{r}_{j}}, 
\quad
& \text{if} \quad \mathcal{C}^{r}_{\text{miss}} = \emptyset,
\\
\beta^r_{\text{miss}} = \beta^{r}_{i} = \frac{{\sum}_{{j} \in {\mathcal{K}}_{r}} \mathds{1}^{r}_{j}}{(1 + {\sum}_{{j} \in {\mathcal{K}}_{r}} \mathds{1}^{r}_{j})^2},
\quad
& \text{if} \quad \mathcal{C}^{r}_{\text{miss}} \neq \emptyset.
\end{cases}
\end{align}
Under these assignments, all aggregation weights continue to satisfy the normalization constraint in \eqref{eq:aggregation_optimization_2 constraint}.

\section{Supplementary Experimental Results}

\subsection{Convergence Trends under Transient and Intermittent Failures}\label{section:supplementary transient intermittent}

\begin{figure}[h]
\begin{minipage}[h]{1\linewidth}
\centering
\includegraphics[width= 3.5 in ]{8legend.pdf}
\end{minipage}
\begin{minipage}[h]{1\linewidth}
\centering
\subfigure[MNIST, i.i.d.]{
\includegraphics[width= 1.6 in ]{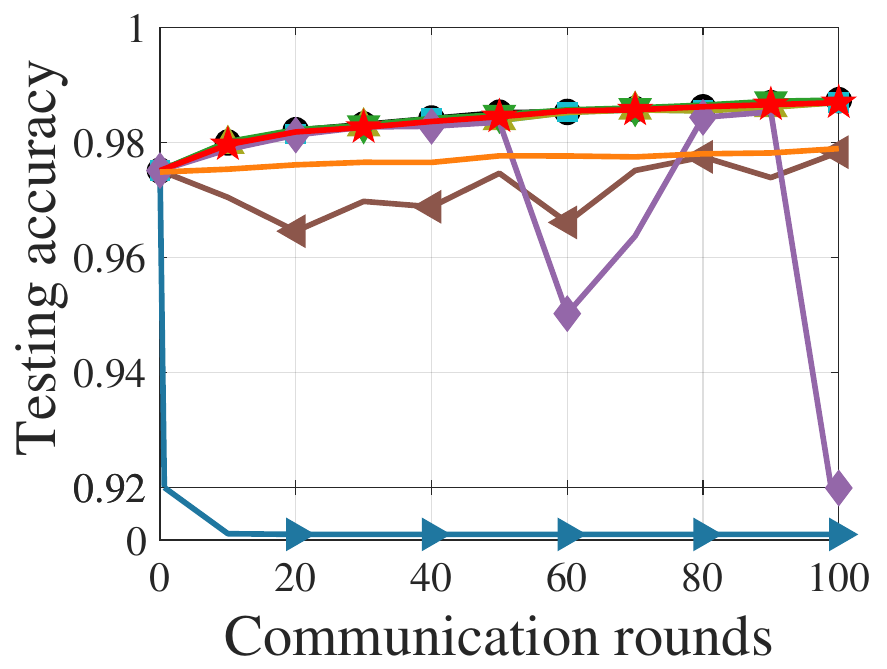}\label{fig:mnist transient iid}}
\subfigure[CIFAR-10, i.i.d.]{
\includegraphics[width= 1.6 in ]{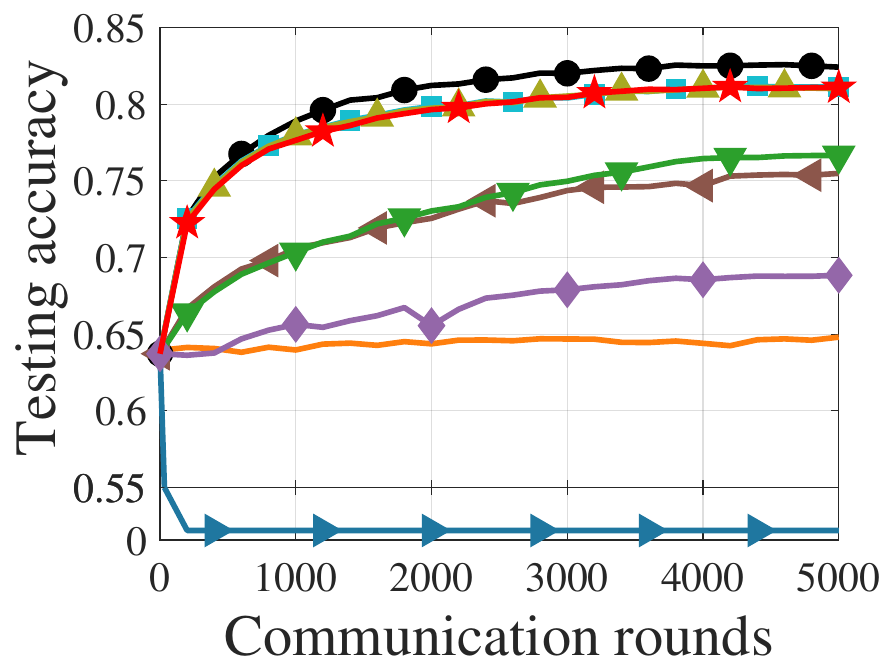}\label{fig:cifar10 transient iid}}
\subfigure[CIFAR-100, i.i.d.]{
\includegraphics[width= 1.6 in ]{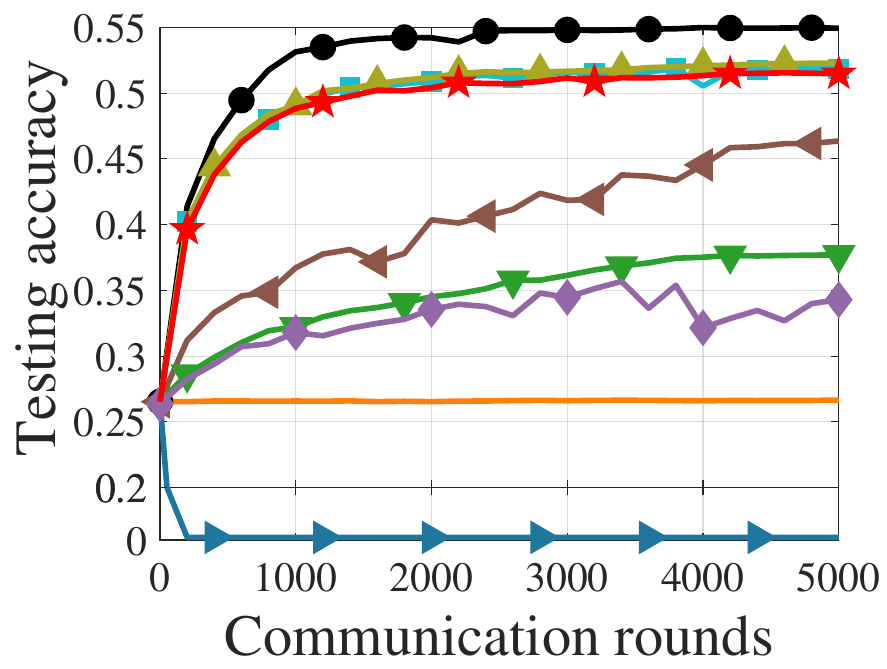}\label{fig:cifar100 transient iid}}

\subfigure[MNIST, non-i.i.d.]{
\includegraphics[width= 1.6 in ]{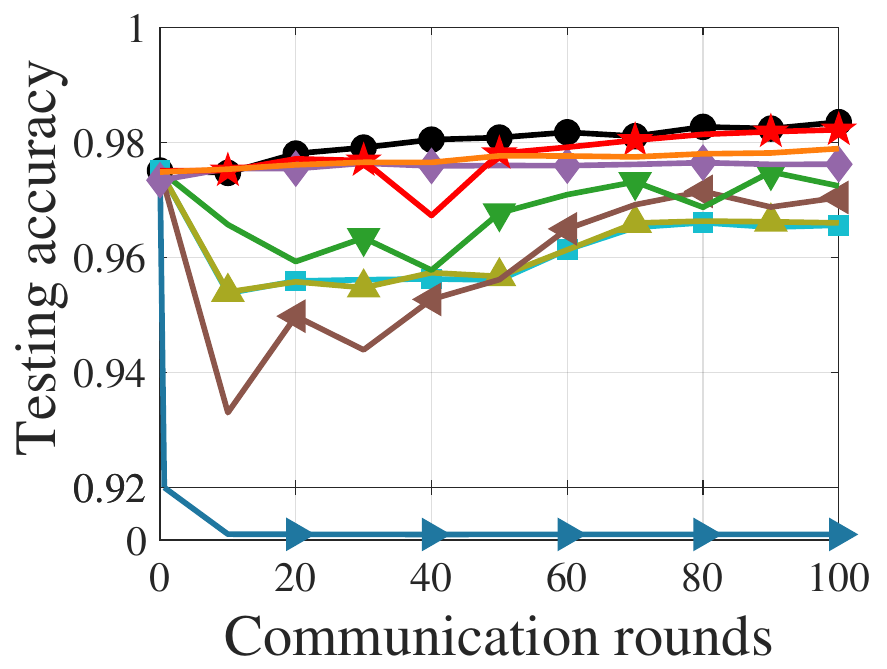}\label{fig:mnist transient noniid}}
\subfigure[CIFAR-10, non-i.i.d.]{
\includegraphics[width= 1.6 in ]{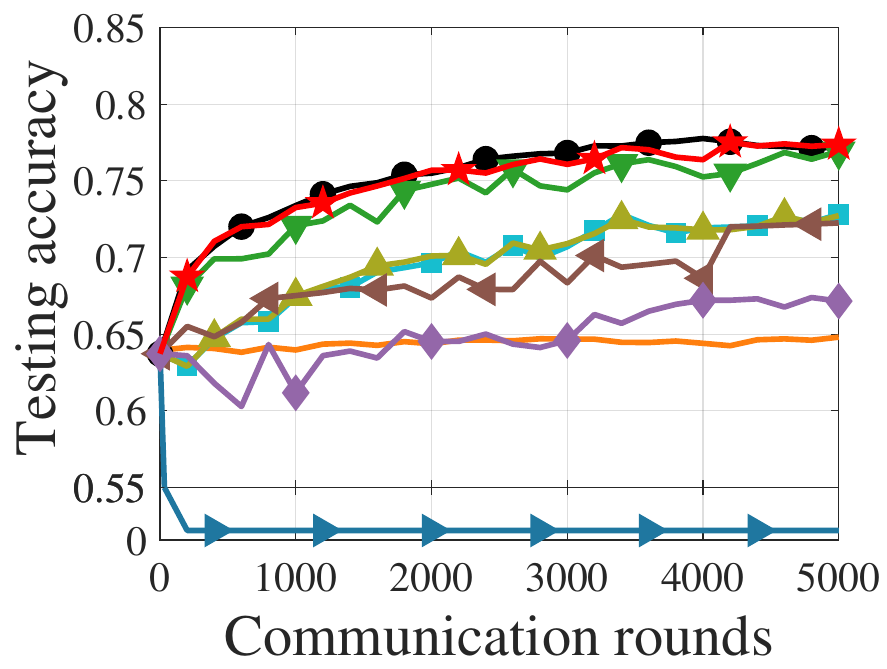}\label{fig:cifar10 transient noniid}}
\subfigure[CIFAR-100, non-i.i.d.]{
\includegraphics[width= 1.6 in ]{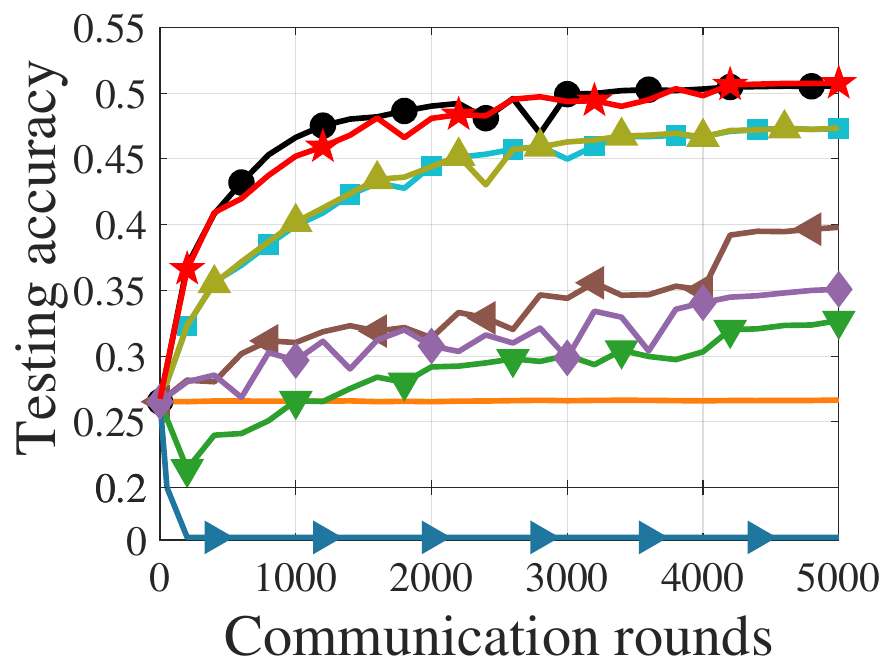}\label{fig:cifar100 transient noniid}}
\caption{
Convergence trends of different \gls{fft} strategies under transient failures ($K=20$, full-parameter fine-tuning).}
\label{fig:performance diff dataset transient full-parameter}
\end{minipage}
\end{figure}

\begin{figure}[h]
\begin{minipage}[h]{1\linewidth}
\centering
\includegraphics[width= 3.5 in ]{8legend.pdf}
\end{minipage}
\begin{minipage}[h]{1\linewidth}
\centering
\subfigure[MNIST, i.i.d.]{
\includegraphics[width= 1.6 in ]{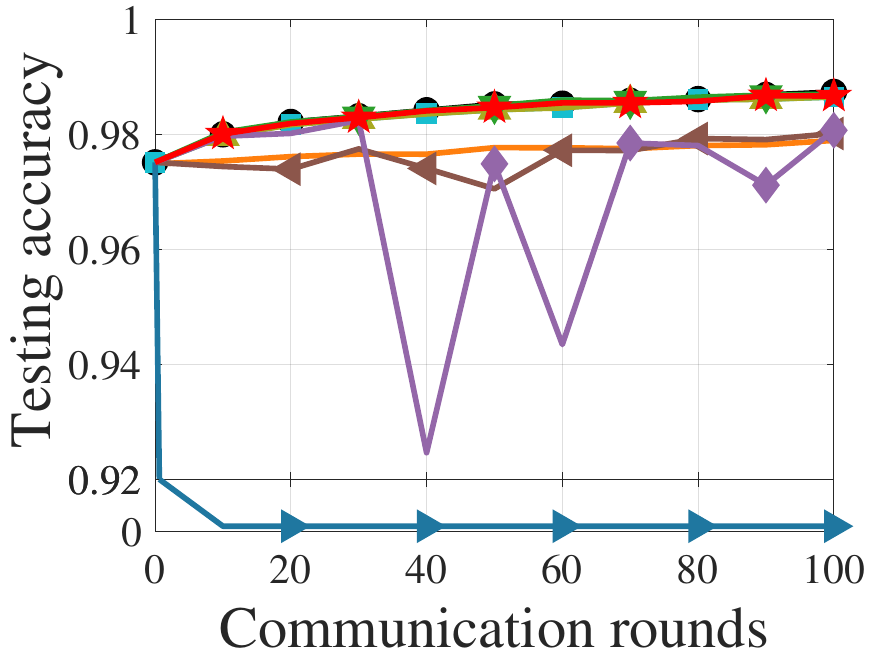}\label{fig:mnist intermittent iid}}
\subfigure[CIFAR-10, i.i.d.]{
\includegraphics[width= 1.6 in ]{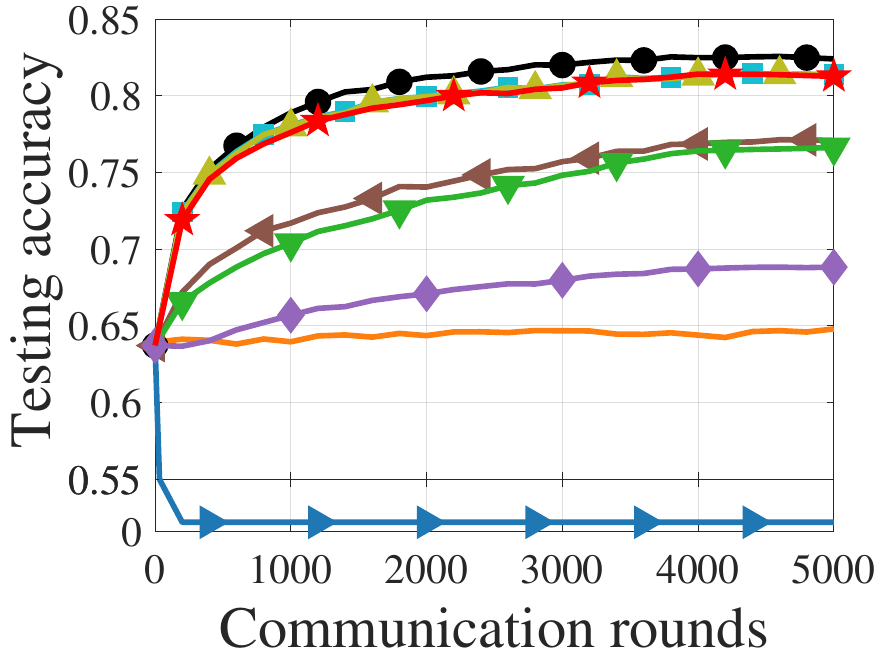}\label{fig:cifar10 intermittent iid}}
\subfigure[CIFAR-100, i.i.d.]{
\includegraphics[width= 1.6 in ]{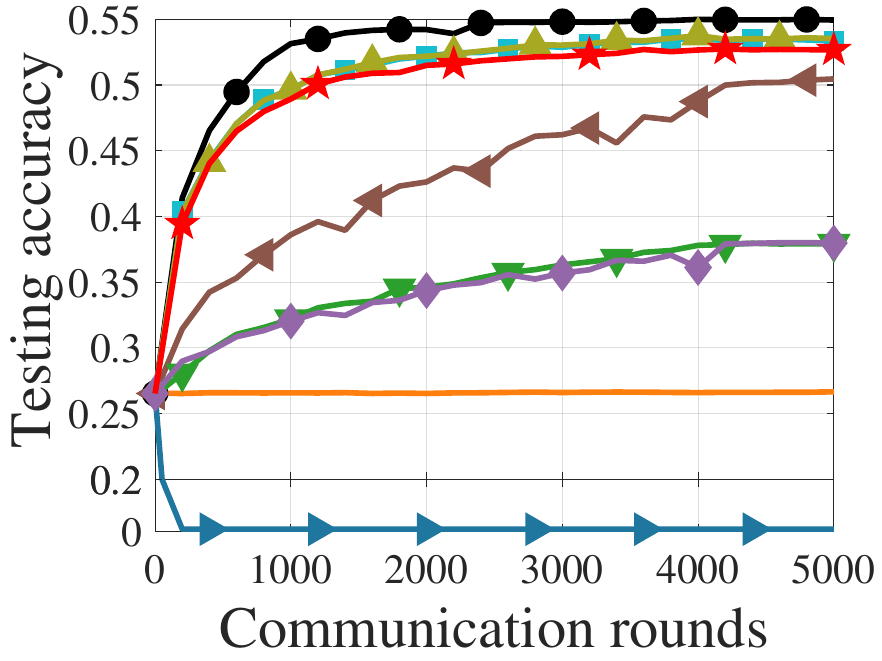}\label{fig:cifar100 intermittent iid}}

\subfigure[MNIST, non-i.i.d.]{
\includegraphics[width= 1.6 in ]{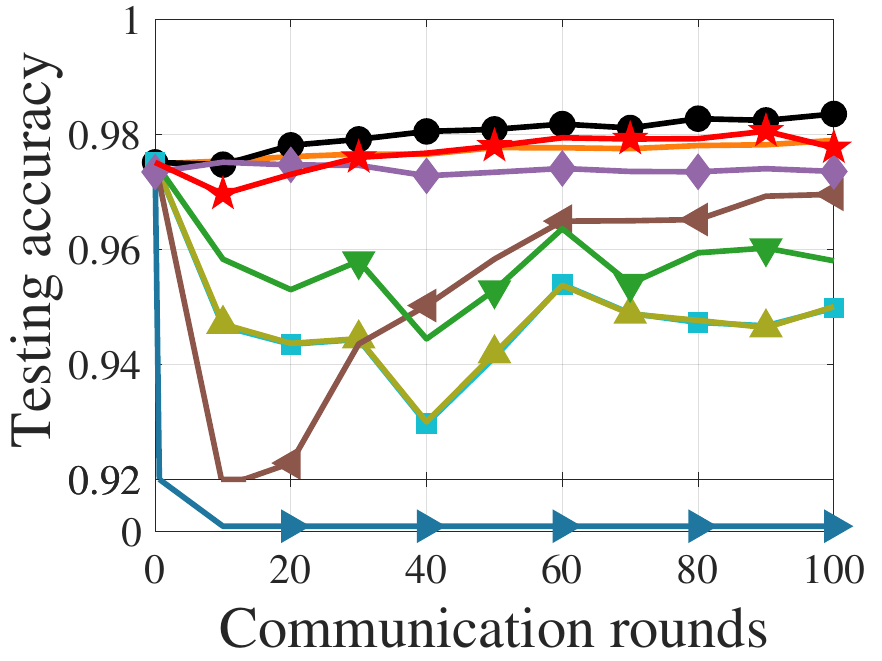}\label{fig:mnist intermittent noniid}}
\subfigure[CIFAR-10, non-i.i.d.]{
\includegraphics[width= 1.6 in ]{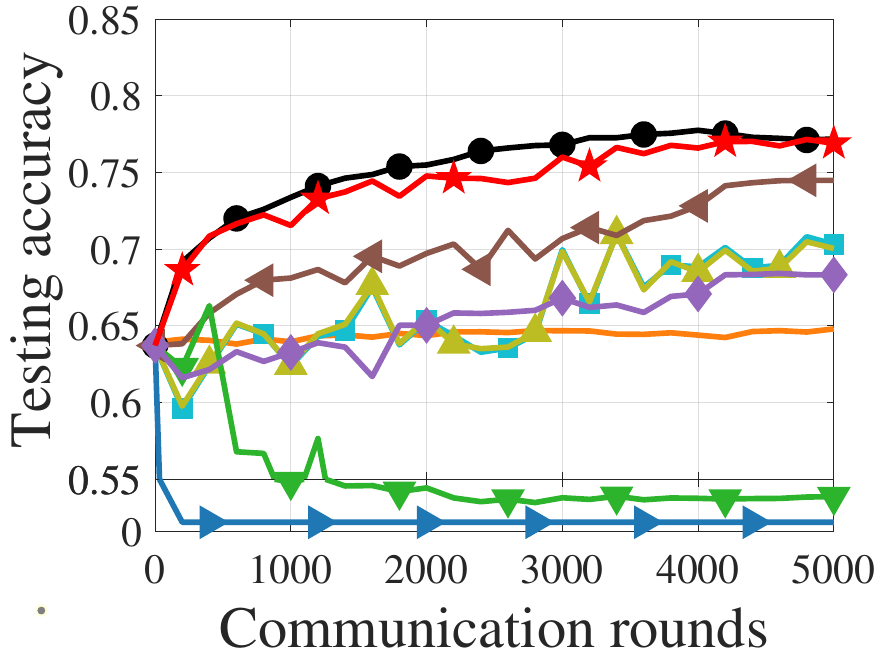}\label{fig:cifar10 intermittent noniid}}
\subfigure[CIFAR-100, non-i.i.d.]{
\includegraphics[width= 1.6 in ]{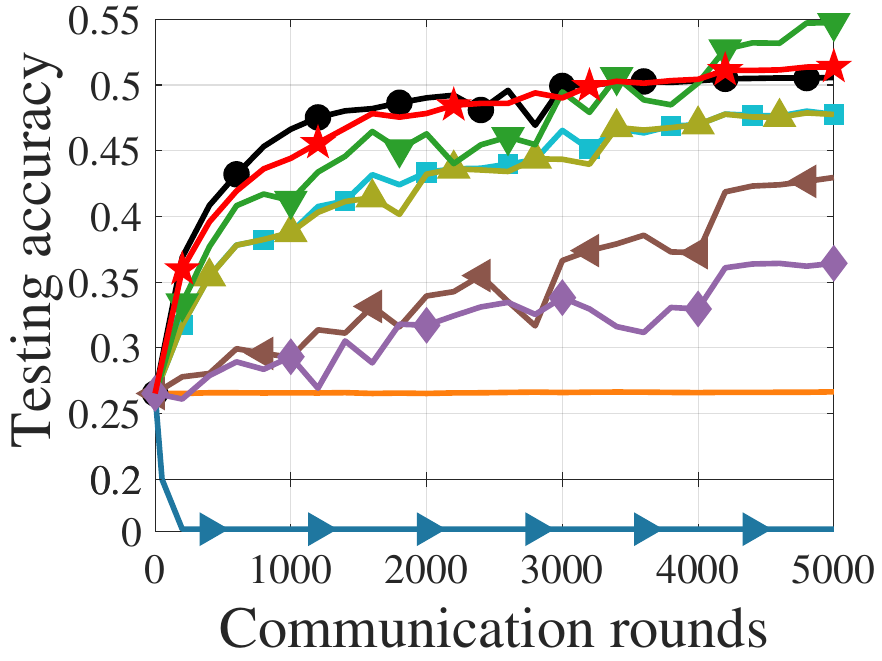}\label{fig:cifar100 intermittent noniid}}
\caption{
Convergence trends of different \gls{fft} strategies under intermittent failures ($K=20$, full-parameter fine-tuning).}
\label{fig:performance diff dataset intermittent full-parameter}
\end{minipage}
\end{figure}

Fig.~\ref{fig:performance diff dataset transient full-parameter} illustrates the convergence trends of different \gls{fft} strategies under transient connection failures across MNIST, CIFAR-10, and CIFAR-100 for both i.i.d.\ and non-i.i.d.\ data settings.
Fig.~\ref{fig:performance diff dataset intermittent full-parameter} presents the corresponding results under intermittent failures.

\subsection{Convergence Trends under Partial Participation}\label{fig:supp performance partial participation}

\begin{figure}[h]
\begin{minipage}[h]{1\linewidth}
\centering
\includegraphics[width= 3.5 in ]{8legend.pdf}
\end{minipage}
\begin{minipage}[h]{1\linewidth}
\centering
\subfigure[MNIST, non-i.i.d.]{
\includegraphics[width= 1.6 in ]{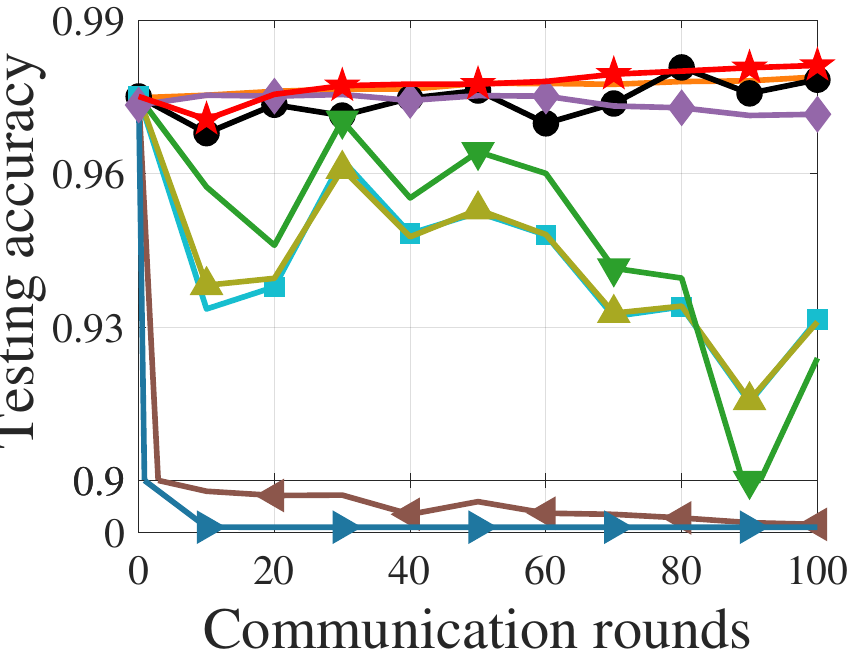}\label{fig:mnist mixed noniid partial}}
\subfigure[CIFAR-10, non-i.i.d.]{
\includegraphics[width= 1.6 in ]{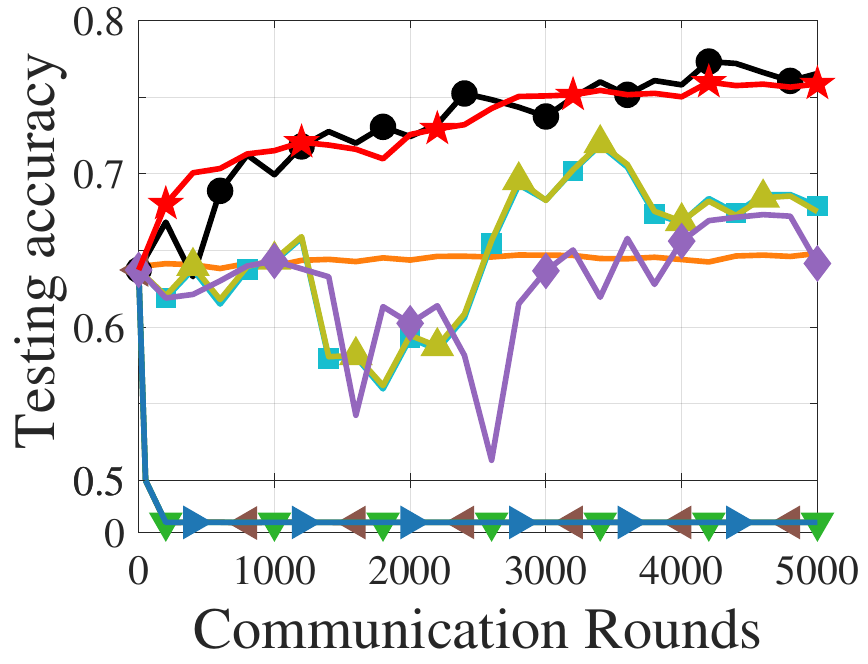}\label{fig:cifar10 mixed noniid partial}}
\subfigure[CIFAR-100, non-i.i.d.]{
\includegraphics[width= 1.6 in ]{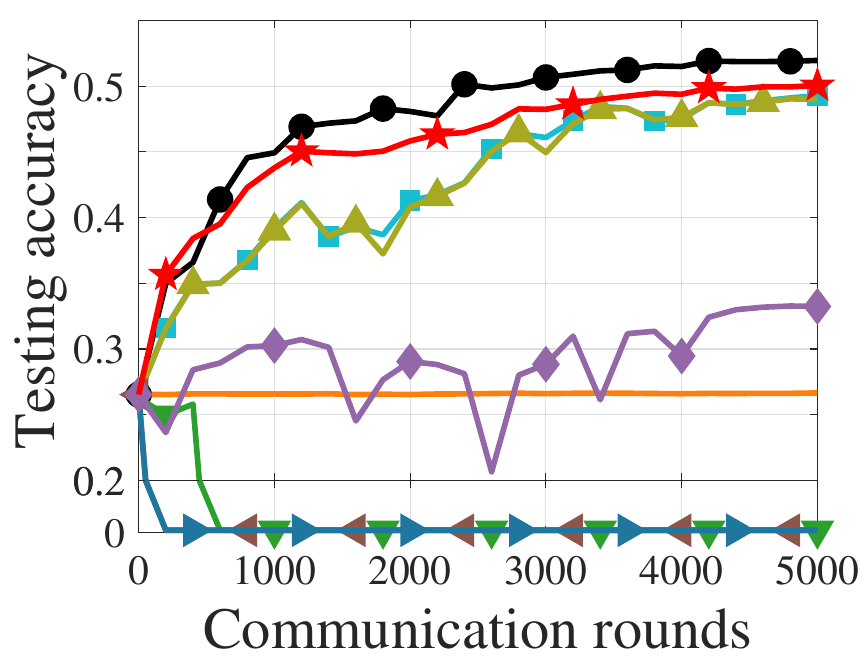}\label{fig:cifar100 mixed noniid partial}}
\caption{
Convergence of different \gls{fft} strategies under full-parameter fine-tuning with partial participation ($K=10$, mixed failures).
}
\label{fig:performance diff dataset mixed noniid partial}
\end{minipage}
\end{figure}

Fig. \ref{fig:performance diff dataset mixed noniid partial} compares the convergence trends of different \gls{fft} strategies under partial participation.

\end{document}